\documentstyle{espart}
\begin{document}
\def \G {\Gamma}
\def\half{{1\over 2}}
\def \km {{~\rm km~ s}^{-1}{\rm Mpc}^{-1}}
\def \d {\delta}
\def \e {\eta }
\def \l {\Lambda}
\def \o {\Omega}
\def \ee {({\eta\over \eta _0})}
\def \g {\Gamma}
\def \n {\nu}
\def \nn {\noindent}
\def \m {\mu}
\def \dm {{\Delta M\over M}}
\def \der {({\delta\rho\over \rho})_H}
\def \dr {{\delta\rho/ \rho}}
\def \dt {{\delta T/ T}}
\def \tik {\langle T_{ik}\rangle}
\def \ie {{\em i.e.~~}}
\def \lleq {\lower0.9ex\hbox{ $\buildrel < \over \sim$} ~}
\def \ggeq {\lower0.9ex\hbox{ $\buildrel > \over \sim$} ~} 

\def\tablerule{\noalign {\vskip3truept\hrule\vskip3truept}}

\def \jetp {Sov. Phys. JETP, }
\def \jetpl {JETP Lett., }
\def \pismajetp {Pisma v Zh. Eksp. Teor. Fiz.,}
\def \sal {Sov. Astron. Lett., }
\def \ap {ApJ, }
\def \apl {ApJ, }
\def \aps {ApJS, }
\def \pd {Phys. Rev. D, }
\def \prl {Phys. Rev. Lett., }
\def \pl {Phys. Lett., }
\def \np {Nucl. Phys., }
\def \mnras {MNRAS, }
\def \aa {A\&A, }
\def \cqg {Class. Quantum Grav.}

\def\beq{\begin{equation}}
\def\eeq{\end{equation}}
\def\ber{\begin{eqnarray}}
\def\eer{\end{eqnarray}}

\newcommand{\sq}{\lower.25ex\hbox{\large$\Box$}}

\begin{frontmatter}
\title{The Case for a Positive Cosmological $\l$-term}
\author{Varun Sahni}
\address{IUCAA -- Inter--University 
Centre for Astronomy \& Astrophysics,
\\ Post Bag 4,  Ganeshkind, Pune 411007, India}
\and
\author{Alexei Starobinsky}
\address{Landau Institute for Theoretical Physics, Kosygina 2, 
Moscow 117334, Russia}

\begin{abstract}
Recent observations of Type 1a supernovae indicating an
accelerating universe have once more drawn attention to the 
possible existence, at the present epoch, of a small positive 
$\l$-term (cosmological constant). In this paper we 
review both observational and theoretical aspects of a small cosmological
$\l$-term. We discuss the current observational situation
focusing on cosmological tests of $\l$ including the age of the universe,
high redshift supernovae,
gravitational lensing, galaxy clustering and the
cosmic microwave background.
We also review the theoretical debate surrounding $\l$: 
the generation of $\l$ in models with spontaneous symmetry breaking and
through quantum vacuum polarization effects -- mechanisms 
which are known to give rise to a 
large value of $\l$ hence leading to the `cosmological constant problem'. 
More recent attempts to generate a small
cosmological constant at the present epoch using either field theoretic
techniques, or by modeling a dynamical $\l$-term
by scalar fields are also extensively discussed. Anthropic
arguments favouring a small $\l$-term are briefly reviewed.
A comprehensive bibliography of recent work 
on $\l$ is provided.
\end{abstract}
\end{frontmatter}
\vfill\eject

\tableofcontents
%\newpage
\section{Introduction}
\label{sec:lambda}

Recent years have witnessed a resurgence of interest in the possibility
that a positive $\l$-term (a cosmological constant) may dominate the total 
energy density in the universe. Interest in the cosmological constant stems 
from several directions:

(i) Observations of high redshift Type Ia supernovae appear to suggest 
that our universe may be accelerating with a large fraction of the 
cosmological density in the form of a cosmological $\l$-term. 
\footnote{For a different interpretation of the supernovae data
%in the context of 
%an inhomogenoeus universe, varying speed of light cosmology
%or varying gravitational constant $G$ can be found in
see \cite{barrow99,celerier99,garcia99}.}
When combined 
with observations of the cosmic microwave background (CMB), an approximately
flat 
Friedmann-Robertson-Walker (FRW) cosmological model with 
total energy density ($\o_m + \o_\l \simeq 1$) is suggested, in agreement
with predictions of the simplest versions of the inflationary scenario
of the early universe (sections \ref{sec:sn}, \ref{sec:mbr}).

(ii) Most dynamical estimates of the amount of clustered matter yield a 
conservative upper limit $\o_m \lleq 0.3$. In addition, theoretical modelling 
of structure formation based on the cold dark matter model (CDM) with 
$\o_m = 1$ has failed to match up with observations at a quantitative
level. By contrast, a flat low density CDM$+\l$ universe with $\o_m 
\simeq 0.3$ and $\o_\l \simeq 0.7$, and with an approximately flat 
(or, Harrison-Zeldovich-like, $n_S\approx 1$) initial Fourier spectrum of 
scalar (adiabatic) inhomogeneous metric and density perturbations agrees 
remarkably well with a wide range of observational
data ranging from large and intermediate angle CMB anisotropies to
observations of galaxy clustering on large scales. Since an 
approximately flat initial spectrum of adiabatic perturbations is also
precisely what simplest variants of the inflationary scenario predict,
the positive $\l$-term removes a necessity in any complications of
the inflationary scenario (which might be required if the 
universe was found to be open).

(iii) At a theoretical level, a cosmological constant $\l = 8\pi G\rho_{vac}
/c^2$ is predicted to arise out of zero-point quantum vacuum fluctuations
of fundamental scalar, spinor, vector and tensor fields (see section 
\ref{sec:lambda2}). 
Although a theoretically predicted value of $\rho_{vac}$ usually appears to be
much larger than current observational
limits, there is no generic known mechanism which
will set the value of $\l$ to precisely zero either on the basis of 
symmetry arguments
or by dynamical means. \footnote{The value of $\l$ can, of course, 
be set to zero by hand by adding
suitable counterterms to the bare (infinite) value of $\l$ in the Lagrangian. 
This method, however, amounts to a rather ad-hoc adjustment of parameters and 
cannot be regarded as being `generic' (see section \ref{sec:lambda2}).}
Some recent attempts to generate a small $\l$ at the present 
epoch either through vacuum polarization and particle creation effects
or by means of dynamically evolving scalar fields 
are discussed in section \ref{sec:small}.

Although none of the above arguments can by themselves be regarded 
as conclusive evidence
for a cosmological constant, the growing body of work on the subject,
combined with a possible deep relationship between a small cosmological
constant today and a large cosmological term 
driving inflation at an early epoch, suggests that the case for a positive
cosmological constant be taken seriously. In this paper we attempt to review
some aspects of the cosmological constant issue emphasizing both 
theoretical as well as observational aspects.
For earlier reviews on the subject the reader is referred to
Zeldovich (1968), Weinberg (1989)
and Caroll, Press and Turner (1992).

From the physical point of view, a $\l$-term represents a new
type of dark non-baryonic matter, completely unknown from laboratory
experiments.
Its difference from another type of dark non-baryonic
matter that has been already introduced in cosmology for almost two decades
from observations of gravitational clustering is essentially that matter
described by the $\l$-term is, (a) not gravitationally clustered at all
scales at which we see clustering of baryons and dustlike dark matter, and
(b) has a strongly negative effective pressure ($P < 0,~|P| \sim \rho c^2$).
Thus, remarkably, by investigating the behaviour of the present universe
we are studying novel fundamental physics. Extragalactic astronomy and
cosmology once more become a driving force for new insights in physics ! 

\section{The Cosmological Constant revisited}
\label{sec:history}

In 1917, only a few years after introducing the field equations of
the General theory of relativity (GR), Einstein proposed adding a 
`cosmological constant' to these equations which were modified to
\beq
R_{ik} - \half~g_{ik}R - \Lambda g_{ik} = \frac{8\pi G}{c^4} T_{ik}.
\label{eq:lam1}
\eeq
The main motivation behind introducing the cosmological constant appears 
to have
been Einstein's belief that the equations of General Relativity should be 
compatible with
Mach's principle. 
Einstein was fascinated by the arguments of philosopher/scientist
Ernst Mach. Mach was concerned about the notion of absolute motion 
which prevailed in Newtonian mechanics.
He postulated that all the matter in the universe including the
distant stars provided
a `background' against which
motion could be measured and that unless there was a material
background which  served as a reference frame, it was meaningless to talk of
rest or motion in any absolute sense (Mach 1893).
Einstein  proposed incorporating Mach's principle
into the general theory of relativity by suggesting a solution of the 
equations (\ref{eq:lam1}) in which the universe
was static and closed on itself, much like the closed two dimensional surface 
of a balloon. A static solution of (\ref{eq:lam1}) is possible to construct
since, as shown
in section \ref{sec:static}, 
a positive cosmological constant introduces a repulsive force which
can counterbalance the attractive force of gravity leading to 
the `static Einstein universe'. This universe
has a finite spatial volume with no boundaries, furthermore
the total mass in such a
universe is related directly to its (finite) volume (section 
\ref{sec:closed}).
A low mass universe has a small volume, and an empty universe has no
volume at all !
%As the mass in the universe is decreased its spatial volume shrinks
%till, in the limiting case an empty universe possesses no volume.
The static Einstein universe thus incorporates Mach's principle 
since it demonstrates that
without matter there
can be no space against which background inertial effects can be measured.
%(since there is no universe) and Mach's principle is satisfied !!

It should be borne in mind that in 1917 the idea of the Milky Way being an
island universe was widely believed in, and the notion of the existence of
other
galaxies had not yet been firmly established. 
All this was about to change however, when
in the early 1920s Slipher's work showed that light from several
spiral nebulae (later re-christened galaxies) was redshifted, 
a fact that could be
explained by the Doppler effect if these nebulae/galaxies were moving away from 
us.\footnote{It is interesting that the same year that 
Einstein introduced the cosmological term
$\Lambda$, de Sitter presented solutions of (\ref{eq:lam1}) with $T_{ik} = 0$,
$\Lambda > 0$, which had both static and
dynamic features.
%book(see Sec. \ref{sec:desit} for a discussion of the de Sitter metric).
Intriguingly,
although the space-time coordinatization originally introduced by de Sitter
was static \cite{desit}, namely
$ds^2 = \cosh^{-2}{Hr}\lbrack dt^2 - dr^2 - H^{-2} \tanh^2{Hr}
(d\theta^2 + \sin^2\theta d\phi^2\rbrack$,
it allowed for a linear redshift-distance relation, 
since $\Gamma^r_{tt} \neq 0$ in the above metric 
%book (\ref{eq:desori}), 
resulting in the motion
of test bodies by virtue of the geodesic equation
$\frac{d^2x^i}{ds^2} + \Gamma^{i}_{kl}\frac{dx^k}{ds}\frac{dx^l}{ds} = 0$
($\Gamma^{i}_{kl}$ is the affine connection). 
This effect was pointed out by Weyl (1923)
and later used by Eddington to interpret Slipher's observations 
in the context of de Sitter's static universe \cite{wein89}.}
In 1922, about five years after Einstein had proposed his static solution, 
%several years before Hubble's observational results 
%were published, 
Aleksander Friedmann constructed a matter
dominated expanding universe without a cosmological constant.
%The conclusive discovery by Hubble (1929) of a linear expansion law relating
%redshift to distance in an expanding universe led Einstein
The possibility that the universe may be expanding led Einstein
to abandon the idea of a static universe and, along with it, the cosmological
constant. 
%(Einstein would later call $\Lambda$ his `biggest blunder'.)
In a 1923 letter to Weyl,
Einstein is quoted as saying \cite{pais82}
`` If there is no quasi-static
 world, then away with the cosmological term !'' 
%dominated expanding universe without a cosmological constant which thereafter
The conclusive discovery by Hubble (1929) of a linear expansion law relating
redshift to distance 
%in an expanding universe made 
made Friedmann models the standard geometrical framework 
within which Hubble's discoveries
were subsequently interpreted 
\cite{wein89,zn83,jvn_cup,peeb93}.

Introduced, then discarded, the cosmological constant
staged several comebacks, the first having to do with the realization that
the static Einstein universe was unstable and, if perturbed, could either
expand or contract. In 1927 Lemaitre constructed an expanding model which
originated from such an asymptotically static state in the distant past.
The Lemaitre model had a long age and has frequently been reinvoked 
whenever the age constraints (associated with high values of $H_0$) get
too tight for standard FRW models (section 
\ref{sec:age}). The Lemaitre model
was also discussed in the early 1960s
when observations appeared to show an excess of quasi-stellar objects (QSO's)
near the redshift $z \simeq 2$. It was felt that a universe which 
`hesitated' or `loitered'
near the quasi-static state at $z \sim 2$ for a sufficient amount of time 
would naturally explain an
abundance of objects at that redshift. 
%the number density of 
%cosmological objects does
%not appear to peak at any particular redshift, instead arguments for a 
Present arguments for a positive cosmological constant are 
associated with observations of high
redshift supernovae which indicate $\o_\l = \l/3H^2 \sim 0.7$ 
\cite{perl98a,perl98b,riess98},
and from cosmological simulations of structure formation which also appear to
favour a positive cosmological constant \cite{kgb93,os95,cen98}.
In the next section
we shall qualitatively analyze solutions of the Einstein equations with a 
non-zero cosmological constant in a 
Friedmann-Robertson-Walker (FRW) universe following the original path
taken by Eddington and Lemaitre.

\section{FRW Cosmological models with $\Lambda \neq 0$.}
\label{sec:static}
A homogeneous and isotropic universe is
characterized by the Friedmann-Robertson-Walker line element 
\begin{equation}
ds^2 = c^2dt^2 - a^2(t)\left({dr^2\over 1 - \kappa r^2} + r^2 d\theta^2 +
r^2\sin^2\theta d\phi^2\right)  ~~~~\kappa = 0, ~\pm 1
\label{eq:lam1a}
\end{equation}
In this metric the Einstein equations (\ref{eq:lam1}) with
matter in the form of a perfect fluid acquire the 
following simple
form
\ber
3(\frac{\dot a}{a})^2 &=& 8\pi G\rho + \Lambda c^2 - 
3\frac{\kappa c^2}{a^2}\label{eq:lam2a},\\
\frac{\ddot a}{a} &=& -\frac{4\pi G}{3}(\rho + 3P/c^2) + \frac{\Lambda c^2}{3}.
%&\equiv & -\frac{4\pi G}{3}\gamma\rho + \frac{\Lambda c^2}{3}
\label{eq:lam2}
\eer
Equation (\ref{eq:lam2}) can be recast to look like the equation of 
motion of a point
particle on
the surface of a sphere of radius $R \equiv a$ and mass $M$, setting $c = 1$
we obtain
\begin{equation}
\ddot{R} = -\frac{GM}{R^2} + \frac{\Lambda}{3}~R.
\label{eq:9}
\end{equation}
The total `gravitating mass'
$M = \frac{4\pi}{3}R^3(\rho + 3P)$ reflects the fact that
`pressure carries weight' in Einstein's theory of gravity.
From (\ref{eq:9}) we find that 
a particle on the sphere feels both attractive 
and repulsive forces.
The force of repulsion $F_{rep} = \frac{\Lambda}{3}~R$
is caused by the cosmological constant and increases 
with distance if $\l > 0$. (For negative $\l$ this becomes a
force of `attraction', formally
resembling the force of confinement
between quarks which binds them within the nucleus.)

The opposite signs of the forces of attraction and repulsion in (\ref{eq:9})
allow for a large number of new solutions
to the Einstein equations. As pointed out in the previous section,
Einstein himself used the repulsive effect of the cosmological constant to
balance the attraction of matter resulting in a static closed universe
which Einstein felt was in agreement with Mach's principle. 
A quantitative analysis of solutions to (\ref{eq:lam2a}) \& (\ref{eq:lam2}) can 
be gained by eliminating $\rho$ in these equations and combining them into a 
single equation for the evolution of the scale factor in the presence of a 
$\Lambda$-term
\beq
2\frac{\ddot a}{a} + (1 + 3w)\lbrack \frac{{\dot a}^2}{a^2} +
\frac{\kappa c^2}{a^2}\rbrack - (1 + w)\Lambda c^2 = 0
\label{eq:lam2b}
\eeq
which is also valid if $\Lambda$ is a function
of time (\ie if $T_{ik} = \Lambda (t) g_{ik}$). 
(We have assumed that matter has an equation of state $P = w\rho c^2$.)
A comprehensive quantitative
analysis of (\ref{eq:lam2b}) has been carried out in \cite{felten86} for a 
cosmological constant,  and in \cite{coop98}
for a time varying cosmological 
term $\Lambda (t)$. For our purpose it will be sufficient
to note that the qualitative behaviour of the universe
in the presence of a cosmological term which is 
either constant or time varying, 
can be understood very simply by rewriting
(\ref{eq:lam2a}) in the suggestive form (we assume $c=1$ for simplicity)
\beq
\half {\dot a}^2 + V(a) = E 
\label{eq:lam3}
\eeq
where 
\beq
V(a) = - (\frac{4\pi G}{3}\rho a^2 + \frac{\Lambda a^2}{6}),
~~~
E = - \frac{\kappa}{2}.
\label{eq:lam4}
\eeq
Since $\rho = \rho_0(a_0/a)^{3(1+w)}$, we find, substituting $w=0$ for dust
\beq
V(a) = - (\frac{A}{a} + \frac{\Lambda a^2}{6})
\label{eq:lam4a}
\eeq
where $A = \frac{4\pi G}{3}\rho_0a_0^3$.
(We assume for simplicity that matter is pressureless so that $w = 0$, however
the qualitative analysis given below remains valid for matter
possessing more general equations of state.)
Equation (\ref{eq:lam3}) reminds one of classical motion with conserved 
energy $E$ in a one dimensional potential $V(a)$ whose generic form is shown in
Figure \ref{fig:pot} for $w = P/\rho \geq 0$.
From the form of $V(a)$ several things can be said about the behaviour of the
expansion factor $a(t)$. 
We shall first examine the case $\kappa = 1$ ($E < 0$)
since it provides us with the
largest variety of qualitatively different solutions to the Einstein equations.

\subsection{Closed universe models ($\kappa = 1$).}
\label{sec:closed}

\begin{figure}
\vskip 1.0cm
\centerline{
\psfig{file=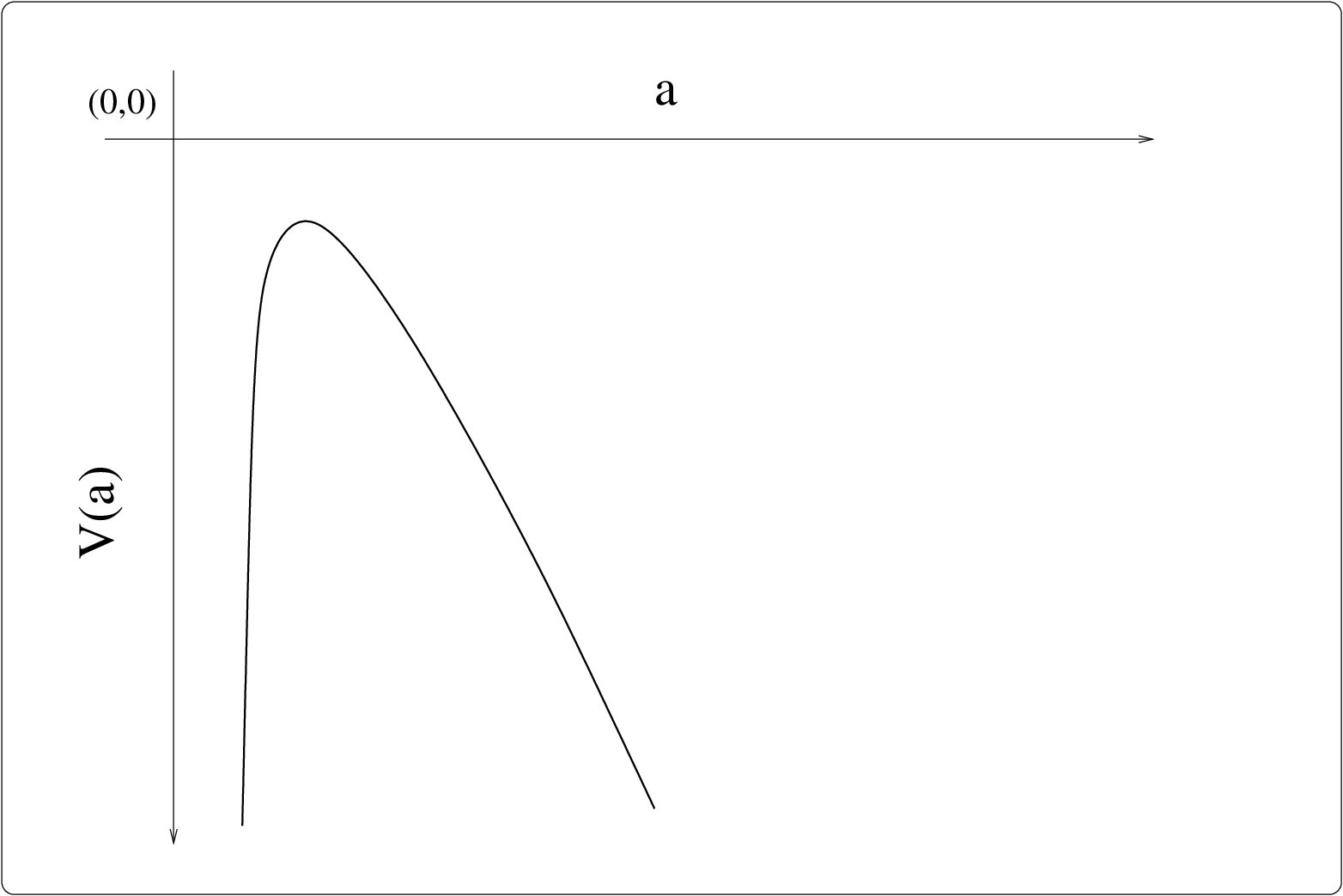,width=10cm,angle=-90}}
\caption{\footnotesize 
The `effective potential' $V(a)$ describing the expansion of the
universe in the presence of matter and a cosmological constant (see equation 
(\ref{eq:lam3})).
The large variety of solutions to the Einstein equations can be analyzed
by studying the kindered problem of the
motion of a particle moving under the
influence of the potential $V$.}
\bigskip
\hrule
\medskip
\label{fig:pot}
\end{figure}

%The situation with greatest variety 
%arises if $\kappa = 1$ so that $E < 0$. 
Consider a particle moving with
negative total energy under the influence of the potential $V(a)$ 
shown in fig. (\ref{fig:pot}),
then the following situations arise (the one dimensional particle coordinate 
is equivalent to the value of `$a$' --  the expansion factor.)

(i) {\em Oscillating models}:
The particle moves from left to right (starting from $a = 0$)
but with insufficient energy to surmount the potential barrier. 
Consequently the expansion factor $a(t)$ first
increases then decreases describing a universe which, after expanding, contracts 
into a singularity. Such models are called {\em oscillating models of the first 
kind} \cite{jvn_cup}. 

(ii) {\em Bouncing models}:
The particle moves from right to left (starting from 
$a = \infty$) again with insufficient energy to surmount the potential,
in this case $a(t)$ first decreases then increases and the universe
rebounds after collapsing without ever reaching a singular state.
Such models are called {\em bouncing models} or
{\em oscillating models of the second kind}, 
an example of such a model is provided by the complete de Sitter space-time
\begin{equation}
ds^2 = c^2 dt^2 - H^{-2}\cosh^2{(Ht)}\lbrack d\chi^2 +
\sin^2\chi(d\theta^2 + \sin^2\theta d\phi^2)\rbrack
\label{eq:de4}
\end{equation}
where $-\infty < t < \infty, 0 \leq \chi \leq \pi , 0 \leq \theta \leq \pi,
0 \leq \phi \leq 2\pi$.

(iii) {\em Static Einstein Universe (SE)}: 
The particle is placed at the top of the potential with exactly zero
kinetic energy: ${\ddot a} = {\dot a} = 0$. This situation, describes the
static Einstein universe.
Setting
${\ddot a} = {\dot a} = 0, \kappa = 1$ in (\ref{eq:lam2a}) and
assuming for simplicity
that
matter is pressureless ($w = 0$) we obtain
\beq
\Lambda_{crit} = \frac{4\pi G}{c^2}\rho_m = \frac{1}{a_0^2}
\eeq
which relates the value of the cosmological constant to the density of matter
{\em and} the curvature of space. The volume and mass of a SE universe are
respectively $V = 2\pi^2 a_0^3$, $M = V\rho_m = 2\pi^2 a_0^3\rho_m$. As a 
result $M = (\frac{\pi c^2}{2G})a_0$, and one finds $\lim_{a_0 \rightarrow 0} M 
\simeq 0$,
\ie the mass of the static Einstein universe decreases as its radius shrinks
to zero, consequently a static empty universe simply cannot exist !
This feature of SE found favour with the proponents of Mach's
principle as discussed in section \ref{sec:history}.

(iv) {\em Loitering Universe}:
The static Einstein universe is clearly unstable: small fluctuations can 
make it either contract or expand (these correspond to tiny perturbations
of a particle located at the hump of $V(a)$ in fig. (\ref{fig:pot})
which cause it to roll either towards the left
($a \rightarrow 0$)
or towards the right ($a \rightarrow \infty$).
Based on this observation, an interesting new model of the universe was proposed
by Eddington and Lemaitre in which the value of $\Lambda$ was kept slightly 
larger than $\Lambda_{crit}$. In this case the universe begins from the Big 
Bang,
approaches the static Einstein universe and remains
close to it for a substantial
period of time before re-expanding \cite{edd30,lema27}.
(If $\Lambda < \Lambda_{crit}$ the universe will contract instead of expanding.) 
The quasi-static or {\em loitering}
phase, during which the universe remains close to $a \simeq a_0$, has
several appealing features not present in models which expand monotonically
\cite{sfs92}:
(i) density perturbations grow at the exponential (Jeans) rate
$\delta \propto \exp{\sqrt{4\pi G\rho}~t}$ and not at the weaker rate
$\delta \propto t^{2/3}$ characteristic of an Einstein-de Sitter universe;
(ii) a prolonged quasi-static phase results in an older universe,
ameliorating the `age' problem which can arise in matter dominated flat
cosmologies 
if the value of the Hubble parameter turns out to be large (see section 
\ref{sec:age}).

Interest in loitering models rose dramatically in the late 1960's
when observations suggested the existence of an excess of quasars near
redshift $z_l \simeq 2$. To explain these observations the Lemaitre model
with a quasi-static (loitering) phase at $z_l\simeq 2$ was invoked
\cite{pss67,shklovsky67,rowan68}.
(Loitering at $z_l$ arises if the cosmological constant exactly balances
$\rho_m$ leading to the relation: $(1 + z_l)^3 = \Omega_\Lambda/\Omega_m$,
where $\Omega_\Lambda = \Lambda/3H^2$.
A decaying cosmological constant will lead to loitering at higher values of
$z_l$ which has certain advantages from the standpoint of current observations
\cite{sfs92}.)

(v) {\em Monotonic Universe}: 
The particle approaches the potential from the left ($a = 0$) with
sufficient energy to surmount it and travel on towards $a \rightarrow \infty$. 
In such a situation the scale factor will have an inflection point
at ${\ddot a} \simeq 0$, $\dot a > 0$. By adjusting 
initial conditions so that the particle remains close to the hump of the 
potential for a sufficiently long duration,
one recovers the `loitering' models
discussed in (iv).

(vi) {\em Nonsingular Oscillating model}:
Another cosmological model deserving mention consists of a form of matter which
behaves as a $\Lambda$-term when the universe is small, as the universe expands
the $\Lambda$-term decays into either radiation or matter.
The energy density in such a model can be phenomenologically described by
$8\pi G\rho = \Lambda/(1 + \Lambda a^p/\alpha)$,
so that
$\lim_{a \rightarrow 0}~8\pi G\rho \simeq \Lambda, ~~~
\lim_{a \rightarrow \infty} ~8\pi G\rho \simeq \alpha/a^p$, $p = 3, 4$
for matter and radiation respectively.
The potential $V(a) = -\frac{4\pi G}{3}\rho a^2$ associated with this model
has a broad minimum
which leads to a non-singular oscillatory motion of the expansion factor
$a(t)$. This toy model is interesting since it exhibits an infinite number of
expansion and contraction cycles without ever becoming singular.

\begin{figure}
\centerline{
\psfig{file=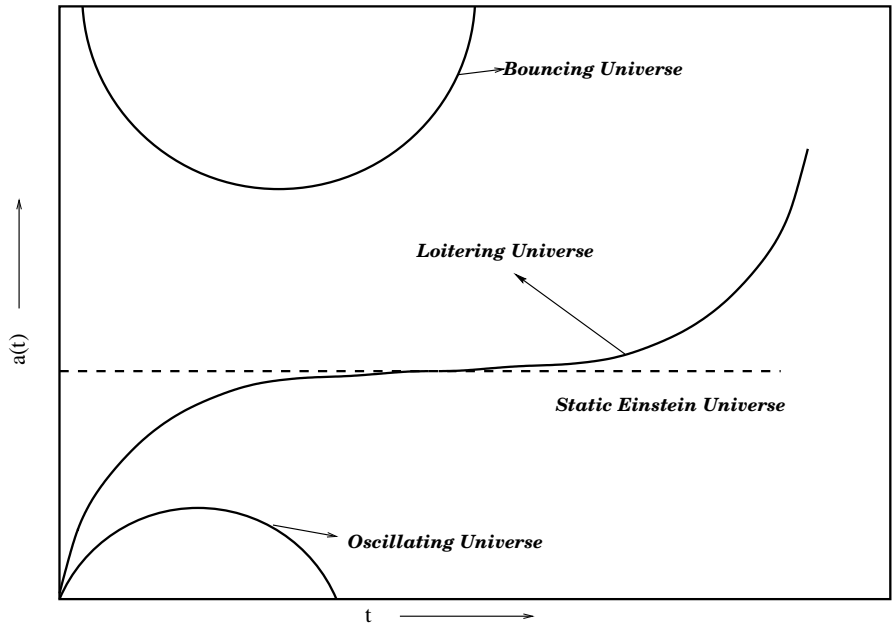,width=12cm,angle=-90}}
\caption{\footnotesize
Four distinct possible solutions of the Einstein equations with
a cosmological constant are schematically
shown for a closed universe ($\kappa = +1$).
(Incidentally none of these solutions arise if $\kappa = 0, -1$.)}
\label{fig:loiter}
\end{figure}

(vii) Other possibilities not shown in Figure (\ref{fig:loiter})
include `asymptotic models' in which
the universe asymptotically approaches or moves away 
from the static Einstein universe. 
The reader is referred to \cite{felten86,jvn_cup} for 
a more quantitative discussion of these issues.

Although the above discussion referred to cosmological models
filled with matter having non-negative pressure and a
cosmological constant, it is 
easy to show that the qualitative behaviour of the
universe described in (i) -- (vii) remains valid, if we generalize
the definition of the $\Lambda$-term to include any form of matter which
violates
the {\em strong energy condition} so that $\rho_{\Lambda} + 3P_{\Lambda} < 0$
\cite{sfs92}.

\include{psfig}

\subsection{Spatially open and flat cosmological models.}
\label{sec:open}

The preceding discussion referred to closed universe models for which 
$\kappa = 1$ and $E < 0$.
For flat and open models ($\kappa = 0, -1$) the total energy is non-negative
$E \geq 0$ and motion in the potential $V(a)$ becomes unbounded, since a 
particle always has sufficient energy to surmount the potential barrier 
in figure (\ref{fig:pot}).
As a result the expansion factor $a(t)$ shows monotonic behaviour,
starting from the singular point at $a = 0, t = 0$ and increasing without bound
as $t \rightarrow \infty$.
For $\l > 0$ the universe 
passes through an inflection point
at which the expansion of the
universe changes from deceleration $(\ddot a < 0$)
to acceleration $(\ddot a > 0$) (from (\ref{eq:lam2a}) \& (\ref{eq:lam2}) it can 
be shown that this
usually occurs at a redshift when $\l$ is still not dominating the
expansion dynamics of the universe; see section \ref{sec:sn}).

In the important case when the universe is 
spatially flat and contains pressureless matter (dust) and
a positive cosmological constant, the expansion factor
has the exact analytical form:
\beq
a(t) \propto \left(\sinh {{3\over 2}}
\sqrt{{\Lambda\over 3}}ct\right)^{2/3}
\label{eq:lam5}
\eeq
which interpolates smoothly between the matter dominated epoch in the past
($a \propto t^{2/3}$) and an inflationary epoch in the future 
($a \propto e^{\sqrt{\frac{\l}{3}}t}$). Equation (\ref{eq:lam5})
will be used later, when we examine some observational aspects of a 
universe with a cosmological constant in Section \ref{sec:obs_lambda}

Finally, oscillating, bouncing and loitering models, as well as the static
Einstein universe, are
clearly absent in flat and open FRW models. 

\section{Observational consequences of a cosmological $\Lambda$-term}
\label{sec:obs_lambda}

Arguments favouring $\Lambda > 0$ at the present epoch essentially stem
from four sets of observations:

(i) {\em The age issue:} A high value of the Hubble constant 
$H_0 \sim 80 {\rm km/sec/Mpc}$ predicts a short age of the
universe which is incompatible with the ages of the oldest stars (12 - 16 Gyr)
unless the universe is open ($\o_m < 0.1$) or flat and $\l$ dominated
$\o_m + \o_\l = 1$.
The appeal of this argument has somewhat decreased following 
recent Hipparcos parallax measurements indicating a lower value 
$H_0 \leq 67 ~{\rm km/sec/Mpc}$ and also a lower age for globular clusters:
$11.5 \pm 1.5$ Gyr. Still, recent observations of
old galaxies at high redshifts 
are extremely difficult to accommodate 
within the framework of a flat matter dominated cosmology unless
the Hubble parameter is very small ($H_0 \lleq 45$ km/sec/Mpc; 
section \ref{sec:age}).

(ii) {\em Structure formation:} The standard COBE normalized
cold dark matter model of
structure formation with $\o_m = 1$ appears to be in serious conflict with 
observations.
The situation may be remedied if the universe is flat, with most of matter
smoothly distributed in the form of a cosmological constant and only
a small fraction $\o_mh \simeq 0.2$ in clustered matter.
(Here $h$ is the Hubble constant in units of 100 km/s/Mpc).
Studies of the abundance and evolution of clusters of galaxies and of lensing
by clusters also appear to favour a low density universe
(section \ref{sec:clusters}).

(iii) {\em Baryon excess in clusters:} In a spatially flat universe with
$\o_m = 1$ the mass fraction in 
baryons in the Coma cluster is expected to greatly
exceed nucleosynthesis bounds 
leading to what has been called 
the `baryon catastrophe'. The mass fraction in baryons can be kept in agreement 
with nucleosynthesis constraints only if $\o_m h\simeq 0.16$ \cite{white93}
($\o_m$ includes contribution from baryons and clustered dark matter).
Agreement with the inflationary scenario which strongly favours a 
spatially flat universe then suggests that the remaining
mass might be in the form of a cosmological constant.

(iv) {\em High redshift supernovae and the cosmic microwave background:} 
Preliminary results 
from this rapidly advancing field of cosmology 
suggest that the universe may be accelerating universe with a dominant
contribution to its energy density coming in the form of 
cosmological $\l$-term.
These results, when combined with CMB anisotropy observations on intermediate
angular scales, strongly
support a flat universe
$\o_m + \o_\l = 1$ with $\o_\l \sim 0.6 - 0.7$ 
(sections \ref{sec:sn} \& \ref{sec:mbr}).

In the first half of this paper we shall briefly review
the present observational
status of the cosmological constant referring the reader to the original papers
and earlier reviews \cite{cpt92,cohn98} for more details.

\subsection{$H_0, q_0$ and the Age of the Universe}
\label{sec:age}

The quest for understanding the geometry of our universe has been one
of the central aims of cosmology since the 1960s
and Alan Sandage in 1970 even described the whole of observational
cosmology as being a ``search for two numbers''. The first of these
numbers -- the Hubble parameter
$H_0 = ({\dot a}/a)_0$, provides us with measure of the observable
size of the universe and its age. The second $q_0 = - H_0^{-2}({\ddot a}/a)_0$
is called the deceleration parameter and
probes the equation of state of matter and the cosmological density parameter.
In the presence of a cosmological constant,
\beq
q_0 = \frac{\o_m}{2} - \o_\l.
\label{eq:decel1}
\eeq
In a critical density universe with $\o_m + \o_\l = 1$, the deceleration
parameter
\beq
q_0 = \frac{3}{2}\o_m - 1,
\label{eq:decel2}
\eeq
consequently a critical density universe will accelerate if $\o_m < 2/3$.
The observational quest for $q_0$ showed that evolutionary effects play
a dominant role in this important quantity and for a while it was felt that
it may be virtually impossible to disentangle the true cosmological
`signal' for $q_0$ from evolutionary `noise'. Recent years however have
witnessed an important turnaround with the development of new and
more powerful techniques 
which are either less sensitive to evolutionary effects
or for which evolutionary effects are better understood.

In the next section, we shall consider several promising cosmological
tests which could shed light on the composition of the universe and its
geometrical properties. These tests include
gravitational lensing, the use of high redshift
supernovae as calibrated standard candles,
and the angular size-redshift relation.
Before we do that however, we shall turn our attention to another fundamental
quantity which has traditionally played an important role in constraining
cosmological models -- the age-redshift relation.

The presence of a $\Lambda$-term leads to an increase in the age of the universe 
with far-reaching observational consequences. To appreciate this let us 
first consider the critical density Einstein-de Sitter universe with
$a \propto t^{2/3}$, so that
\beq
t_0 = \frac{2}{3}H_0^{-1}.
\label{eq:age1}
\eeq
The value of $H_0$ therefore serves to determine the age uniquely in a spatially
flat matter dominated universe. Moderately high values $H_0 \geq 75 ~$km 
s$^{-1}$ Mpc$^{-1}$
result in an age for the universe which is smaller than the ages of the
oldest globular clusters making an Einstein-de Sitter universe with a high
value of $H_0$ difficult to reconcile with observations. The situation can
be remedied if we live in an open universe. Assuming for simplicity that
the universe is empty (a good approximation if $\Omega_m \leq 0.2$)
we get $a \propto t$ so that 
\beq
t_0 = H_0^{-1}.
\label{eq:age2}
\eeq
Combining (\ref{eq:age1}) and (\ref{eq:age2}) we get 
$\frac{2}{3}H_0^{-1} \leq t_0 < H_0^{-1}$ for matter dominated cosmological 
models
with $\Omega_m \leq 1$. 
(A longer age $t_0 > H_0^{-1}$ can be achieved in the 
presence of a cosmological constant.)
An open universe though older, nevertheless has two difficulties associated with 
it: the first is related to the growth of density perturbations which slow
down considerably in an open universe leading to large primordial 
fluctuations in
the Cosmic Microwave Background which may be difficult to reconcile
with observations (assuming standard adiabatic fluctuations with
scale-invariant initial spectra). 
The second is related to the Omega problem: 
a low Omega universe requires extreme fine tuning of
initial conditions, which some find to be an unattractive feature of
open/closed models.

Let us now consider a more general situation in which the universe
has a $\Lambda$-term in addition to normal matter.
A closed form expression for the age of the universe in spatially flat
models is given by \cite{kolbt90}
\beq
t_0 = \frac{2}{3H_0}\bigg \lbrack\frac{1}{2~\o_\l^{1/2}}
\log{\frac{1 + \o_\l^{1/2}}{1 - \o_\l^{1/2}}}\bigg\rbrack
%H = \sqrt{\frac{\Lambda}{3}}~\coth{(\frac{3}{2}\sqrt{\frac{\Lambda}{3}} t)}
\label{eq:age3}
\eeq
where $\Omega_\Lambda = \Lambda/3H_0^2 = 1 - \o_m$.

\begin{figure}[htb]
\centerline{
\psfig{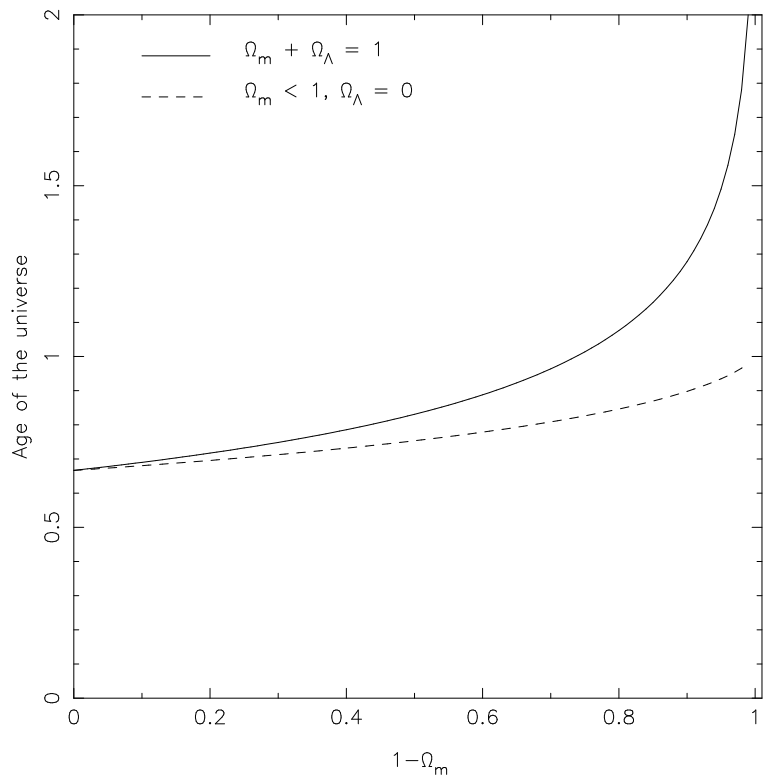}}
\caption{\footnotesize
The age of the universe (in units of $H_0^{-1}$)
is shown as a function of $1 - \Omega_m$
for (i) flat models with a cosmological constant
$\Omega_m + \Omega_\Lambda = 1$
(solid line), and
(ii) for open cosmological models $\Omega_m < 1$ (dashed line).}
\bigskip
\hrule
\medskip
\label{fig:age}
\end{figure}

In Figure \ref{fig:age}
we show the present age of a universe consisting of matter and
a cosmological constant and parametrized in terms of the variables
$\Omega_\Lambda$ and $\Omega_m$. 
We find that the age of a flat
universe with $\o_\l = 1 - \o_m$ is always greater than that of an open universe
for identical values of $1 - \o_m$. Additionally $t_0$ can exceed $H_0^{-1}$ if 
$\o_\l {\lower0.9ex\hbox{ $\buildrel > \over \sim$} } ~0.74$.
%which can be inverted to obtain a $t(H)$ relation analogous to 
%(\ref{eq:age1}), (\ref{eq:age2}).

No exact forms for $t(H)$ are available for a time dependent $\Lambda$-term.
To study this and other cases, it is useful to express the Hubble parameter
as a function of the cosmological redshift $z$. This can easily be
done for a general
multicomponent universe consisting of several non-interacting matter
species characterized by equations of state $P_{\alpha} =
w_{\alpha}\rho_{\alpha}$, for which the Hubble parameter can be written as
\begin{equation}
H(z) = H_0 h(z) = H_0(1 + z) \left[ 1 - \Omega_{total} + \sum_{\alpha}
\Omega_{\alpha}(1 + z)^{\gamma_{\alpha}}\right]^{\half}. \label{eq:age4}
\end{equation}
where $\Omega_{total} = \sum_{\alpha}\Omega_{\alpha}$,
$\gamma_\alpha = 1 + 3w_\alpha$ and $1 + z = a_0/a(t)$
is the cosmological redshift parameter.

Let us assume that the universe, in addition to matter and radiation,
consists of
a decaying $\l$-term modelled by 
a fluid with equation of state $P_X = (m/3 - 1)\rho_X$ so that
$\l = \l_0(a_0/a)^m$,
$m \leq 2$.
The dimensionless Hubble parameter $h(z)$ then becomes
\beq
h(z) = \left[ (1 - \Omega_{total})(1 + z)^2 + \Omega_m(1 + z)^3 + 
\Omega_\Lambda (1 + z)^m \right]^{\half}
\label{eq:age6} 
\eeq
%$\Omega_\Lambda = \Lambda_0/3H_0^2$ and 
where $m = 0$ corresponds to a cosmological {\em constant}, 
and we neglect the presence of radiation.
In a spatially flat universe $\Omega_{total} = \Omega_\Lambda + \Omega_m = 1$ 
(the present value of $\l$ is therefore given by
$\Lambda_0 = 3H_0^2\lbrack 1 - \Omega_m\rbrack$). 
A useful relationship between the cosmological time parameter $t$ and the
cosmological redshift $z$ can be obtained by differentiating
$1 + z = a_0/a(t)$ with respect to time, so that $dz/ dt = - H(z)(1 + z)$.
This leads to the following 
completely general 
expression for the age of the universe at a redshift $z$ 
\beq
t(z) = H_0^{-1}\int_z^\infty \frac{dz'}{(1 + z')h(z')}.
\label{eq:age7}  
\end{equation}
with $h(z)$ supplied by either (\ref{eq:age4}) or (\ref{eq:age6}).

A running debate over the previous decade or so has centered around whether 
or not the
universe has an `age problem', \ie on 
whether matter dominated cosmological models
are substantially younger than their oldest constituents (which happen to be 
metal poor old globular cluster stars). 
A key role in this controversy is played by
the Hubble parameter, whose present value is known to within an uncertainty of
about two. Higher values of $H_0$ clearly give rise to a younger universe
whereas lower values lead to an older one. 
%On the other hand, the most reliable
%lower limits on the ages of the oldest stars within globular clusters give
%values $12 - 14$ Gyr, and there appears to be a convergence of opinion that
%Pre-Hipparcos age estimates of the oldest stars within globular clusters
%gave typical values between 
%$15 - 18$ Gyr. A high value of $H_0 \sim 75 \km$
%favoured by many \cite{pea91,freed94,pierce94},
%combined with the above age estimates is clearly incompatible with a 
%flat $\Omega_m = 1$ cosmology, and perhaps 
%also with open models with $\Omega_m > 0.1$
%(see figure {\ref{fig:age}). 
%Such large values of $H_0$ if subsequently confirmed,
%will lend strong support to the presence of a cosmological constant
%or some other form of `repulsive gravity'
%(or cast doubts on our ability to model stellar interiors).

At the time of writing lower values 
$H_0 {\lower0.9ex\hbox{ $\buildrel < \over \sim$} } ~65 \km$ are
strongly supported by observations, especially in the light of 
new parallax measurements made by the {\em Hipparcos} satellite for Cepheid
stars, which has led to a reanalysis of distances to globular clusters and 
consequently of their
age estimates\footnote{An important indicator of the absolute age of a
globular cluster star
is its luminosity when it leaves the main sequence. Since luminosity is
related
to
distance (to the star) ages of globular clusters are very sensitive to
distance
callibrators.} which have 
dropped to $11.5 \pm 1.5$ Gyr \cite{chaboyer98,krauss98}.
(Low values of $H_0$ are also suggested from an analysis of the 
Sunyaev-Zeldovich effect from X-ray emitting clusters \cite{hughes98},
from Type Ia supernovae\cite{riess98,perl98b} 
and from Cepheids observed by the HST.)
Lower values of $H_0$ reconcile 
matter dominated flat models with the revised ages of 
globular clusters \cite{krauss98}
and with limits from nucleochronology 
which indicate $t_0 \geq 7.8$ Gyr \cite{chaboyer98}.
%(The latter are    
%less restrictive but in principle can be 
%more reliably known than the former \cite{cpt92}).
Low values of $H_0$ combined with an
absence of stellar systems with ages greatly exceeding 20 Gyr also argue
against large values of $\Lambda$, 
since from Fig \ref{fig:age} we see that $\Omega_\Lambda
{\lower0.9ex\hbox{ $\buildrel > \over \sim$} } ~0.85$ suggests an age $t_0 
{\lower0.9ex\hbox{ $\buildrel > \over \sim$} } ~24$ Gyr 
(if $H_0 = 50$km/sec/Mpc).
Finally the recent supernovae based measurements of Perlmutter et al. (1998b)
suggest a best-fit age of the universe $t_0 \simeq 14.9 ~(\frac{63}{H_0})$ Gyr
for a spatially flat universe with $\o_m \simeq 0.28$ and $\o_\l \simeq
0.72$.

The above arguments were largely limited to the {\em present}
age of the universe. Ages of high redshift objects at $z > 1$ provide
crucial information about the age of the universe at that redshift 
\cite{kenn96}. The existence of at least two high redshift galaxies having
an evolved stellar population and hence an old age sets very severe
constraints on a flat matter dominated universe 
\cite{dunlop96,krauss97,dunlop98,peacock98,alcaniz99}. For instance the radio
galaxy 53W091 at $z = 1.55$ discovered by Dunlop et al. (1996) is reported to
be at least 3.5 Gyrs old. The age of a spatially flat matter dominated
universe at a redshift $z$ is easily obtained from (\ref{eq:age7}) to be
\beq 
t(z) = \frac{2}{3H_0} (1+z)^{-\frac{3}{2}}.
\eeq
Consequently the discovery of 53W091 can be accommodated within an $\o_m = 1$,
CDM model only if the Hubble parameter is uncomfortably small \cite{krauss97}
$H_0 \lleq 45 \km$.
However both open and flat $\l$-dominated models
alleviate the age problem for 53W091.

At even higher redshifts, recent work \cite{ytk98}
aimed at age-dating a high redshift QSO at
$z = 3.62$ using delayed iron enrichment by Type Ia supernovae as a 
cosmic clock, sets a lower bound of 1.3 Gyr on the age of the universe
at that redshift. This discovery can be accommodated within 
a spatially flat cosmology only if $\o_m + \o_\l = 1$ (low density open
models with $\o_m \leq 0.2$ are also permitted).
However, the age dating of stellar populations requires complex modelling
and although both open and flat $\l$-dominated models are clearly
favoured by current observations, more work needs to be done before 
matter dominated flat models are excluded on the basis of age arguments alone.
\footnote{It may be 
appropriate
to mention that models with a cosmological constant
may {\em never be singular} and therefore could possess an infinite age as
demonstrated by the `bouncing models' in Fig. \ref{fig:loiter}.
However the value of the cosmological constant in such models
is several orders of magnitude larger than permitted by current observations.
The relevance of such models is therefore likely to be
limited to the very early universe and will not affect the
age problem discussed here.}

%\section{Gravitational lensing}
%Supernovae and the value of $\l$.}
\subsection{The luminosity distance and gravitational lensing}
\label{sec:lens}

Before proceeding to discuss possible constraints on $\o_\l$ from gravitational
lensing in this section and high redshift supernovae in the next,
let us introduce a quantity which plays a crucial role in these discussions,
namely the {\em luminosity distance} $d_L(z)$
 upto a given redshift $z$.
Consider an object of absolute luminosity ${\cal L}$ located at a coordinate 
distance $r$ from an observer at $r = 0$. Light emitted
by the object at a time $t$ is received by the observer at $t = t_0$, 
$t$ and $t_0$ being related by the cosmological redshift 
$1 + z = a(t_0)/a(t)$. The 
luminosity flux reaching the observer is
\beq
{\cal F} = \frac{\cal L}{4\pi d_L^2},
\label{eq:age8}
\eeq
where $d_L$ is the luminosity distance to the object \cite{jvn_cup}
\beq
d_L = a(t_0)r(1 + z).
\label{eq:age9}
\eeq
The luminosity distance $d_L$ depends sensitively
upon both the spatial curvature and the
expansion dynamics of the universe. To demonstrate this we
determine $d_L$ using the expression for the coordinate distance $r$
obtained by setting $ds^2 = 0$ in (\ref{eq:lam1a}),
resulting in
\beq
\int_0^{r}{dr'\over\sqrt{1 - \kappa r'^2}} = \int_{t}^{t_0}{ dt\over a(t)}
= \eta_0-\eta
\label{eq:age10}
\eeq
which gives
\ber
r &=& \sin ~(\eta_0-\eta)~~~~~~~(\kappa = +1)\nonumber\\
&=& ~~~~~~~\eta_0 - \eta~~~~~~~~(\kappa = 0)\nonumber\\
&=& \sinh~(\eta_0-\eta) ~~~~~(\kappa = -1)\nonumber\\
\label{eq:age11}
\eer
where
$\eta = \int_0^t cdt/a(t)$.

Furthermore, since $dz/dt = - (1 + z)H(z)$, we get
\begin{equation}
\eta_0 -\eta = \int_t^{t_0}{c dt \over a(t)} = \frac{c}{a_0H_0}\int_0^z 
{dz'\over h(z')}
\label{eq:age11c}
\end{equation}
where $h(z) = H(z)/H_0$ is defined in (\ref{eq:age4}), and, 
in a universe with several components
\beq
{\kappa\over a_0^2H_0^2} = \Omega_{total} - 1.
\label{eq:age11b}
\eeq
Substituting 
(\ref{eq:age11b}) and (\ref{eq:age11c}) in (\ref{eq:age9}) we get the 
following expression for the luminosity
distance in a multicomponent universe with a cosmological term \cite{cpt92}
%\begin{eqnarray}
%d_L(z) & =&  {(1+z) H_0^{-1}Sin(\eta_0 - \eta)\over
%|\Omega_{total} - 1|^{\frac{1}{2}}}
%~~~~~~~~~~~~~~(\kappa = +1)\nonumber\\
%d_L(z) & =&  {(1+z) H_0^{-1} Sinh(\eta_0 - \eta)
%\over |\Omega_{total} - 1|^{\frac{1}{2}}}
%~~~~~~~~~~~~(\kappa = -1)\nonumber\\
%d_L(z) &=& (1+z)H_0^{-1}\int_0^z {dz'\over h(z')}
%~~~~~~~~~~~~~~~~~~~~~(\kappa = 0)\nonumber\\
%\label{eq:age11d}
%\end{eqnarray}
\beq
d_L(z) =  {(1+z)cH_0^{-1}\over 
|\Omega_{total} - 1|^{\frac{1}{2}}}S(\eta_0-\eta)
\label{eq:age11d}
\eeq
where
\beq
\eta_0 - \eta = |\Omega_{total} - 1|^{\frac{1}{2}}\int_0^z {dz'\over h(z')},
\label{eq:age11e}
\eeq
and $S(x)$ is defined as follows:
$S(x) = \sin{(x)}$ if $\kappa = 1$ $(\Omega_{total} > 1)$,
$S(x) = \sinh{(x)}$ if $\kappa = -1$ $(\Omega_{total} < 1)$,
$S(x) = x$ if $\kappa = 0$ ~$(\Omega_{total} = 1)$.

\begin{figure}
\centerline{
\psfig{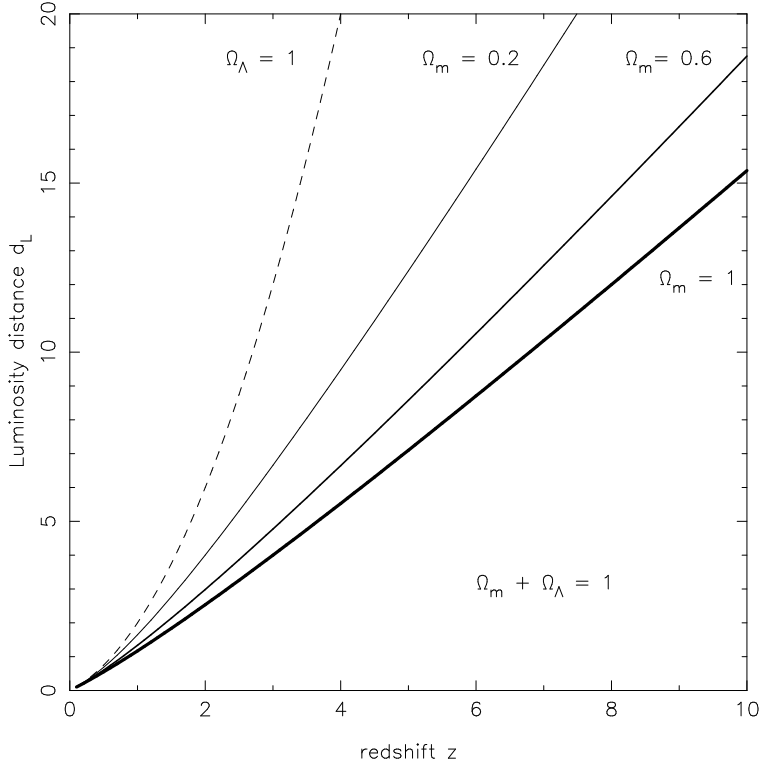}}
\caption{\footnotesize The luminosity distance $d_L$ (in units of $H_0^{-1}$)
is shown as a function of
cosmological redshift
$z$ for flat cosmological models with a cosmological constant $\Omega_m +
\Omega_\Lambda = 1$.
Heavier lines correspond to larger values of $\Omega_m$. For comparison we
also show (dashed line) the angular size in a flat de Sitter
universe ($\Omega_\Lambda = 1$).}
\bigskip
\hrule
\medskip
\label{fig:distance}
\end{figure}

\begin{figure}[htb]
\centerline{
\psfig{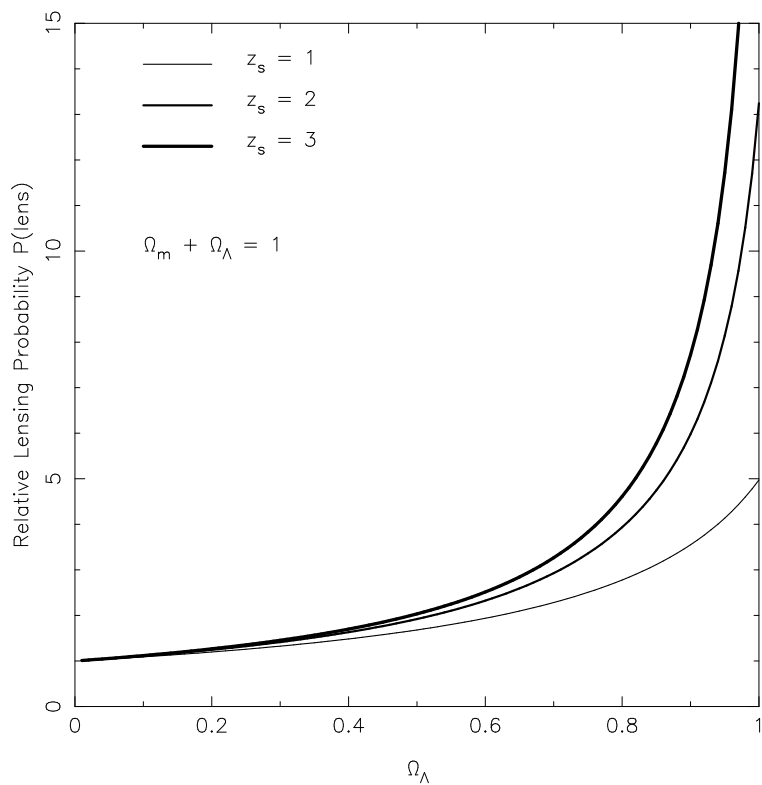}}
\caption{\footnotesize 
The lensing probability $P(lens)$ evaluated relative to the fiducial 
case $\Omega_m = 1$ is shown as a function of
$\Omega_\Lambda$ for flat cosmological models $\Omega_m +
\Omega_\Lambda = 1$. The source redshift is taken at $z_s = 1, 2, 3$
respectively.}
\bigskip
\hrule
\medskip
\label{fig:lens}
\end{figure}

Before we turn to applications, let us consider a simple
example which 
provides us with an insight into the role played by the
luminosity distance $d_L$ in cosmology.
In a spatially flat universe the expression for $d_L$ simplifies considerably,
so that we get for the matter dominated model ($a \propto t^{2/3}$)
\beq
d_L^{MD} = 2c~H_0^{-1}\lbrace (1 + z) - (1 + z)^\half\rbrace,
\label{eq:age12}
\eeq
on the other hand in de Sitter space $(a \propto \exp{(H_0 t)})$
\beq
d_L^{DS} = c~H_0^{-1} z(1 + z).
\label{eq:age13}
\eeq
Comparing (\ref{eq:age12}) and (\ref{eq:age13}) we find $d_L^{DS}(z) > 
d_L^{MD}(z)$,
which means that an object located
at a fixed redshift will appear brighter in 
an Einstein-de Sitter universe than
it will in de Sitter space 
(equivalently in the steady state model).\footnote{For instance a galaxy at 
redshift $z = 3$ will appear 9 times brighter in a flat matter dominated 
universe
than it will in de Sitter space (see Fig \ref{fig:distance}).}
This is also true for a 
two component universe consisting of matter and a cosmological constant
as demonstrated in Fig \ref{fig:distance}. In a 
spatially flat universe the presence of a $\Lambda$-term increases the
luminosity distance to a given redshift, leading to interesting astrophysical
consequences. Since the physical volume associated with a unit redshift interval 
increases in models with $\l > 0$, the likelihood that light from a quasar 
will encounter a lensing galaxy is larger in such models. Consequently
the probability that a quasar is lensed by
intervening galaxies increases appreciably in a $\Lambda$ dominated
universe, and can be used as  a test to constrain the value of 
$\Omega_\Lambda$ \cite{fuku90,fuku91,turner90}. 
Following \cite{fuku92,cpt92,cohn98} we give below the probability of a quasar 
at redshift $z_s$ being lensed relative to the fiducial Einstein-de Sitter
model ($\Omega_m = 1$)
\beq
P({\rm lens}) = \frac{15}{4}\lbrack 1 - \frac{1}{\sqrt{1 + z_s}}\rbrack^{-3}
\int_0^{z_s}\frac{(1+z)^2 dz}{h(z)}\bigg\lbrack 
\frac{d(0,z)d(z,z_s)}{d(0,z_s)}\bigg\rbrack^2
\label{eq:age13a}
\eeq
where $d(z_1,z_2)$ is a generalization of the angular distance 
$d_A = d_L (1 + z)^{-2}$ discussed in Section \ref{sec:angle}:
\beq
d(z_1,z_2) =  {1\over (1 + z_2)
|\Omega_{total} - 1|^{\frac{1}{2}}}S(\eta_{12})
\label{eq:age13b}
\eeq
where
\beq
\eta_{12} = \eta_1-\eta_2 = |\Omega_{total} - 1|^{\frac{1}{2}}\int_{z_1}^{z_2} 
{dz\over h(z)}
\label{eq:age13c}
\eeq
and $S(\eta_{12})$ is defined as follows, 
$S(\eta_{12}) = \sin{(\eta_{12})}$ if $\kappa = 1$ $(\Omega_{total} > 1)$,
$S(\eta_{12}) = \sinh{(\eta_{12})}$ if $\kappa = -1$ $(\Omega_{total} < 1)$,
$S(\eta_{12}) = \eta_{12}$ if $\kappa = 0$ ~$(\Omega_{total} = 1)$.
In Fig \ref{fig:lens} we show the lensing probability $P({\rm lens})$ 
for the spatially flat universe $\Omega_m + \Omega_\Lambda = 1$.
A large increase in the lensing probability over the fiducial $\Omega_m = 1$
value is clearly seen in models with low $\Omega_m$ (high $\Omega_\Lambda$).
(For a broader analysis of parameter space see \cite{cpt92}.)

Turning now to the observational situation,
at the time of writing the best observational estimates give a $2\sigma$ 
upper bound $\Omega_\Lambda < 0.66$ obtained from multiple images of 
lensed quasars
\cite{kochanek93,kochanek96,maoz93}. 
Since radio sources are not plagued by some of the systematic
errors arising in an optical search (notably extinction in the lens galaxy and
the quasar discovery process) a search involving radio
selected lenses can yield useful complementary information to optical
searches \cite{falco98}. 
Recent work by Falco et al (1998) gives $\Omega_\Lambda < 0.73$ which is 
only marginally consistent with 
optical estimates, 
a combined analysis of optical and radio data yields a slightly more
conservative upper bound $\Omega_\Lambda < 0.62$ at the
$2\sigma$ level (for flat universes) \cite{falco98}.
(Constraints on $\o_\l$ from both lensing and Type Ia Supernovae are
discussed in \cite{waga98}; also see next section. An interesting new method of 
constraining $\o_\l$ from weak lensing in clusters is discussed in
\cite{fort98}, also see \cite{bartel98} and section \ref{sec:clusters}.)
Improved understanding of statistical and systematic uncertainties combined
with new surveys and better quality data promise to make gravitational 
lensing a powerful technique for constraining cosmological parameters and
cosmological world models.

%\subsubsection{The luminosity distance and high redshift supernovae}
\subsection{Type Ia Supernovae and the value of $\l$.}
\label{sec:sn}

The luminosity distance also plays a crucial 
role in determining cosmological 
parameters once the absolute brightness of a class of objects is known.
Of considerable importance in this context is
the magnitude-redshift relation which relates the
apparent magnitude $m$ of an object to its absolute magnitude $M$ 
\beq
\mu \equiv m - M = 5\log_{10}{\frac{d_L}{Mpc}} + 25
\label{eq:age14}
\eeq
where $\mu$ is known as the distance modulus.
%One can therefore attempt to constrain
%the form of $d_L$ if both $m$ and $M$ are known
%within reasonable limits. 
Since $d_L$ depends upon the geometry of space
and its material content,
the magnitude-redshift relation (\ref{eq:age14}) can, in principle
be used to
determine $\Omega_\Lambda$ and $\Omega_{total}$ if both $m$ and $M$ are known
within reasonable limits. \footnote{In practice 
(\ref{eq:age14}) must be corrected for effects associated with the
redshifting of light as it travels to us, commonly called the K-correction.
For instance photons being detected using a red filter would originally
have had a `blue spectrum' if the source was located at $z \simeq 1$.
Other possible sources of systematic errors include luminosity evolution,
intergalactic extinction,
Malmquist bias, the aperture correction,
weak lensing etc.
A more complete discussion of these issues can be found in 
\cite{jvn_cup,peeb93}.}

\begin{figure}
\centerline{
\psfig{file=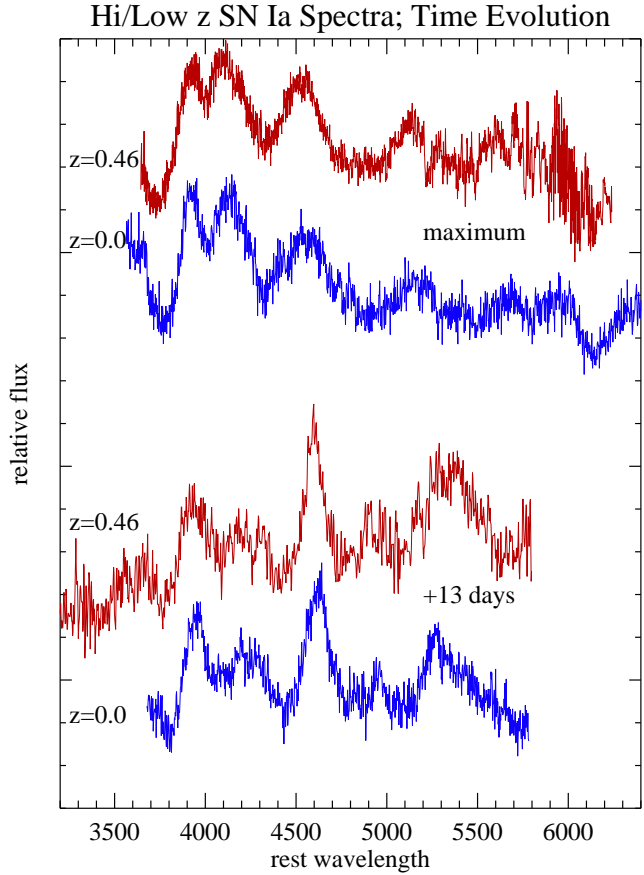,width=10cm}}
\caption{\footnotesize
Spectra of two type Ia supernovae at high ($z = 0.46$) and low ($z = 0.0$)
redshift show
a remarkable similarity; from \cite{schommer99}.}
\bigskip
\hrule
\medskip
\label{fig:sn_spectra}
\end{figure}

The recent discovery that type Ia supernovae may be used as calibrated
standard candles for obtaining estimates of the luminosity distance
$d_L$ through (\ref{eq:age14}) has aroused great interest.
Type Ia supernovae are explosions which arise as a white dwarf star crosses
the Chandrasekhar stability limit while accreting matter from a companion
star
\cite{hoyle60,arnett69,colgate69}. (See \cite{bruno} for a recent review
of Type Ia supernovae.)
The high absolute luminosity of SNe Ia ($M_B \simeq - 19.5$ mag) suggests
that they can be seen out to large distances making them ideal candidates for 
measuring and constraining cosmological parameters \cite{colgate79,branch92}.
Of crucial import to using type Ia supernovae for estimating the luminosity
distance $d_L$ 
has been the observation that: (i) the dispersion in their luminosity at 
maximum light is extremely small ($\lleq 0.3$ mag);
(ii) the width of the supernova light
curve is strongly correlated with its intrinsic luminosity: 
a brighter supernova will have a broader light curve indicative of a more 
gradual
decline in its brightness \cite{phillips93}. 
Both (i) and (ii) reduce the scatter in the absolute luminosity of type Ia
supernovae to $\sim 10\%$ making them excellent standard candles \cite{branch92}. (Remarkably a comparison of the spectra of high and low redshift supernovae
shows no significant differences as illustrated in figure \ref{fig:sn_spectra},
lending support to the standard candle hypothesis.)

\begin{figure}
\centerline{
\psfig{file=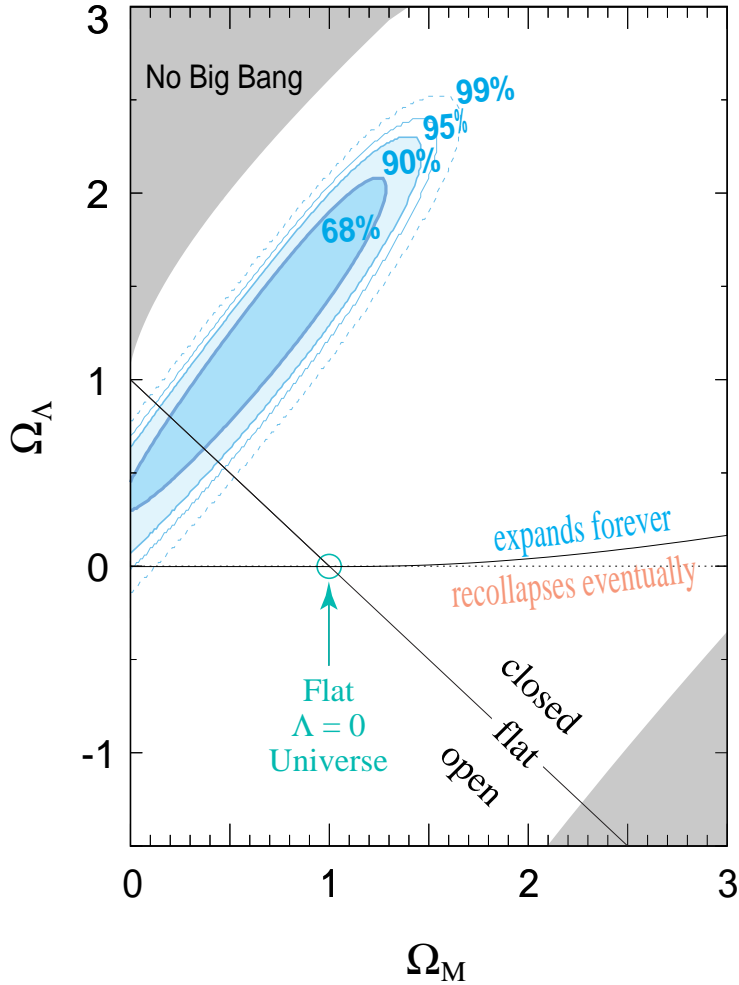,width=10cm}}
\caption{\footnotesize 
Best-fit confidence regions in the $\o_m-\o_\l$ plane obtained
from the analysis of Type Ia high redshift supernovae of 
Perlmutter et al. (1998b).
The upper-left shaded region corresponds to the singularity free 
`bouncing universe' models
discussed in section \ref{sec:closed}.} 
\bigskip
\hrule
\medskip
\label{fig:sn}
\end{figure}

Nearby type Ia supernovae ($z < 0.15$) 
have been used to determine the value of $H_0$
and to calibrate the sample, whereas those further away are used to
obtain reasonable estimates of 
cosmological parameters by minimizing
the $\chi^2$ statistic
\beq
\chi^2(H_0, \o_m,\o_\l) = \sum_i\frac{\lbrace\mu_{p,i}(z_i;H_0,\o_m,\o_\l) - 
\mu_{0,i}\rbrace^2}{\sigma_{\mu0,i}^2}
\eeq
where $\mu_p$ are model dependent `predicted' values of the distance modulus
obtained from (\ref{eq:age11d}) and (\ref{eq:age14}), 
and $\mu_0(z_i)$ are the observed values.
At least two groups -- the Supernova Cosmology Project \cite{perl98a} and
the High-Z Supernova Search Team \cite {riess98}
have been engaged in both finding and
calibrating supernovae at low and high redshifts.
(Recent information
on the supernovae search can be obtained from the web sites \cite{web}.)
At the time of
writing, both groups have analyzed data for several dozen type Ia supernovae
and a consensus seems to be emerging 
that a positive value of $\Omega_\Lambda$ is
strongly preferred. 
For instance, treating type Ia supernovae as standard candles and then
using distance estimates to 42 moderately  high redshift supernovae with 
$z \lleq 0.83$,
Perlmutter et al. (1998b) find that the joint probability distribution of
the parameters $\o_\l$ \& $\o_m$ is well approximated by the relationship
(valid for $\o_m \leq 1.5$)
$$0.8 \o_m - 0.6 \o_\l \simeq -0.2 \pm 0.1.$$ 
The best-fit confidence region in the $\o_m - \o_\l$ plane shown in
(\ref{fig:sn})
appears to favour a
closed universe. However, as we shall see in the next section, when
combined with the results of cosmic microwave experiments, the combined
likelihood of $\o_m,~\o_\l$ peaks near $\o_m + \o_\l \simeq 1$.

 These results provide an interesting insight into the expansion dynamics
of the universe during its recent past. For instance,
a cosmological model which 
passed through an epoch of matter domination 
before the present $\l$ dominated epoch, also passed through an inflexion point
at which the expansion of the
universe changed from deceleration $(\ddot a < 0$)
to acceleration $(\ddot a > 0$). From (\ref{eq:lam2a}) \& (\ref{eq:lam2}) it can
be shown that this
occurred at a redshift when $\l$ was still not dominating the
density of matter in the universe.
For instance from (\ref{eq:lam2}) we find that 
deceleration is succeeded by acceleration at the epoch
\beq
(1 + z_\ast)^3 = 2\frac{\o_\l}{\o_m}.
\label{eq:ast}
\eeq
On the other hand the epoch of equality between $\rho_m$
and
$\l$ occurred at the redshift
\beq
(1 + z_\star)^3 = \frac{\o_\l}{\o_m}
\label{eq:star}
\eeq
where $\o_\l = \l c^2/3H_0^2$.
Substituting the `best-fit' values obtained by Perlmutter et al. (1998b)
for a flat universe
$\o_m \simeq 0.28$, $\o_\l \simeq 0.72$ 
we get $z_\ast \simeq 0.726$ and
$z_\star \simeq 0.37$ so that $z_\star < z_\ast$. From (\ref{eq:decel2}) 
we also get 
$q_0 = -0.58$ for the deceleration parameter, indicating 
an accelerating universe (the combined Sn+CMB data give a slightly larger value
$q_0 \simeq -0.5$). 

Supernovae data can also be used to constrain time dependent $\l$ models
of the kind discussed in section \ref{sec:dynamic}. 
For models with $\l = \l_0 (1+z)^m$ equations (\ref{eq:ast}) and 
(\ref{eq:star}) are generalised to
\beq
(1 + z_\ast)^3 = -(1 + 3w)\frac{\o_\l}{\o_m}.
\eeq
and 
\beq
(1 + z_\star)^3 = \frac{\o_\l}{\o_m}^{-1/w}
\eeq
where $w = m/3-1$ is the (time-independent) equation of state of the $\l$-field.
A combined analysis of gravitational lensing and Type Ia supernovae gives
the best-fit value $\o_m \simeq 0.33$ for a spatially flat universe
\cite{waga98}. Attempts to constrain the decay rate 
result in $0.24 \lleq \o_m \lleq 0.38$
and $m \lleq 0.85$ (at the 68\% confidence level)
which in turn places constrains on the cosmic
equation of state $w = m/3-1 \lleq -0.72$.
The combined supernovae \& lensing data therefore convincingly rule
out a network of tangled
cosmic strings ($w \simeq -1/3$) and strongly favour a cosmological
{\em constant} ($w = -1$).

In the case of
scalar field models with potentials $V(\phi) \sim \phi^{-p}$ and 
$V(\phi) \sim (e^{1/\phi} - 1)$
the scalar field density in a spatially flat universe 
is constrained to lie in the range \cite{ptw99,wang99}
$\o_\phi \in (0.6,0.7)$ and the effective current equation of state is
$w_\phi < -0.6$ (at the 95$\%$ confidence level). 
(Scalar field or `quintessence' models give rise to an equation of 
state which is usually a function of time (equivalently redshift), 
exceptions to this rule are also known 
and are discussed in section \ref{sec:quint}. Some constraints
on a time dependent equation of state have been studied in 
\cite{cooray99,efst99,ht98,saini99}.)

\begin{figure}
\centerline{
\psfig{file=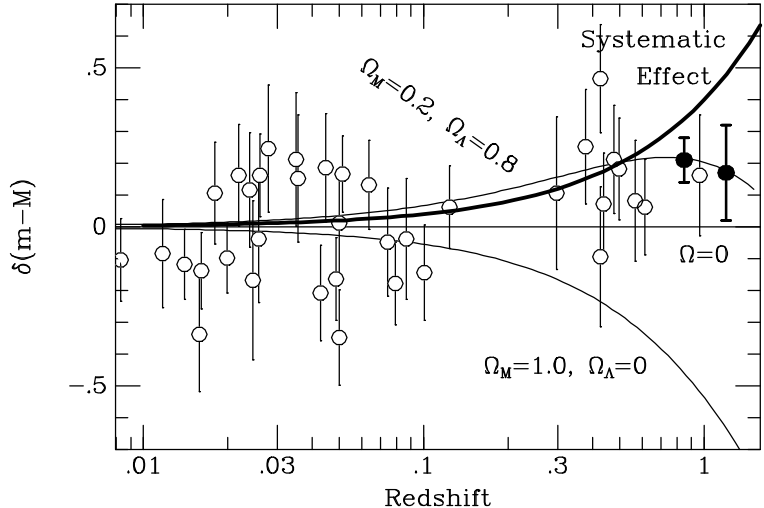,width=10cm}}
\caption{\footnotesize
The effects of a systematic bias in the properties of SnIa 
are illustrated in this figure
from Schommer, Suntzeff and Smith (1999). The upper thin curve corresponds to 
the best fitting $\l$ model and the lower thin curve to the Einstein-de Sitter
universe. The SnIa data are plotted with respect to an empty universe
with $\Omega = 0$. Systematic effects caused by evolution or extinction are
expected to follow the rising heavy curve which is
distinguished from the $\l$ curve showing a downturn at $z \sim 1$.}
\bigskip
\hrule
\medskip
\label{fig:sn_evolution}
\end{figure}

The results obtained by both the Supernova Cosmology Project and
the High-Z Supernova Search Team
team present
the strongest `direct' 
evidence for a non-zero cosmological constant. However much work 
needs to be done both in understanding systematic uncertainties 
as well as Sn Ia properties before the case for a positive 
$\Lambda$ is firmly established.\footnote{An accelerating universe can also
be accommodated within the framework of the Quasi-Steady State Cosmology of
Hoyle, Burbidge and Narlikar (1993).}
One of the major causes of worry faced by 
`standard candles' is source evolution and/or extinction. Clearly if
the observed faintness of the supernova is being caused by physical processes
rather than the accelerated expansion of the universe
 then all bets on $\l > 0$ are off. Fortunately
the effects of both evolution and extinction are likely to monotonically
increase with redshift whereas the effect of 
$\l$ peaks at redshifts $z \sim 1$ and declines thereafter; 
see figure \ref{fig:sn_evolution}.
Consequently observations conducted at redshifts just beyond $z = 1$
combined with lower redshift observations should be able to disentangle
source evolution/extinction from effects due solely to $\l$ \cite{schommer99}.
As we shall show in the next section, 
much stronger constraints on $\o_m$ and $\o_\l$
emerge if we combine the supernovae results with observations of the
cosmic microwave background.

\subsection{Constraints on $\l$ from the cosmic microwave background.}
\label{sec:mbr}

On large angular scales $\theta \ggeq 1^\circ$ photons of the
cosmic microwave background traveling to us from the last scattering surface
probe scales that were causally unconnected at the time of recombination.
\footnote{In matter dominated models 
the horizon at last scattering subtends an angle
$\theta \simeq 1.8^{\circ}
\Omega_m^{1/2} (1000/z_{rec})^{1/2} \simeq 1.8^{\circ}$
for $\Omega_m \simeq 1$ and $z_{rec} \simeq 1000$. In flat $\l$ dominated
models the dependence of $\theta$ on $\o_m$ is much weaker, consequently
$\theta \simeq 1.8^{\circ}$ provides a good approximation for most values of
$\o_m$.}
As a result observations of the CMB anisotropy on large scales provide us
with a very clean probe of the primordial matter fluctuation spectrum
before its distortion by astrophysical processes. On such large scales
the main contribution to the CMB anisotropy comes from the Sachs-Wolfe effect
\begin{equation}
{\delta T\over T} = - \frac{1}
{2}\int_{\eta_{rec}}^{\eta_0}\frac{\partial h_{\alpha\beta}}
{\partial\eta}e^{\alpha}e^{\beta}d\eta,
\label{eq:l55a}
\end{equation}
which relates temperature fluctuations to the integral of the 
variation of the metric
evaluated along the line of sight \cite{sw67}.
The evaluation of (\ref{eq:l55a}) in a flat matter dominated universe
is simplified
by the fact that linearized
the gravitational potential does not
evolve with time, with the result that the above expression reduces to 
\begin{equation}
{\delta T\over T} \simeq {1\over 3}{\delta \phi\over c^2}
\label{eq:l55}
\end{equation}
which relates fluctuations in the CMB to those in the
gravitational potential at the surface of last scattering.
Equation (\ref{eq:l55}) can therefore be successfully used to determine
the amplitude of primordial metric fluctuations with the help of COBE data.
The presence of a cosmological constant however 
causes the linearized gravitational potential to evolve with time,
the full Sachs-Wolfe integral (\ref{eq:l55a})
must therefore be used both to determine and 
normalize the primordial fluctuation spectrum \cite{ks85}.

The CMB temperature distribution can be written as
\begin{equation}
T(\theta, \phi) = T_0\left[1 + \frac{\delta T}{T}
 (\theta, \phi) \right],
 \label{eq:l56}
 \end{equation}
where $T_0$ is the blackbody temperature $T_0 = 2.728 \pm 0.004^\circ K$
\cite{fixsen96}.
$\delta T/T$ can be written in terms of a multipole expansion
on the celestial sphere:
\begin{equation}
\frac{\delta T}{T} (\theta, \phi) =
\sum_{l=2}^\infty~\sum_{m=-l}^l a_{lm} Y_l^m (\theta, \phi),
\label{eq:l57}
\end{equation}
Information pertaining to a particular theoretical model is contained
in the coefficients $a_{lm}$ which are usually assumed to be 
statistically independent and distributed in the manner of a Gaussian
random field with zero mean and variance
\begin{equation}
C_l \equiv \langle|a_{lm}|^2\rangle
\eeq
where the angle brackets indicate
an ensemble average over possible universes.

\begin{figure}[htb!]
\centerline{
\psfig{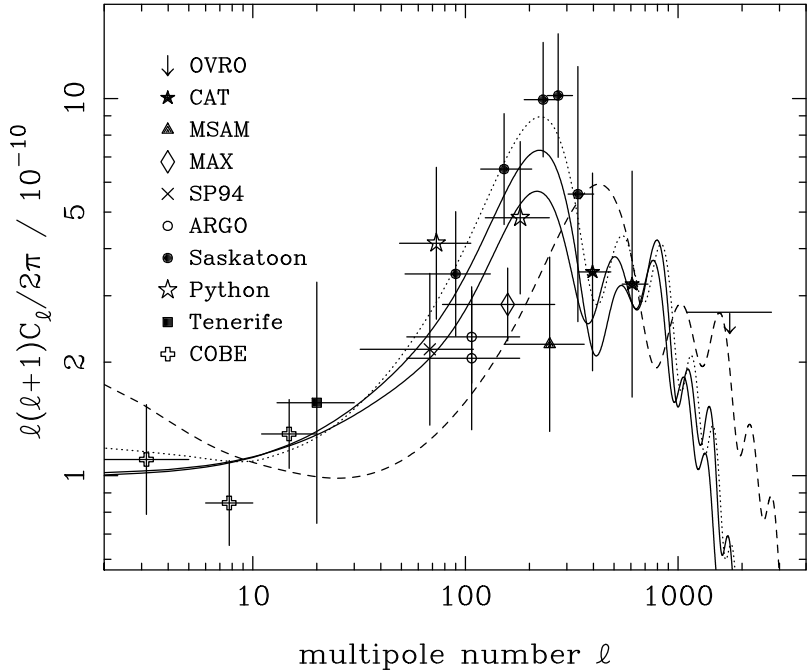}
}
\caption{\footnotesize
The angular power spectrum of the cosmic microwave background is plotted against
the angular wavenumber $l$ (in radians$^{-1}$). The predictions of the following
theoretical models are tested against observations:
(i) The flat $\l$CDM model with parameters
$(\o_\l,\o_m,\o_b,h) = (0.7,0.3,0.05,0.65)$ (dotted line); 
(ii) Flat CDM models with
$(\o_m,\o_b,h) = (1,0.1,0.5)$ and $(\o_m,\o_b,h) = (1,0.05,0.5)$ (solid lines),
the larger $\o_b$ model shows a higher Doppler peak; (iii) Open CDM model
with $(\o_m,\o_b,h) = (0.3,0.05,0.65)$ (broken line). Here $\o_m = \o_{cm}+\o_b$, where $\o_{cm}$ is the cold (non-baryonic) matter component.
For more details,
see Peacock (1999) and Bond et al. (1997).}
\bigskip
\hrule
\medskip
\label{fig:cl}
\end{figure}

The quantity that is directly measured by observations is 
the angular correlation
of the temperature anisotropy
\beq
C(\theta) = \bigg\langle \frac{\delta T}{T}(\hat{n_1})\frac{\delta 
T}{T}(\hat{n_2})\bigg\rangle = \frac{1}{4\pi} \sum_l \bigg\lbrack\frac{l + 
\half}{l(l+1)}\bigg\rbrack
C_l P_l(\cos\theta) W_l
\label{eq:ang}
\eeq
where $\cos\theta = \hat{n_1}\cdot\hat{n_2}$, $P_l$ are Legendre polynomials
and $W_l$ is the filter
function of the experiment used to measure the CMB;
$\langle \rangle$ denote an ensemble average in the case of theoretical
predictions and angular average in the context of observations. 
(The relationship between
$C(\theta)$ and the angular power spectrum $C_l$ is analogous to that between 
the two point correlation function $\xi$ and the matter power spectrum $P(k)$.)

At low multipoles $l \lleq 60$ 
the contribution to $C_l$ is mainly from the Sachs-Wolfe effect due
to scalar density perturbations and (in some models) tensorial gravity waves.
(The value of the tenth multipole provides a convenient choice for normalization
 of the perturbation spectrum \cite{bw97}.)
At large $l > 60$ however, the main contribution to $C_l$ is due to oscillations
in the photon-baryon plasma before decoupling, which leave their
imprint in the CMB at the time of last scattering.
These oscillations give rise
to Doppler peaks in $C_l$ the location of the peak being 
determined by the angle subtended by the sound horizon at the time of 
recombination (see figure \ref{fig:cl}). 
The sound horizon depends upon $\o_{\rm baryon}$ \& $\o_m$ 
whereas the angular diameter 
distance
to the last scattering surface depends upon $\o_\l$, $\o_m$ and the spatial 
curvature of 
the universe. (Both $\o_\l$ and the spatial curvature are extremely small
at the time of last scatter and therefore do not contribute to the
sound horizon. On the other hand, the location of the doppler peak is not
very sensitive to $\o_{baryon}$ provided $\o_{baryon} \ll \o_m + \o_\l$.)

The angular scale corresponding to the first Doppler peak 
is sensitive to both the curvature of the
universe and its matter content, its location 
can therefore be used to place strong constraints on cosmological models.
There are some indications that the first Doppler peak has been measured
near $l \simeq 200$ \cite{boom99}, since $l_{\rm peak} \sim 200 \Omega^{-1/2}$
one obtains $0.85 \leq \o \leq 1.25$ at the 68\% confidence level.
(The height of the peak is related to the baryon fraction in the 
universe and also to the scalar/tensor ratio S/T, 
the larger the baryon density the higher the peak, a small value of S/T 
reduces the peak height. The peak height also depends on the rate of
expansion of the universe and hence on $H_0$ \cite{husug95,jungman96,hu96}; 
for low values
$\o_{\rm baryon} \lleq 0.05$ the peak height decreases if $H_0$ increases,
whereas the reverse is true for a larger baryon fraction.)
In figure \ref{fig:cl} we show the angular power spectrum of the cosmic 
microwave background
for the flat $\l$CDM model with $\o_\l = 0.7$ (dotted line), 
for comparison we also
show spatially flat (solid line) and open (dashed line)
matter dominated models with $\o_m = 1$ and $\o_m = 0.3$ respectively.

\begin{figure}[htb!]
\centerline{
\psfig{file=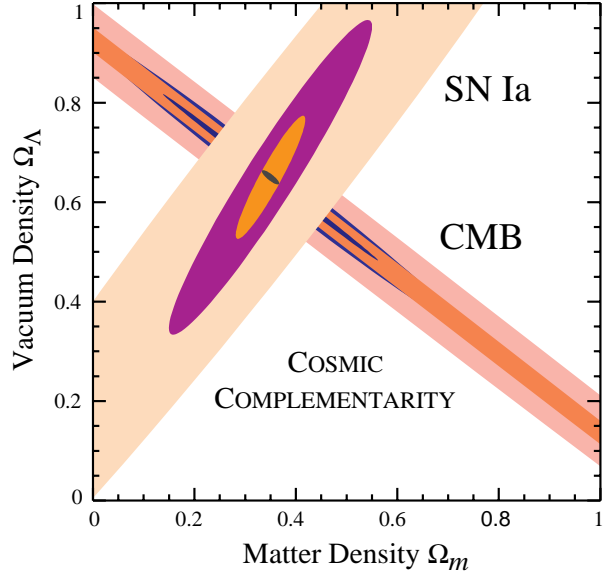,width=8cm}}
\caption{\footnotesize
The `cosmic complementarity' principle is beautifully illustrated
by these best-fit contours obtained using expected data from future supernovae
and CMB experiments. The 68$\%$ confidence regions are shown for three
sets of hypothetical supernovae data
%containing 100, 200 and 400 events
%providing `pessimistic' and `optimistic' prognosis of which are
likely to be recorded in
five years time. The CMB analysis refers to the upcoming MAP and PLANCK
satellite missions. The assumed fiducial model is $\l$CDM with $\o_m = 0.35$,
$\o_\l = 0.65$ and $H_0 = 65$ km. sec$^{-1}$ Mpc$^{-1}$.
One clearly sees that the degeneracy in parameter space from supernovae
observations is almost orthogonal to the degeneracy arising from CMB 
measurements.
For more details
see Tegmark et al. (1998).}
\bigskip
\hrule
\medskip
\label{fig:cmb}
\end{figure}

It should however be pointed out that the CMB alone cannot uniquely
differentiate between two models having identical matter content, perturbation
spectra and with the same angular diameter distance to the last scattering 
surface. Such models will be degenerate in the sense that they will
produce very similar CMB anisotropies \cite{eb98,efst98}.
A degeneracy in parameter space happens to be a common feature of most 
cosmological tests. Fortunately different tests often have complementary 
degeneracies. 
(A degeneracy arises when a result remains unaffected by a specific 
combination of parameter changes.)
For instance the degeneracy in the $\o_m - \o_\l$ plane
from high redshift supernovae tests is almost orthogonal to that
in a CMB analysis. Thus combining Type Ia supernovae measurements 
with the results from CMB experiments can serve to substantially 
decrease the errors on expected values of $\o_m$ and $\o_\l$ as 
illustrated in figure \ref{fig:cmb} and figure \ref{fig:sn_cmb}
\cite{white98,tegmark98,eb98,efst98}.
Since the location of the Doppler peak near $l \simeq 200$ supports 
a spatially flat universe \cite{hancock98,boom99}, 
a combined likelihood analysis of CMB anisotropies from the 1997 test
flight of the BOOMERANG experiment
and Type Ia Supernovae data gives best fit values \cite{boom99}
\beq
%\o_m = 0.25_{-0.12}^{+0.18},~ \o_\l = 0.63_{-0.23}^{+0.17}.
0.2 \leq \o_m \leq 0.45, ~ 0.6 \leq \o_\l \leq 0.85
\eeq
which strongly favour a flat universe with $\o_m + \o_\l \simeq 1$
(also see \cite{efst98,line98,tegmark99}).

\begin{figure}[htb!]
\psfig{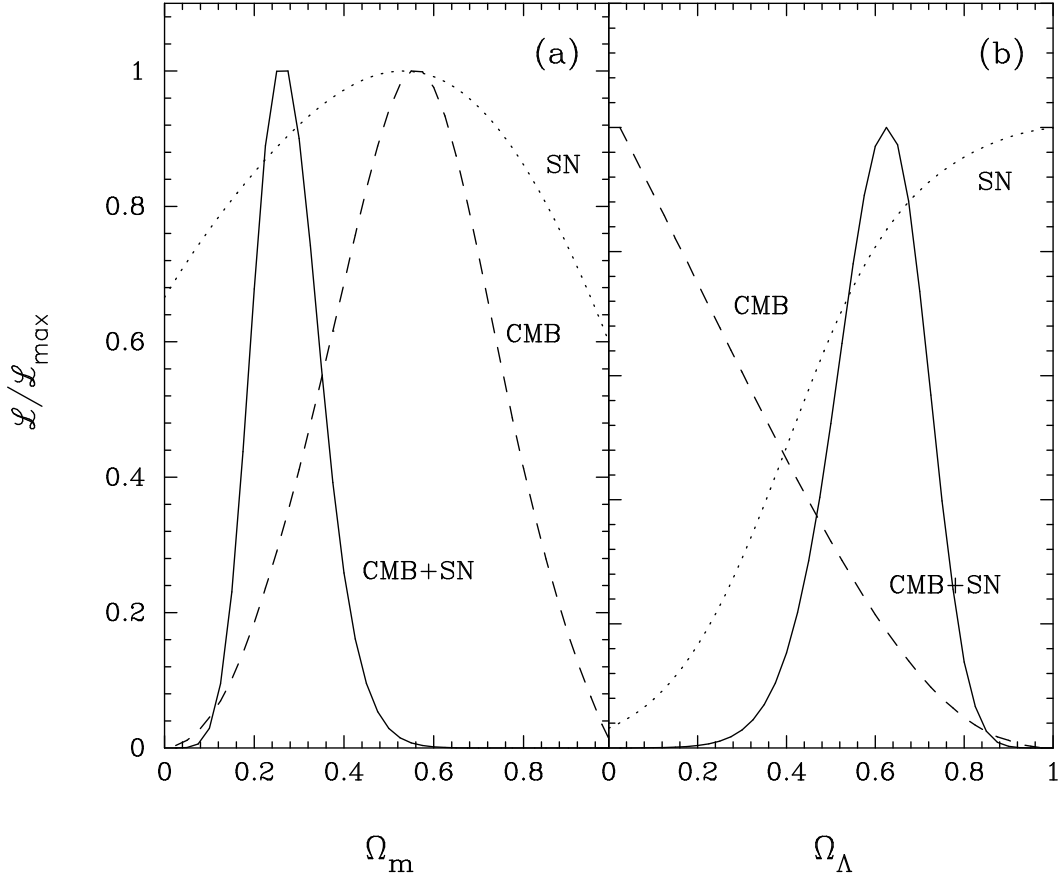}
%hscale=40,vscale=40,hoffset= -50,voffset=210}
\caption{\footnotesize 
%Likelihood contours in the $\o_m-\o_\l$ plane (left) are
%derived using a 
%combined likelihood analysis of CMB and supernovae data. These contours 
%show that the combined CMB+Sn likelihood function is strongly 
%peaked at $\o_m = 0.25$ and $\o_\l = 0.63$ thereby favouring a flat universe 
%(shown by the dotted-dashed straight line).
Marginalized
likelihood functions are shown for SN data alone (dotted lines)
CMB data alone (dashed lines) and the combined SN and CMB data (solid lines).
The CMB+SN likelihood function sharply peaks near $\o_m = 0.25$ and 
$\o_\l = 0.63$ favouring a flat universe with
$\o_m + \o_\l \simeq 1$.
More details may be found in Efstathiou et al. (1998).}
\bigskip
\hrule
\medskip
\label{fig:sn_cmb}
\end{figure}

%This degeneracy can however be broken if one appeals to the magnitude-redshift
%relation for high redshift type Ia supernovae surveyed in the last section
%which provide 
%constraints which in the $\o_\l-\o_m$ plane
%are almost orthogonal to those derived from the CMB 

\subsection{The Angular size - redshift relation.}
\label{sec:angle}
Another potentially sensitive test of models is related to the fact
that the angular size $\Delta\theta$ 
of an extended object $D$ located at a redshift $z$,
depends rather sensitively on the properties of the cosmological
model in which it
is being measured. Knowing the absolute size of an object (e.g. galaxy or
radio source) and the angle subtended by a distribution of such objects in the 
universe, it may be possible (after correcting for projection and evolution 
effects) to say something about the geometry of space and the matter content
of the universe.

It is easy to derive a relationship between $D$ and $\Delta\theta$.
Consider an object of proper length $D$ at a coordinate
distance $r$, and assume for simplicity that the object is aligned
along the $\theta$ axis so that coordinates marking its `top' and `bottom'
are respectively
$(r, \theta_1 + \Delta\theta_1, \phi_1)$ and $(r, \theta_1, \phi_1)$. 
The observer is at $r = 0$.
The proper length of the object can be obtained by setting
$t$ = constant  in the FRW line-element (\ref{eq:lam1a}) 
giving \cite{jvn_cup}
\beq
ds^2 = -D^2 = - a^2(t)r^2\Delta\theta_1^2.
\eeq
As a result we get the following expression for the angle 
subtended by the object at the location of the observer 
\beq
\Delta\theta = \frac{D}{d_A}
%{D\over a(t)r} = {D (1+z)^2\over d_L(z)}
\label{eq:angle2}
\eeq
where $d_A = a(t)r$ is the `angular-size distance'. Since 
$1 + z = a_0/a(t)$ one gets $d_A = d_L (1 + z)^{-2}$, where 
$d_L(z) = a_0r(1+z)$ is the luminosity distance discussed in the previous 
section.
Accordingly (\ref{eq:angle2}) may be rewritten as
\beq 
\Delta\theta = {D (1+z)^2\over d_L(z)}.
\label{eq:angle2a}
\eeq
%where the luminosity distance in FRW models is given in 
%(\ref{eq:age11d}). 
In Fig \ref{fig:angle}  we have plotted the angular size - redshift relation
for flat cosmological models with a cosmological constant. 
(We have used expressions (\ref{eq:age11d}) for the luminosity distance $d_L$ 
and
(\ref{eq:age6}) for the dimensionless Hubble parameter $h(z)$, assuming
a flat universe $\o_{total} = \o_m + \o_\l = 1$.)

\begin{figure}
\centerline{
\psfig{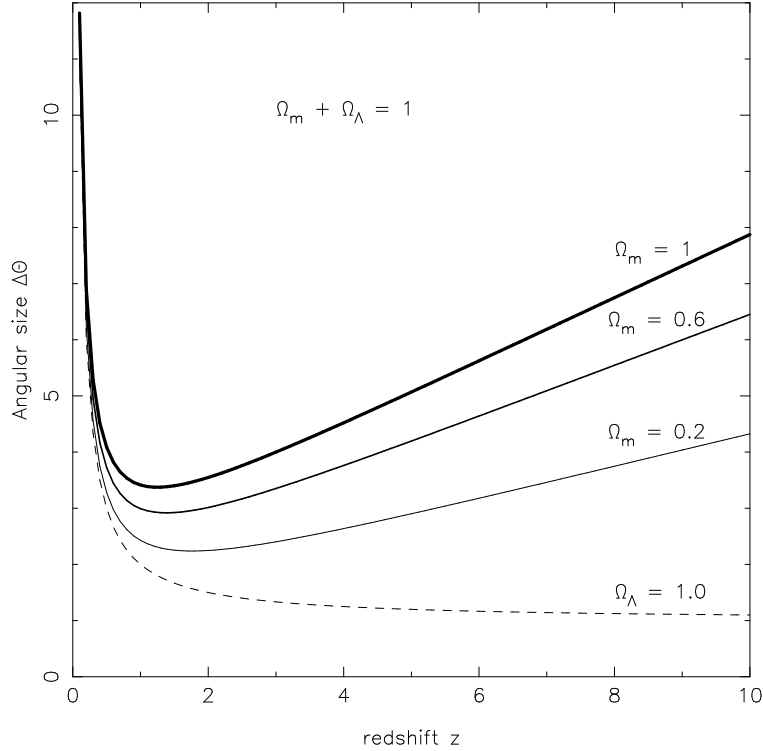}}
\caption{\footnotesize 
The angular size is shown as a function of cosmological redshift
$z$ for flat cosmological models with a cosmological constant $\Omega_m + 
\Omega_\Lambda = 1$. 
Heavier lines correspond to larger values of $\Omega_m$. For comparison
we also show (dashed line) the angular size in a de Sitter
universe ($\Omega_\Lambda = 1$).}
\bigskip
\hrule
\medskip
\label{fig:angle}
\end{figure}

We find that as the object is moved to higher redshifts its 
angular size first decreases (as naively expected) but soon 
begins to increase after passing through a minimum value.
The appearance of a minimum angular size at a given redshift $z_{\rm min}$
is a generic feature
of cosmological models with $\Omega_m > 0$. 
Differentiating (\ref{eq:angle2a}) with respect
to the redshift after substituting for $d_L$ from (\ref{eq:age12}),
and then setting $\delta\Delta\theta/\delta z = 0$ gives
$z_{\rm min} = 1.25$ for the flat matter dominated Einstein-de Sitter universe.
From figure \ref{fig:angle} we find that the location of 
the minimum angular size moves to higher redshifts as 
$\Omega_\Lambda$ is increased,
until in the limiting case $\Omega_\Lambda = 1$ there is no
minimum at all.
(Formally $z_{\rm min} \rightarrow \infty $ in de Sitter space, indicating that 
the angular size of an object decreases monotonically with redshift
without ever reaching a minimum value.)

The suggestion that angular sizes of galaxies could be used to discriminate 
between cosmological models was first made in \cite{hoyle59}.
Curiously the angular size of a typical galaxy at a redshift $z \sim 1$ is
roughly one arc second which is close to the limiting value of the 
angular resolution (`seeing') allowed by the Earth's atmosphere \cite{peeb93}.
Beyond $z \sim 1$ the angular size of an object increases, and if one is
confident that galaxies of a given class at higher redshifts are similar 
in form to their
lower redshift counterparts, then this test can in principle provide a 
powerful means of discriminating between world models especially with
the use of satellite data which can get around the `seeing' limit. 
Other (larger) objects which can be used to probe the angular size-redshift
relation include clusters of galaxies \cite{jvn_cup} and both extended and 
compact
radio galaxies \cite{kapahi89,cham90}. Extended radio sources which include
the twin radio lobes surrounding a radio galaxy can have sizes ranging from
a few kpc to $\sim 1000$ kpc, consequently the typical angular size of such
objects is $\sim 20$ arc seconds which can easily be measured using
ground based techniques. 

However, a word of caution must now be added, 
both clusters and radio galaxies
are prone to strong evolutionary effects which could lead to a change in size
over cosmological epoch. Thus a comprehensive understanding of physical
effects associated with both clusters (subclustering, virialization etc.)
and radio galaxies (evolution of radio lobes and the central engine etc.)
is necessary before
the  angular size-redshift relation
can be used to unambiguously determine cosmological parameters including
$\Omega_\Lambda$.

Recently Kellerman (1993) and Gurvits et al.
(1998) have
studied the angular sizes of
compact radio sources (QSO's and AGN's)
arguing that the central `engine' powering these objects is likely
to be controlled by a limited number of
 physical parameters (mass of central black
hole, accretion rate etc.) and may therefore be subject
to less evolutionary effects
than extended radio sources. On the basis of an analysis of a large number
of  sources spanning a wide redshift range $0.01 < z < 4.73$ these
authors
claim that an increase in the angular size has been detected
which is consistent with $\o_m \sim 1$.
(However working with the same data set as Kellerman (1993),
Kayser (1995) has shown that a significant $\l$ cannot be ruled out.)

%A complementary test which can be used to constrain
%closed world models containing unusual forms of matter such as a cosmological
%constant is related to the angular size in 
%closed cosmological models with antipodes.
An interesting feature of closed universes not present in the flat models 
considered in figure \ref{fig:angle}, or in open models,
is the presence of antipodal points. The presence of
antipodes can contribute to changing the
angular size as well as to the lensing of a source galaxy or quasar 
\cite{gott89}
and therefore provide us with a good means to constrain
closed cosmological models.
A metric describing the closed FRW universe is
\begin{equation}
ds^2 = c^2dt^2 - a^2(t)\lbrack d\chi^2 + \sin^2\chi(d\theta^2 + \sin^2\theta 
d\phi^2)]
\label{eq:angle2b}
\end{equation}
where $0 \le \chi, \theta \le \pi, 0 \le \phi \le 2\pi$. If we assume that 
the observer is located at $\chi = 0$ then the associated antipodal point is at 
$\chi_a = \pi$. Substituting $ds^2 = 0$
we obtain $\chi = \eta \equiv \int dt/a$, where $\eta$ is the conformal time
coordinate. In a matter dominated universe the form for the
expansion factor is $a(\eta)=A(1-\cos\eta)$, $ct=A(\eta-\sin\eta)$,
and the Hubble parameter is given by $H \propto \sin{\eta}/(1 - \cos{\eta})$
where $0 \leq \eta \leq 2\pi$. Thus a light ray from the antipodal point
$\chi_a = \pi$ reaches $\chi = 0$ at the time of maximum expansion $\eta = \pi$
 (corresponding to $H = 0$). Consequently in a matter dominated 
closed universe, light from an antipodal
 point can {\em never} reach an observer during the expanding phase 
(when $H > 0$). This situation changes when one considers a closed universe
with a cosmological term. In this case the universe is not obliged to 
recollapse and one can observe an antipodal point during the
expansion epoch.
The location of antipodal points can be derived from the following
considerations: 
from (\ref{eq:age11d}) we find for a closed
universe 
\beq
d_L(z) =  {(1+z) H_0^{-1}Sin(\eta_0 - \eta)\over
|\Omega_{total} - 1|^{\frac{1}{2}}}
\eeq
where 
\beq
\eta_0 - \eta = |\Omega_{total} - 1|^{\frac{1}{2}}\int_0^z {dz'\over
h(z')}
\eeq
since $d_A = d_L (1+z)^{-2}$ it follows that for 
\beq 
\eta_0 - \eta = n\pi, ~~ d_A \simeq 0 
\label{eq:antipode}
\eeq
%$d_A \simeq 0$ and
and, from (\ref{eq:angle2}), $\Delta\theta \rightarrow \infty$,
\ie the angular size of
an object located close to one of the antipodal points (\ref{eq:antipode})
can become very large. Consequently such an object will appear to us to
be extremely bright even if located at a high redshift !
The presence of `normal' galaxies
and quasars, as well as gravitationally lensed objects out to redshifts 
$\simeq 4.92$ set a lower limit on the antipodal 
redshift $z_a (\Omega_\Lambda, \Omega_m) > 4.92$ which can be used to
constrain the cosmological parameter pair $(\Omega_\Lambda, \Omega_m)$
 in a closed universe \cite{durrer90,lahav91,coop98}.
(Multiple images of a source object 
located further away than
the antipodal redshift $z_a$ are very difficult to form \cite{gott89}.)
Since the supernovae analysis prima-facie appears to favour a closed universe,
antipodal constraints may be used to further narrow down the allowed range in
parameter space. (The somewhat exotic possibility that gamma 
ray bursts may derive some 
of their excess luminosity from being located close to an antipode in a closed
Universe has been considered in \cite{saini99}.)

%\section{Clusters of galaxies and the value of $\o_m$.}
\subsection{Clusters of galaxies and the Large Scale Structure of the Universe}
\label{sec:clusters}

Observations of large scale structure 
indicate that the model which comes closest to
explaining most observational features of galaxy clustering is
$\l$CDM, a
model containing a cosmological constant in addition to baryons and cold
dark matter \cite{kgb93,os95}. 
Parameters of this model which agree well with observations
are $\o_\l$h$^2
\simeq 0.33$, $\o_b \simeq 0.02$, $\o_m \simeq 0.3$,
where h = $H_0/100$ is the Hubble parameter in units of 100 km/sec/Mpc.
(Setting $h = 0.7$ gives $\o_\l = 0.68$.)

There are several reasons as to why the presence of 
$\l$ improves the performance of the standard cold dark
matter model.
The first is related to the fact that in a spatially flat universe
linearized density perturbations
grow at a slower rate in the presence of  $\l$ than in its absence. 
(The growth
rate is however faster
than that in an open universe for identical values of $1-\o_m$.)
This changes the initial normalization of the density field since
the linearized gravitational potential now becomes time-dependent,
which affects the Sachs-Wolfe integral
discussed in section \ref{sec:mbr}.
The slow down in the rate of growth also affects the abundance of very
massive objects (clusters and superclusters) some of which may have
formed only relatively recently and would therefore feel the presence of 
long wavelength modes still in the linear regime \cite{cen98}.
A small value of $\o_m$ (alternatively, a large value of $\o_\l = 1-\o_m$)
also affects the matter 
power spectrum in $\l$CDM models which is strongly influenced by the
epoch of matter radiation equality. 
This effect is incorporated in the {\em shape parameter}\footnote{The 
shape parameter is so named because it affects the shape of the Power spectrum
$P(k)$, which interpolates between the asymptotic regimes \cite{sc95}
$P(k) \propto k$ for $k \rightarrow 0$ and 
$P(k) \propto k^{-3}\log^2{k}$ for $k \rightarrow \infty$. The maximum value
of $P(k)$ occurs near $k \sim d_{eq}^{-1}$.}
$\Gamma = \o_m h$: 
a small value of $\o_m$ leads to a larger value of the horizon at 
matter-radiation equality $d_{eq} \simeq 16 /(\Gamma h)$ Mpc
and hence to more long wavelength power in the fluctuation spectrum
$P(k) = \langle |\delta_k|^2 \rangle$. 
Both open models and $\l$CDM models show better agreement
with galaxy clustering data on large scales \cite{esm90}, the `best fit'
value of $\Gamma$ being
$\Gamma \simeq 0.25$.

An independent estimate of $\o_m$ is provided by the peculiar velocities of 
galaxies in our neighborhood (on scales $\sim 10-100$ h$^{-1}$ Mpc).
The results of a joint estimate from velocity flows and supernovae gives the
most likely values $\o_m \simeq 0.5$ and $\o_\l \simeq 0.8$, thereby
favouring an approximately flat universe \cite{dekel99}.

A low value of $\o_m$ is also indicated by studies of clusters of galaxies.
Clusters of galaxies have traditionally been powerful probes of cosmological
structure formation scenario's. 
The masses of rich clusters can be estimated using three independent methods:
the velocity dispersion
of member galaxies, the cluster X-ray temperature due to hot intracluster
gas and strong gravitational lensing of background galaxies by the cluster.
All three methods provide an estimate of the cluster mass which ranges from
$10^{14}$ to $10^{15}$ h$^{-1}$ M$_\odot$ for the mass located 
within the central $1.5$h$^{-1}$
Mpc. region of a cluster \cite{bahcall99}. 
The resulting median mass-to-light ratio for rich clusters is
$M/L_B \simeq 300 \pm 100$h $M_\odot/L_\odot$, which when
integrated over the full
range of luminous matter in the universe 
gives an estimate for the density parameter 
$\o_m = 0.2 \pm 0.1$. 

A low value of $\o_m$ is also indicated by a 
study of baryonic matter 
within clusters.
In a detailed study of the composition of the Coma cluster which included
estimates of the baryonic mass fraction provided by X-ray emitting gas 
and virial measurements of its total mass,
White et al (1993)
showed that the ratio of baryonic matter to total mass 
$\o_bh^{3/2}/\o_m = 0.07 \pm 0.03$. As a result the baryonic mass fraction
greatly exceeds nucleosynthesis constraints $\o_bh^2 = 0.015 \pm 0.005$
if $\o_m = 1$, leading to a `baryon catastrophe'. However no catastrophe
occurs if $\o_mh^{1/2} = 0.21 \pm 0.12$ 
since the value of $\o_b$ is now small enough to be acceptable by
nucleosynthesis constraints \cite{os95}. This result therefore is strongly 
supportive of either an open universe or one that is $\l$ dominated and flat,
so that $\o_m = 1 - \o_\l \ll 1$.

Observations of cluster abundances can be used to
provide good estimates of $\sigma_8$ --
the average root-mean-square mass fluctuation in a sphere of
radius $8$h$^{-1}$ Mpc. The best-fit value of $\sigma_8$ consistent with
present day cluster abundances is $\sigma_8 \simeq 0.5 ~\o_m^{-0.5}$.
This value gives a measure of the clustering amplitude on small scales
and therefore 
can be used to normalize the power spectrum of density perturbations.
A complementary method of normalization is provided by large angle
CMB anisotropies measured by COBE. Taken together the $\sigma_8$ normalization
on small scales and the COBE normalization on large scales ($\sim 1000$ Mpc.)
provide very useful constraints on the cosmological parameters $\o_m, \o_\l,
\o_B$, on the biasing parameter $b = \delta_{lum}/\delta_{dark}$  and on the 
`primordial tilt' in the power spectrum $|\delta_k|^2 \propto
k^n$ which can be shown to lie in the range $|1-n| \lleq 0.2$ \cite{os95}.

\begin{figure}
\centerline{
\psfig{file=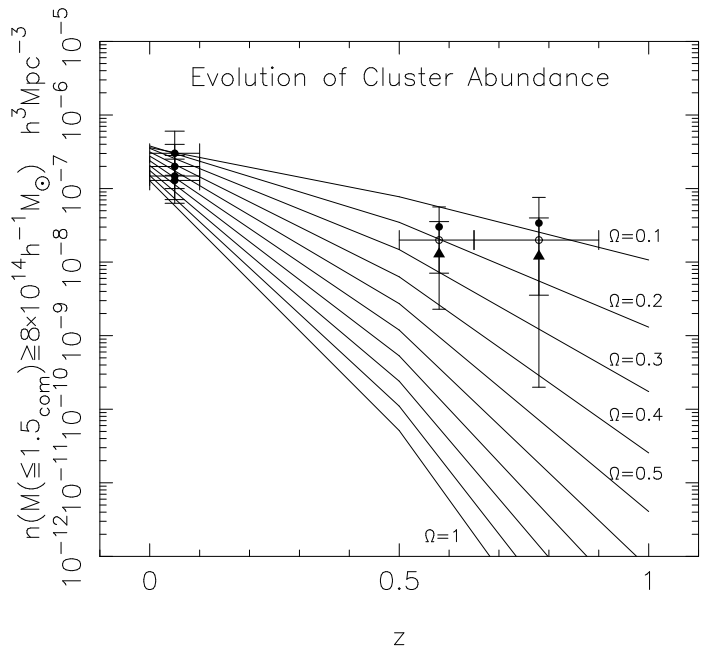,width=10cm}}
\caption{\footnotesize The observed and expected cluster abundance is shown as a 
function of redshift for massive clusters with $M_{cl} \ggeq 8\times 10^{14} 
M_\odot$ located within the Abell radius of $1.5$h$^{-1}$Mpc.
The curves show the expected cluster abundances in CDM models
with different $\o_m$. Figure courtesy of Neta Bahcall (1999).}
\bigskip
\hrule
\medskip
\label{fig:clusters}
\end{figure}

A potentially powerful method for discriminating between different cosmological  
models is provided by the abundance of rich clusters of galaxies measured
at high redshifts. 
The presence of large amounts of X-ray emitting gas in many rich clusters
provides us with a useful observational tool with which
to probe cluster mass.
Observations of galaxy clusters are then 
matched against theoretical models which model cluster formation and evolution
using either Press-Schechter techniques or N-body/Hydro-simulations
\cite{ob92,henry97,fbc97,bb97,cmv97,bahfan98,eke98,vl98}.
As discussed earlier the growth of long wavelength
perturbations which are still in the linear regime, is significantly slower in 
low density models (both with and without a cosmological constant) than in a
critical density $\o_m = 1$ universe. This leads to dramatic differences in
the redshift dependence of the rich cluster abundance in cosmological models:
rich clusters are much rarer
at high redshifts in an $\o_m = 1$ universe than they are in a low density
universe (see figure \ref{fig:clusters}). For instance,
whereas almost all massive clusters with $M \sim 10^{15}M_\odot$ 
are expected to have formed by $z \sim 0.5$ in a low density universe,
only a small fraction ($< 10\%$)
of the present day $10^{15}M_\odot$ clusters would
have been in place by $z \sim 0.5$ in an $\o_m = 1$ 
universe \cite{gioia97,cen98,vl98}. 
The existence of three massive clusters in the 
redshift range $z \sim 0.5 - 0.9$
has therefore been viewed as a difficulty for
the standard cold dark matter model with $\o_m = 1$ 
for which $10^{-3}$ rich clusters are expected at $z > 0.5$ 
\cite{gioia97,bahfan98,bahcall99}.
It must be noted however that large uncertainties in
both the observational data (only a few very massive clusters have been 
reliably observed at high $z$) and in our theoretical understanding of rich
clusters, makes it difficult at present to place unambiguous constraints 
on the values
of $\o_m$ and $\o_\l$ \cite{vl99}. 
It is hoped that better quality data from satellite
launches planned for the immediate future (XMM) and more accurate modelling of 
large scale structure will improve the situation significantly in the near 
future.

Constraints on the abundance of rich clusters
also come from {\em arcs} caused
by the strong gravitational lensing of extended background sources (galaxies,
radio sources) by foreground clusters.
Since clusters act as gravitational lenses for background sources, the
larger number of clusters at early epochs in (i) open,
low $\Omega_m$ models, and (ii) flat, high $\Omega_\Lambda$ models, relative to
(iii) flat $\Omega_m = 1$ models leads to a greater abundance of arcs in
both (i) and (ii) relative to (iii).
An estimate by Bartelmann et al. (1998) based on numerical
simulations of large scale structure, has shown that
an order of magnitude more arcs are predicted in flat models with
$\Omega_m \simeq 0.3$, $\Omega_\Lambda \simeq 0.7$ (${\cal N}_{\rm arcs}
\sim 280$) than in the flat $\Omega_m = 1$ model (${\cal N}_{\rm arcs}
\sim 36$). In open models this effect is even more dramatic
(${\cal N}_{\rm arcs} \sim 2400$ for $\Omega_m \simeq 0.3, \Omega_\Lambda = 0$).
However, impressive as these results are,
the absence of a comprehensive data base for arcs
and uncertainties
in the modelling of galaxy clusters
makes it difficult to attempt to constrain theoretical models
on the basis of observations at present. (Bartelmann et al. (1998)  
however make a case for
a low density universe by arguing that the observed number of arcs in the 
EMSS arc survey extrapolated to the full sky is 1500 - 2300, which is close to
what one observes for low density models in their numerical simulations.)
Both observational data sets and the theoretical modelling of clusters
are likely to improve significantly in the near future
giving this method potentially great importance in the ongoing
`quest for $\Lambda$'.

Finally, the Lyman-$\alpha$ forest which populates the spectra of quasars
provides a potentially powerful means of discriminating between rival models
of structure formation and in probing the presence of a cosmological $\l$-term
at intermediate redshifts $0 \leq z \leq 5$ \cite{cen98,dwein98,hui98}.

\section{Theoretical issues: Vacuum fluctuations and the Cosmological constant}
\label{sec:lambda2}
Having summarised the observational evidence for a cosmological $\l$-term
let us now turn our attention to some theoretical implicationsi of $\l$.

A turning point in our understanding of the cosmological constant occurred when
Zeldovich (1968), intrigued by $\Lambda$-based
cosmological models presented to explain
an excess of quasars near redshift $\sim 2$, showed
that zero-point vacuum
fluctuations must have a
Lorentz invariant form $P_{vac} = -\rho_{vac}c^2$,
equivalently $T_{ik}^{vac} = \Lambda g_{ik}$,
\ie the vacuum within the quantum framework
had properties identical to those of a
cosmological constant.

Let us review this situation beginning
with an oscillator consisting of a single 
particle of mass $m$ moving under the influence of a potential $V = \half kx^2$.
At the classical level one expects the lowest energy state to be associated with
the particle at rest at $x = 0$, so that the total energy vanishes: 
$E = T + V = 0$. 
Thus, within the classical framework, the vacuum can be viewed as 
a state having zero energy and momentum.
However when viewed in terms of quantum mechanics the situation changes,
the uncertainty relation preventing the particle (wave function) from 
simultaneously
having a fixed location ($x = 0$) and a fixed velocity ($T = 0$). As a result,
the ground state energy of the oscillator is finite and is given by 
$E = \half \hbar\omega$, where $\omega = k/m$. 
Turning now to quantum theory, it is
well known that after secondary quantization
a classical field can be looked upon as an ensemble of oscillators each with
frequency $\omega(k)$. The net `zero-point energy' of this field is 
$E = \sum_k\half \hbar\omega(k)$. 
Thus the uncertainty
relation endows the vacuum with both energy and 
pressure!

The existence of zero-point
vacuum fluctuations
has been spectacularly demonstrated by the Casimir effect.\footnote{The Casimir 
effect arises because vacuum fluctuations satisfy the 
quantum mechanical wave equation and hence are sensitive to boundary conditions. 
As shown by Casimir (1948) the presence of two flat parallel conducting plates
at a separation $l$, alters the distribution of electromagnetic 
field modes existing in the vacuum,
resulting in  an attractive force per unit area between
the plates: $F = - \hbar c\pi^2/240 l^4$ which is of vacuum origin.
The Casimir effect has been experimentally measured by Spaarnay (1957) 
and others \cite{bd82,tabor69,wein89,milton98}.}
The vacuum energy associated with zero-point fluctuations
is formally infinite and results
in a `cosmological constant problem' for the universe \cite{wein89}. 
Because of the importance of this result we shall
%we shall verify the 
%`reality' of zero-point fluctuations by 
perform a simple calculation aimed at evaluating
the zero-point energy associated with a quantized scalar field in flat 
space-time. (The reader is referred to Birrell \& Davies 1982 for a discussion 
of 
quantization of higher spin fields.)

Consider the action defined in flat four dimensional space-time 
\footnote{Following Landau \& Lifshitz (1975) we use Latin
indices to describe space-time coordinates, so that i,j,k... = 0,1,2,3;
and Greek indices to describe spatial coordinates:
$\alpha,\beta,\gamma ..$ = 1,2,3.}
\begin{equation}
{\cal{S}} = \int{\cal L}(x)d^4x
\label{eq:c1}
\end{equation}

where ${\cal L} (x)$ is the Lagrangian density for a massive scalar field
\begin{equation}
{\cal L} = \half(\eta^{ij}\Phi_{,i}\Phi_{,j} - m^2\Phi^2)
\label{eq:c2}
\end{equation}
propagating in flat space-time with metric $\eta_{ij}$.

The variational principle $\delta_\Phi {\cal S} = 0$ gives the Klein-Gordon
equation

\begin{equation}
(\sq + m^2)\Phi = 0
\label{eq:c3}
\end{equation}
where $\sq \equiv \eta^{ik}\partial_i\partial_k$.

%To quantize the system we treat the field $\Phi$ as an operator 
%\begin{equation}
%\Phi(x) = \int d^3k\left[a_k\phi_k({\bf x},\eta)
%+ a_k^{\dagger} \phi_k^*({\bf x},\eta)\right]
%\label{eq:c4}
%\end{equation}
%where $a_k,~a_k^{\dagger}$ are annihilation and creation operators
%$[a_k,a_{k'}^{\dagger}] = \delta_{kk'}$, defining the vacuum state
%$a_k|0\rangle = 0$ , \forall k$.
%An orthonormal set of solutions to (\ref{eq:c3}) is
%\begin{equation}
%\phi_k = \frac{1}{\sqrt{(2\pi)^3 2\omega_k}}\exp{(i{\bf k}{\bf x} - i\omega_k 
%t)}
%\label{eq:c5}
%\end{equation}
%where $\omega_k^2 = k^2 + m^2$.
%
%We shall find it more convenient to work with the discrete analogue of 
%(\ref{eq:c5}) in which the field modes are defined using 
%periodic boundary conditions on a three dimensional
%torus of side $L$ so that
To quantize the system we treat the field $\Phi$ as an operator
\beq
\Phi(x) = \sum_{\bf k}\lbrack a_k\phi_k({\bf x},\eta)
+ a_k^{\dagger} \phi_k^*({\bf x},\eta) \rbrack
\label{eq:c4}
\eeq
where $a_k,~a_k^{\dagger}$ are annihilation and creation operators
$[a_k,a_{k'}^{\dagger}] = \delta_{kk'}$, defining the vacuum state
$a_k|0\rangle = 0$, $\forall k$.
An orthonormal set of solutions
defined using periodic boundary conditions on a three dimensional
torus of side $L$ is given by \cite{bd82}
\ber
\phi_{\bf k} &=& \frac{1}{\sqrt{2L^3\omega}}\exp{(i{\bf k}{\bf x} - i\omega_k 
t)}\nonumber\\
k_j &=& \frac{2\pi n_j}{L}, ~~~ n_j \in I
\label{eq:c5}
\eer
where $\omega_k^2 = k^2 + m^2$, and 
the field modes have been normalized using 
\beq
(\phi_{\bf k}, \phi_{\bf k'}) = \delta_{{\bf k}{\bf k'}} 
\eeq
where
\beq
(\phi_1,\phi_2) = -i\int\lbrack\phi_1\partial_t\phi_2^* - 
\phi_2^*\partial_t\phi_1\rbrack d^3x.
\eeq

Consider next, the energy-momentum tensor
\begin{equation}
T_{ij} = \Phi_{,i}\Phi_{,j} - \half\eta_{ij}\eta^{kl}\Phi_{,k}\Phi_{,l}
+ \half m^2\Phi^2\eta_{ij}
\label{eq:c6}
\end{equation}
where $T_{00}$ defines the energy density
\beq
T_{00} = \half ({\dot\Phi}^2 + \partial_\mu\phi\partial^\mu\Phi + m^2\Phi^2)
\eeq
and $T_{0\alpha}$ the momentum density
\beq
T_{0\alpha} = \frac{\partial\Phi}{\partial t}\frac{\partial\Phi}{\partial 
x^\alpha}, ~~~\alpha = 1,2,3.
\eeq
Substituting from (\ref{eq:c4}) \& (\ref{eq:c5}) into (\ref{eq:c6}) one obtains 
for the Hamiltonian $H$
\begin{equation}
H \equiv \int T_{00} d^3x = \half\sum_{\bf k} 
(a_{\bf k}^{\dagger}a_{\bf k} + a_{\bf k} a_{\bf k}^{\dagger})\omega_{\bf k}
\label{eq:c7}
\end{equation}
which can be further simplified using the commutation relation 
$[a_k,a_{k'}^{\dagger}] = \delta_{kk'}$ to
\begin{equation}
H = \sum_{\bf k}(a_{\bf k}^\dagger a_{\bf k} + \half)\omega_{\bf k}.
\label{eq:c7a}
\end{equation}

A similar operation on the momentum density yields \cite{bd82}
\beq
P_\alpha \equiv \int T_{0\alpha} d^3 x = \sum_{\bf k} a_{\bf k}^{\dagger}a_{\bf 
k} k_\alpha, ~~~ \alpha = 1,2,3.
\label{eq:c8}
\eeq
Inspecting expressions (\ref{eq:c7}) and (\ref{eq:c8}) for
the Hamiltonian $H$ and the momentum operator $P_\alpha$ we find, for the 
expectation value of
these quantities in the vacuum state $|0\rangle$ 
\begin{equation}
\langle 0|{\bf P}|0\rangle = 0, ~~~~~~~~~~~~~~ \langle 0|{H}|0\rangle = 
\half \sum_{\bf k} \omega_{\bf k}.
\label{eq:c9} 
\end{equation}

Transforming the sum $\sum_k$ to an integral we get 
\beq
\half \sum_{\bf k} \omega_{\bf k} = \half (\frac{L}{2\pi})^3\int\omega({\bf k})
d^3 k = \frac{L^3}{4\pi^2}\int_0^\infty\sqrt{k^2 + m^2}~k^2 dk.
%\half \sum_{\bf k} \omega_{\bf k} = \half
%c~(\frac{L}{2\pi\hbar})^3\int\omega({\
%bf k})
%d^3 k = \frac{c~L^3}{4\pi^2\hbar^3}\int_0^\infty\sqrt{k^2 + m^2c^2}~k^2
%dk
\label{eq:c9b}
\eeq
%the resulting energy density of the vacuum is
%\beq
%\epsilon_{vac} \equiv \langle T_0^0\rangle_{vac} 
%= \frac{c}{4\pi^2\hbar^3}\int_0^\infty\sqrt{k^2 + m^2c^2}~k^2 dk
%\label{eq:c9bb}
%\eeq
From (\ref{eq:c9}) \& (\ref{eq:c9b}) we find that the energy density of 
zero-point vacuum fluctuations is dominated by
ultraviolet divergences which diverge as $k^4$ when $k \rightarrow \infty$.
The vacuum state therefore has zero momentum and infinite energy !
(In terms of Feynman diagrams the energy density of zero-point fluctuations 
is associated with a one-loop vacuum graph, see figure (\ref{fig:loop}).)

Within the framework of Newtonian gravity and either classical or quantum 
mechanics,
an infinite (or very large)
vacuum energy does not cause serious problems since interaction 
between particles is governed not by the absolute value of the potential 
energy $V$, but
by its gradient $\nabla V$. 
As a result one can always redefine $V' \rightarrow V + V_0$
so that the minimum of $V'$ has zero net energy. The situation changes 
dramatically
when we view the vacuum within the framework of general relativity. 
A central tenet of the general theory of relativity is that the gravitational
force couples to {\em all} forms of energy through the Einstein equations
$G_{ik} = \frac{8\pi G}{c^4}T_{ik}$. Therefore if the vacuum has energy then 
it also
gravitates ! In order to probe this effect further one needs to know the
equation of state possessed by the vacuum energy, equivalently  the form of its
energy momentum tensor $T^{ik}_{\rm vac}$.
This question was answered by
Zeldovich (1968) who showed that the vacuum state 
had to have a Lorentz-invariant form, one that was left unchanged by 
a velocity transformation and hence appeared the same to all observers. 
This requirement is exactly satisfied by the 
equation of
state $P = -\rho$ possessed by the cosmological constant, since the relation
$T_{ik} = \Lambda g_{ik}$ is manifestly Lorentz-invariant.
\footnote{Zero-point fluctuation
s are usually regularized by `normal ordering' -- a rather ad hoc procedure
which involves the substitution $a_ka_{k}^{\dagger} \rightarrow a_{k}^{\dagger}a
_k$ in (\ref{eq:c7}). In curved space-time a single regularization is not
enough to rid  $\langle T_{ik}\rangle$ of all its divergences. 
Three remaining `infinities' must be
regularized, leading to the renormalization of additional terms in
the one-loop effective Lagrangian for the gravitational field, which, in an FRW
universe becomes:
${\cal L}_{\rm eff} = \sqrt{-g}
\lbrack \l_\infty + R/16\pi G_\infty + \alpha_\infty R^2 +
\beta_\infty R_{ij}R^{ij}\rbrack$. Renormalization of the first term
$\l_\infty \rightarrow 0$ corresponds to
normal ordering. The presence of the second term $R/16\pi G_\infty$, led Sakharo
v to postulate that the gravitational field might be `induced' by one-loop
quantum effects in a curved background geometry, since one could recover
the ordinary Einstein action by renormalizing the `bare' value $G_\infty$ to its
observed value: $G_\infty \rightarrow G_{\rm obs}$
\cite{sakharov68}. Thus {\em both} the cosmological constant $\l$ and
the gravitational constant $G$ may be induced by quantum effects.
The remaining two terms in
${\cal L}_{\rm eff}$ give rise to vacuum polarization effects and have been
extensively discussed in the literature \cite{bd82,gmm80}.}

All fields occurring in nature contribute an energy density to the vacuum and
expressions analogous to (\ref{eq:c9}) for bosons can also be derived for
fermions. Since fermionic 
creation, annihilation operators anti-commute this leads to
\beq
\langle 0|{H_f}|0\rangle = - \half \sum_{\bf k} \omega_{\bf k}.
\label{eq:c10}
\eeq
Comparing (\ref{eq:c10}) with (\ref{eq:c9}) we find that the 
zero-point energy of fermions is equal and opposite 
to that of bosons (having identical mass).

The advent of supersymmetry in the 1980s, incorporating a 
fundamental symmetry between bosons and fermions, led to the hope that the
cosmological constant problem would finally be resolved, since the one-to-one
correspondence between bosons and fermions in such theories was expected to
lead to
cancellation between bosonic and fermionic infinities \cite{zumino75}. 
However supersymmetry
is expected to exist only at very high energies/temperatures. 
At low temperatures
such as those existing in the universe today, supersymmetry is broken.
One might therefore expect the cosmological constant to vanish in the early 
universe only to reappear later 
%at the time of SUSY breaking so that
%$\rho_{\rm SUSY} \sim M_{\rm SUSY}^4 \sim (10^3 {\rm GeV)^4$.
%It is interesting that on the logarithmic scale the value of $\rho_{\rm SUSY}$
%lies midway between the Planck value $\rho_{Pl} \sim (10^{18} GeV)^4$ and
%the observed value of the vacuum energy $\rho_0 \sim (10^{-3} eV)^4$.
%This might indicate that the present value of the cosmological constant
%is provided by a theory in which the effective energy scale of the vacuum 
%was given by $M_{\rm SUSY}^2/M_{Pl} \sim 10^{-3} eV$.
, when the universe has cooled sufficiently.
For instance the QCD vacuum is expected to generate a cosmological constant
of the order of $\Lambda_{QCD}^4 \sim 10^{-3}$GeV$^4$ which,
though considerably smaller than the Planck scale $\sim 10^{76}$GeV$^4$,
 is still many
orders of magnitude larger than the observed value $\rho_{vac} \sim 10^{-47}$
GeV $^4$. 
Thus the
cosmological constant problem re-emerges to haunt the present epoch !

Although the cosmological constant problem remains unresolved, an important
aspect of Zeldovich's work was that it demonstrated 
a firm physical mechanism for the generation of a
%which could give rise to a 
cosmological constant. Later work, mostly associated with 
inflationary model-building, further strengthened this idea by
showing that an effective cosmological
constant could arise due to
diverse physical processes including symmetry breaking,
vacuum polarization in curved space-time, higher dimensional `Kaluza-Klein' 
theories etc. Some of these developments have been reviewed in \cite{linde90}.
%book and in section \ref{sec:inflation}.

\section{The cosmological constant and spontaneous symmetry breaking}
\label{sec:ssb}

An important development
in our understanding of the `vacuum energy'
was associated with the phenomenon
of symmetry breaking in the electroweak Weinberg-Salam model. 
Consider the
scalar field action
\beq
S = \int{\sqrt{-g}{\cal L}} d^4x
\eeq
where ${\cal L}$ is the Lagrangian density
\beq
{\cal L} = \half ~g^{ij}\partial_i\phi\partial_j\phi - V(\phi)
\label{eq:inf2a}
\eeq
and the scalar field potential has the form
\beq
V(\phi) = V_0 -\half \mu^2\phi^2 + \frac{1}{4}\lambda\phi^4.
\label{eq:inf2}
\eeq

This particular form of the potential (illustrated in Fig. \ref{fig:ssb})
endows the system with some interesting properties. For instance since the
symmetric state $\phi = 0$ is unstable
($V''(\phi) < 0$) 
the system settles in the ground state $\phi = +\sigma$
or $\phi = -\sigma$, where $\sigma = \sqrt{\mu^2/\lambda}$ thus breaking the
reflection symmetry $\phi \leftrightarrow - \phi$ present in the Lagrangian.
The energy momentum tensor $T_{ik}$ of a scalar field with Lagrangian density
${\cal L}$
is given by
\begin{equation}
T_{ik} = \phi_{,i}\phi_{,k} - g_{ik}{\cal L}.
\label{eq:inf4}
\end{equation}
Assuming $\phi$ to be homogeneous and time-independent one finds the ground 
state energy-momentum tensor to be
\beq
T_{ik} = g_{ik} V(\phi = \sigma),
\eeq
the vacuum state therefore has precisely
the form of an effective cosmological constant
$T_{ik} = g_{ik} \l_{eff}$ where
$\l_{eff} = V(\phi = \sigma) = V_0 - \mu^4/4\lambda$.
Setting $V_0 = 0$, results in a {\em negative} cosmological term
$\l_{eff} = - \mu^4/4\lambda$.
Substituting parameters arising in the electroweak theory results in a
lower limit on the value of the vacuum energy density \cite{wein89}
$\rho_{vac} = |\l_{eff}|/8\pi G = 10^6{\rm GeV}^4$, which is almost $10^{53}$ 
times
larger than current observational upper limits on the cosmological constant
$\rho_{vac,0} = \l_0/8\pi G \sim 10^{-29}{\rm g/cm}^3 \simeq 10^{-47}{\rm GeV}^4
$.
Clearly in order not to violate observational bounds today, one must set
$V_0 \simeq \mu^4/4\lambda$ so that $\l_{eff} \sim
\l_0$. An interesting feature of this `regularization' of the cosmological
constant is that, while drastically reducing the value of
the cosmological constant today
it simultaneously generates a large cosmological constant
$\sim V_0$ during an early epoch
before symmetry breaking, thereby giving rise to the possibility of 
inflation !
The cosmological constant problem therefore 
presents us with a dilemma: it is certainly good to have a large cosmological 
constant during an early epoch so as to resolve -- via inflation -- the
horizon and flatness problems and possibly generate seed
fluctuations for galaxy formation. However one must
simultaneously ensure that the value of $\l$ today is small
so as not to conflict with observations. As we have seen,
in models with SSB this dual requirement
of `{\em large $\l$ in the past + small $\l$ at present}' results
in an enormous fine tuning of initial conditions.

\begin{figure}
\centerline{
\psfig{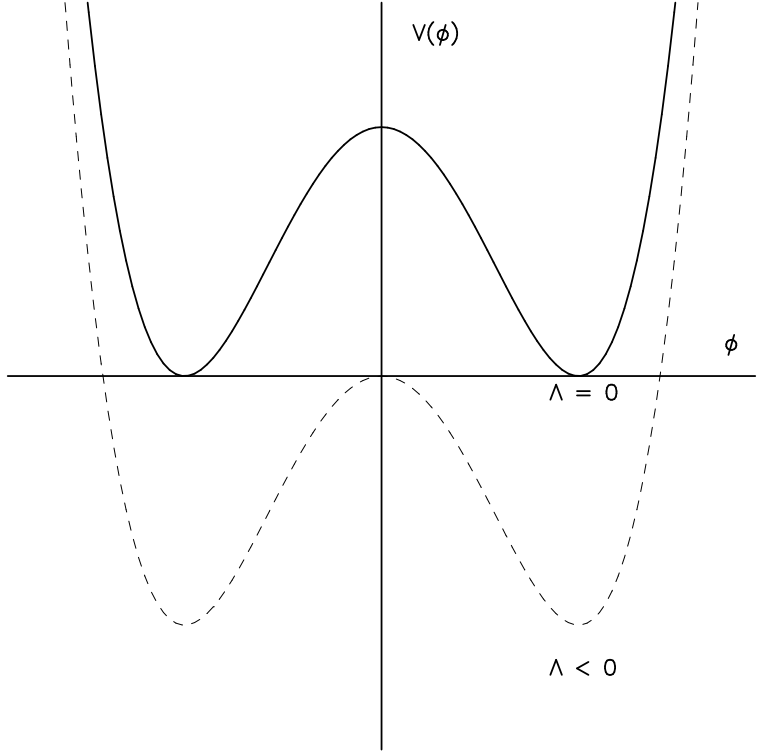}}
\caption{\footnotesize 
The `Mexican top-hat' potential describing spontaneous symmetry 
breaking shown: before (dashed) and after (solid) the 
cosmological constant has been `renormalized'.}
\bigskip
\hrule
\medskip
\label{fig:ssb}
\end{figure}

\section{Mechanisms for generating a small current value of $\l$.}
\label{sec:small}

As we saw in the last two sections, 
the vacuum associated with both one-loop quantum effects and models with
spontaneous symmetry breaking, 
has properties identical to those of a 
cosmological constant. 
There is one problem however, in the case of zero-point fluctuations, the
vacuum density turns out to be
infinite leading to an
infinitely large cosmological term and
resulting in a {\em cosmological constant problem} for cosmology
(see section \ref{sec:lambda2}). Assuming that the ultraviolet divergences
responsible for the cosmological constant problem can be
cured by (hitherto unknown) physics occurring near the Planck scale,
one gets a finite but very large value $$\rho_\Lambda = 
\Lambda c^2/8\pi G \simeq \rho_{Pl} = c^5/G^2\hbar \sim 5\times 10^{93}~
\hbox{\rm g~cm}^{-3},$$ 
where $\rho_{Pl}$ is the Planck density.
%which is almost 123 orders of magnitude
%larger than observational upper limits on $\l$.
On the other hand, as we saw earlier,
recent observations of the luminosities of high redshift
supernovae combined with CMB results give the 
following value for the dimensionless density in $\l$
$$\o_\l = \rho_\l/\rho_{cr} \equiv \frac{\l c^2}{3 H_0^2} \simeq 0.7$$ 
where $\rho_{cr} = 3H^2/8\pi G = 1.88\times
10^{-29}$h$^2$ g/cm$^3$ (see sections \ref{sec:sn} \& \ref{sec:mbr}), 
which leads to $\rho_\l \simeq
\rho_{Pl}\times 10^{-123}$, \ie the value of the cosmological constant
today is almost 123 orders of magnitude smaller than 
the Planck density !

As we have shown in section \ref{sec:ssb}, 
a large (negative) value of the vacuum energy also arises in models with
spontaneous symmetry breaking. In this case, the fine tuning involved 
in matching the present value of $\l$ to observations depends upon
the symmetry breaking scale, and ranges from 1 part in $10^{123}$ for the
Planck scale, to 1 part in $10^{53}$ for the electroweak scale.

Clearly the question begging an answer is:
which physical processes can generate a small value for 
$\l$ today
without necessarily involving a delicate fine tuning of initial
conditions?
Although no clear cut answers are available at the time of writing
(it may even be that a very small $\l$ may demand completely new
physics) some avenues which could lead us to interesting
answers
will be
explored in this section.

\subsection{A Decaying Cosmological Constant?}
\label{sec:decay}

One method of resolving the dilemma between a very large cosmological
constant (predicted by field theory) and an extremely small one
(suggested by observations)
with obvious cosmological advantages
is to make the cosmological term time-dependent.
An initially large
cosmological term would give rise to
inflation, ameliorating
the horizon and flatness problems and (possibly) seeding
galaxy formation. The subsequent slow decay of $\Lambda (t)$ would enable
a small present value
$\Lambda(t_0)$ to be reconciled with observations
suggesting $\Omega_\l \sim 0.7$.
(A time dependent cosmological term of course arises in inflationary models
and during cosmological phase transitions, but in such cases the 
post-inflationary decay of the cosmological term is very rapid.)

The first proposal for dynamically reducing the cosmological constant
was made by Dolgov (1983) who considered
a massless non-minimally coupled scalar field having the Lagrangian density
\beq
{\cal L} = \half (\phi^{,l}\phi_{,l} -\xi R\phi^2)
\eeq
and the resulting equation of motion
\beq
\sq\phi + \xi R\phi = 0,
\label{eq:c110}
\eeq
where $R$ is the scalar curvature and $\xi$ the coupling to gravity.
Considering the special case of a homogeneous scalar field,
the Einstein equations become
\beq
3H^2 = \Lambda + 8\pi G(\rho_\phi + \rho_{matter})
\label{eq:c12}
\eeq
where
\beq
\rho_\phi = \half{\dot\phi}^2 + 3\xi H^2\phi^2 + 6\xi H\phi\dot\phi
\label{eq:c11}
\eeq
is the scalar field energy density.
The scalar field equation (\ref{eq:c110}) reduces to
\beq
\ddot{\phi} + 3\frac{\dot a}{a}\dot{\phi} + 6\xi~\bigg\lbrack \frac{\ddot a}{a}
+(\frac{\dot a}{a})^2\bigg\rbrack\phi = 0.
\label{eq:c13}
\eeq
Dolgov made the discovery that, for negative values of $\xi$,
the scalar field is unstable: its
energy density
$\rho_\phi$ becomes large and  negative compensating for the
cosmological constant in (\ref{eq:c12}), so that the resulting {\em
effective} cosmological constant
rapidly decays to zero. Let us demonstrate this by examining
the Einstein equation (\ref{eq:c12}) which together with the scalar field
equation (\ref{eq:c13})
defines a pair of nonlinear differential equations determining
the behaviour of the scale factor $a(t)$ and the scalar field $\phi (t)$.
The term $3\xi H^2\phi^2$ in $\rho_\phi$ can be carried over into the 
left hand side of 
(\ref{eq:c12}) resulting in
\beq
3H^2 \simeq \frac{\Lambda}{1 - 8\pi G\xi\phi^2} + ...
\label{eq:c14}
\eeq
As Dolgov demonstrated,
$\phi(t)$ grows with time if $\xi < 0$, so that
the effective cosmological constant
$\Lambda_{eff} = \Lambda/(1 + 8\pi G|\xi|\phi^2)$ decreases.
The late time behaviour of $a(t)$, $\phi(t)$ obtained
by solving (\ref{eq:c12} - \ref{eq:c13}) with $\rho_m \ll \rho_{\phi}$
has the asymptotic form \cite{dolgov83,ford87}
$$a \propto t^q, ~q ={1\over 2}+{1\over 4|\xi|}, ~\phi \propto t.$$
As a result, $\lim_{t \rightarrow \infty} \Lambda_{eff} \rightarrow 0$
\ie the cosmological term vanishes at late times.

Unfortunately this mechanism cannot be used in real universe.
The first problem with this approach is that the very mechanism which 
decreases the cosmological constant also quenches the effective gravitational 
constant, since from (\ref{eq:c12}),
\beq
G_{eff} = \frac{G}{1 - 8\pi G\xi\phi^2} \rightarrow 0 ~~~ {\rm as} ~~
t \rightarrow \infty .
\eeq
As a result, the effective gravitational constant becomes noticeably
time-dependent: $\dot G_{eff}/G_{eff} = -2/t \sim - 10^{-10}$ yr$^{-1}$,
which strongly contradicts upper limits from Viking radar ranging~
\cite{hel83} and lunar laser ranging experiments~\cite{wnd96}.
Another problem is that such screening of $\l$ is still not sufficient.
The remaining part of $\l$ remains of the order of the Ricci tensor
all the time, while we need it to be much less than the Ricci tensor
during the matter dominated epoch to obtain sufficient growth of
scalar perturbations. Finally, $\Omega_m \ll 1$ during this regime. 

An extension of this method to higher spin fields (massless 
vector and tensor) can remove the first drawback by making a cancellation of
the cosmological constant possible while keeping the
gravitational coupling constant time-independent \cite{dolgov97}.
However, the other difficulties (especially, the second one) remain.
This shows that it is not easy to explain the observed $\l$-term
by a cancellation mechanism. Still this aesthetically attactive
possibility should be investigated further (some variants of the early 
Dolgov mechanism are discussed in \cite{barr87,wein89}).

\subsection{A small value of $\l$ from `fundamental physics'.}
\label{sec:polar}

Zeldovich (1968),
having demonstrated that the energy density of the 
vacuum was infinite at the one-loop level, 
suggested that after the removal of 
divergences, the `regularized' vacuum polarization contributed by
a fundamental particle
of mass $m$ would be described
by the expression 
\beq
\rho_\l \sim \frac{Gm^2}{\hbar c}m~(\frac{mc}{\hbar})^3.
\label{eq:zeld}
\eeq
One can arrive at this result by means of the
following argument: the vacuum consists of virtual particle-antiparticle
pairs of mass $m$ and separation $\lambda = \hbar/mc$. Although the
regularized
self-energy of these pairs is zero, their gravitational interaction is
finite and results in the vacuum energy density
$\epsilon_{vac} \equiv \rho_{vac}c^2 \sim \frac{Gm^2}{\lambda}/\lambda^3 =
Gm^6c^4/\hbar^4$ corresponding to (\ref{eq:zeld}).
(In terms of Feynman diagrams this corresponds to the energy associated
with the two-loop vacuum graph shown in figure \ref{fig:loop}.)
Substituting $m \rightarrow m_e (m_p)$ we find that the electron (proton)
mass gives too small (large) a value for $\rho_\l$.
On the other hand,
the pion mass gives just the right value~\cite{kard97}\footnote{The 
large difference
between $\rho_\l$ obtained using (\ref{eq:zeld}) for the proton and its 
observed
 value prompted Zeldovich to suggest
that Fermi's weak interaction constant $G_F$ might play a role in
determining the vacuum energy, so that 
\beq
\rho_\l \simeq \frac{GG_Fm^8c^5}{\hbar^7}.
\eeq
Although this leads to some improvement,
$\rho_\l$ for the proton is still
several orders of magnitude larger than its observed value.}
\beq
\rho_\l = \frac{1}{(2\pi)^4}~\rho_P~(\frac{m_\pi}{M_P})^6 \simeq
1.3\times 10^{-123}\rho_P=6.91\times 10^{-30}~g\,cm^{-3}~.
\eeq

A small value of $\l$ can also
be derived from dimensionless fundamental constants
of nature using purely numerological arguments.
For instance, the
fine structure constant $\alpha \equiv e^2/\hbar c \simeq 1/137$
when combined with the Planck scale $\rho_P$, suggests the 
relation~\cite{star98c}
\beq
\rho_\l = \frac{\rho_P}{(2\pi^2)^3}e^{-2/\alpha} \simeq
1.2\times 10^{-123}\rho_P=6.29\times 10^{-30}~g\, cm^{-3}~.
\eeq
Or, when expressed in terms of $\o_\l = \frac{8\pi G\epsilon_\l}{3H_0^2}$
we get $\o_\l$h$^2$ = 0.335,
in excellent agreement with observations.
In principle, $\alpha$ could be some other fundamental constant, such as the
`string constant' associated with superstring theory, which might enter 
into exponentially small
expressions for $\l$ of this type.

There are other (equally speculative) ways in which a small vacuum energy
may be connected to `fundamental physics'. For instance in some 
particle physics models
the scale of supersymmetry breaking is rather low and occurs near the
electroweak scale $M_{\rm SUSY} \simeq 10^3$ GeV \cite{anton98}.
The present value $\rho_{\rm vac} \sim (10^{-3} eV)^4$ might 
therefore be `explained'
by a theory in which the effective energy scale of the vacuum 
was given by $\rho_{\rm vac} \sim M_X^4$ where
$M_X = M_{\rm SUSY}^2/M_{Pl} \simeq 10^{-3} eV$.

\begin{figure}
\psfig{file=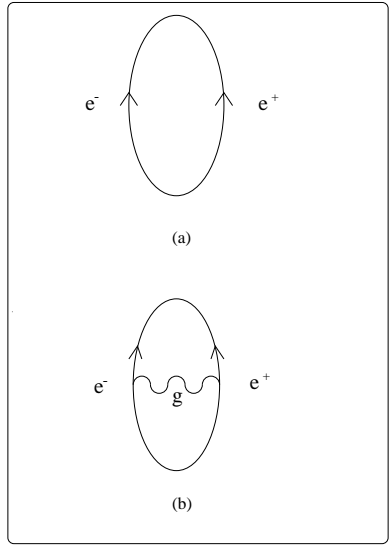,width=6cm}
\caption{This figure shows the 
one-loop (a) and two-loop (b) vacuum diagrams which
contribute towards the vacuum energy density discussed in sections
\ref{sec:lambda2} \& \ref{sec:polar} respectively.}
\bigskip
\hrule
\medskip
\label{fig:loop}
\end{figure}

\subsection{Generating a small 
cosmological constant from Inflationary particle production.}
\label{sec:pp}

A novel means of generating a small $\l$ at the present epoch was
suggested by Sahni \& Habib (1998). 

Massive scalar fields in
curved spacetime satisfy the wave equation
\begin{equation}
\lbrack\sq + \xi R + m^2\rbrack\Phi = 0
\label{eq:pp1}
\end{equation}
where $R$ is the Ricci scalar and $\xi$ parametrizes the coupling to
gravity. 
%$\xi = 0,~1/6$ corresponding to minimal and conformal coupling
%respectively. 
In a spatially flat FRW universe the field variables
separate so that $$\Phi_k = (2\pi)^{-3/2}\phi_k (\eta) \, e^{-i{\bf
k}\cdot{\bf x}}$$ for each wave mode.
The comoving wavenumber $k = 2\pi a/\lambda$ where
$\lambda$ is the physical wavelength of scalar field quanta. Defining
the conformal field $\chi_k=a\phi_k$ and substituting $R = 6{\ddot a}/a^3$
into Eq. (\ref{eq:pp1}) leads to
\begin{equation}
\ddot \chi_k + [k^2 + m^2a^2 - (1 - 6\xi){\ddot a}/a]\chi_k = 0,
\label{eq:pp2}
\end{equation}
where differentiation is carried out
with respect to the conformal time $\eta = \int dt/a$.
Equation (\ref{eq:pp2}) closely resembles the
one dimensional Schr\"{o}dinger equation in quantum mechanics
\beq
\frac{\hbar^2}{2m}\frac{d^2\Psi}{dx^2} + \lbrack E - V(x)\rbrack \Psi = 0.
\label{eq:pp3}
\eeq
Comparing (\ref{eq:pp3}) and (\ref{eq:pp2}) we find that 
the role of the ``potential
barrier in space'' $V(x)$ is  played by the  time dependent term 
$V(\eta) = - m^2a^2 + (1 -
6\xi){\ddot a}/a$ which may be thought of as a 
``potential barrier in time'' \cite{grish75,ss92,grish88}.
(The form of the barrier is shown in
Fig. \ref{fig:pp} assuming that 
inflation is succeeded by radiative and matter dominated eras.)
In quantum mechanics the presence of a barrier leads to particles being 
reflected and transmitted so that $\Psi_{in}(x) = \exp{(ikx)} + 
R(k)\exp{(-ikx)}$ in the incoming region, and
$\Psi_{out}(x) = T(k)\exp{(ikx)}$ in the outgoing region.
Similarly, the presence of the time-like
barrier $V(\eta)$ will lead to particles moving forwards in time as well as
backwards, after being reflected off the barrier. 
The scalar field at late times
will therefore not be in its vacuum state $\phi_k^+$
but will be described by a linear
superposition of positive and negative frequency states
\begin{equation}
\phi_{out}(k,\eta) = \alpha\phi_k^+ + \beta\phi_k^-.
\label{eq:a5}
\end{equation}
The role of reflection and transmission coefficients $R, T$
is now played by the
Bogoliubov coefficients  $\alpha, ~\beta$ which quantify particle production and vacuum polarization effects and are obtained by matching `in modes'
during inflation with `out modes' defined during the radiation or matter
dominated eras.

Due to the existence of space-time curvature, positive and negative
frequencies can be defined only in the limiting case of small wavelengths,
$\lim_{k \rightarrow \infty} \phi_k^{\pm} \simeq \frac{1}{\sqrt{2k}a}\exp {(\mp
ik\eta)}$,
for which effects of curvature can be neglected.
%$|\alpha|^2 - |\beta|^2 = 1$ by flux conservation.
The value of $\alpha, \beta$ is obtained by matching modes corresponding to
the `out' vacuum with those of the `in' vacuum just after inflation.
(The `in' and `out' vacua are defined during inflation and radiation/matter
domination respectively.)

\begin{figure}[tb]
\centerline{
\psfig{file=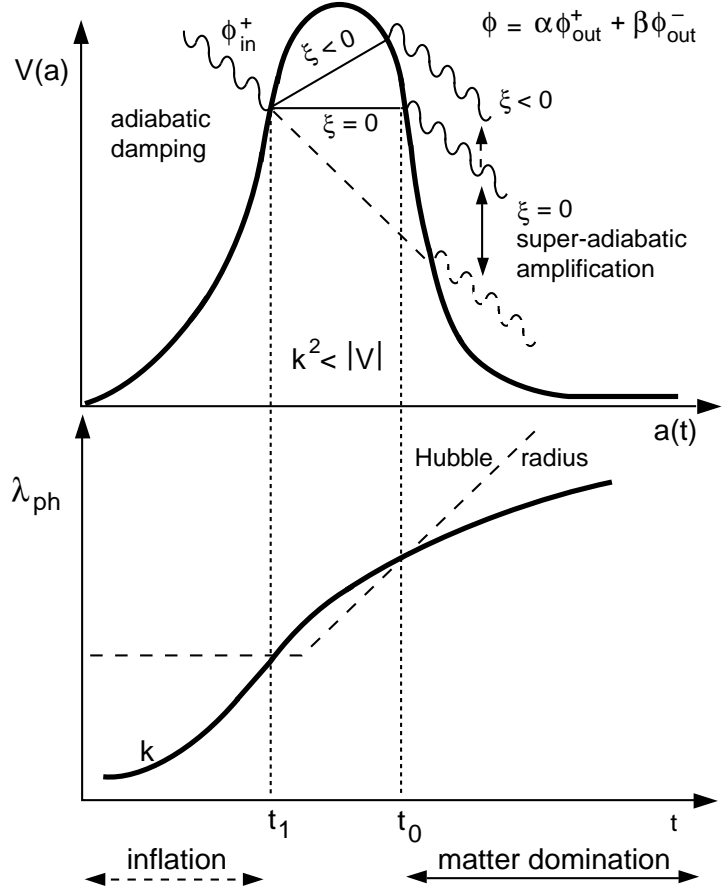,width=10cm}}
\caption{\footnotesize
The process of
super-adiabatic amplification of zero-point fluctuations (particle
production) is illustrated. The amplitude
of modes having wavelengths smaller than the Hubble radius decreases
conformally with the expansion of the universe, whereas that of
larger-than Hubble radius modes freezes (if $\xi = 0$) or grows with
time ($\xi < 0$).
Consequently, modes with $\xi \le 0$ have their
amplitude super-adiabatically amplified on re-entering the Hubble
radius after inflation (from Sahni \& Habib 1998) (the case $\xi = 0$
also describes quantum mechanical production of gravity waves
in a FRW model \cite{grish75}.)}
\bigskip
\hrule
\medskip
\label{fig:pp}
\end{figure}

The net effect of particle creation and vacuum polarization is quantified
by the vacuum expectation value of the energy-momentum tensor 
$\langle T_{ik}\rangle$. For $\xi < 0, |\xi| \ll 1$ and $m/H \lleq 1$
the leading order contribution to 
$\langle T_{ik}\rangle$ is given by
\beq
\langle T_{ik}\rangle \simeq - \xi (R_{ik} - \half g_{ik}R)
\langle \Phi^2\rangle + \half
g_{ik}m^2
\langle \Phi^2\rangle + .....
\label{eq:t0}
\eeq
We immediately see that the first term is simply proportional to the
Einstein tensor and
the second has the covariant form usually associated with a cosmological
constant (\ie $T_{ik} = g_{ik} \Lambda$). Substituting for $\langle T_{ik}
\rangle$ in the semiclassical Einstein equations
\beq
R_{ik} - \half g_{ik}R  = - 8\pi G(T_{ik} + \langle
T_{ik}\rangle)~,
\label{HF}
\eeq
we find
\beq
3H^2 = 8\pi G (\rho_m + \rho_{vac})
\label{eq:t01}
\eeq
where
\ber
\rho_{vac} \equiv \langle T_{00}\rangle &\simeq&
3\xi H^2 \langle \Phi^2\rangle + \half m^2\langle \Phi^2\rangle\label{eq:t00}\\
\langle \Phi^2\rangle &=& \frac{1}{2\pi^2}\int dk k^2
|\phi_{out}(k,\eta)|^2.
\label{eq:t00a}
\eer
The term proportional to $H^2 \langle \Phi^2\rangle$ in (\ref{eq:t00}) may be 
absorbed into the left hand side of
(\ref{eq:t01}) leading to 
\begin{equation}
3H^2 \simeq 8\pi \bar G\big\lbrack \rho_m + \frac{1}{2}m^2
\langle\Phi^2\rangle\big\rbrack
\label{eq:ab8}
\end{equation}
where $\bar G \simeq G/(1 + 8\pi G|\xi|\langle\Phi^2\rangle)$ is the
new, time dependent gravitational constant. (Observational bounds on the
rate of change of $\bar G$
set the constraint $|\xi| \ll 1$.)
As shown in \cite{sahni98} for $\xi < 0$ 
the value of $\langle \Phi^2\rangle$ can be very large, so that
$\bar G \simeq
1/(8\pi|\xi|\langle\Phi^2\rangle$) and
\begin{eqnarray}
\Lambda_{eff} \equiv & 8\pi\bar
G\langle T_{00}\rangle \simeq m^2/2|\xi|,\nonumber\\
\Omega_\l
\equiv & \Lambda_{eff}/3H^2 \simeq \frac{1}{6|\xi|}(m/H)^2
\end{eqnarray}
We therefore find that 
the energy density of created particles defines an {\em effective
cosmological constant} which 
can contribute significantly to the total density of the
universe at late times leading to $\o_m + \o_\l \simeq 1$ \cite{sahni98,sahni99}.

However, it should be noted that this result was obtained in the
Hatree-Fock (or. semiclassical gravity) approximation (\ref{HF}) which
is not exact in considerations of a single quantum field, since metric
and field fluctuations may significantly deviate from their {\it rms}
values. So, further study of this problem using stochastic methods
(similar to those used in stochastic inflation~\cite{st82,st86,vil83} and 
stochastic reheating after inflation~\cite{kls97}) is desirable.
Other field theoretic aspects including the possibility that non-perturbative 
vacuum effects could lead to an effective equation of state today like that of 
radiation plus a cosmological constant have been examined in 
\cite{parker99a,parker99b,parker99c}.

%\section{Dynamical $\l$ models.}
\section{Phenomenological models of a dynamical $\l$-term.}
\label{sec:dynamic}

\subsection{Overview.}

Having, under the pressure of observational evidence and an aesthetical desire 
to keep the inflationary scenario of the early universe as simple as possible,
admitted the existence of a constant positive $\l$-term, it is natural to
take a step beyond Einstein's original hypothesis and consider the
additional possibility 
that the $\l$-term is not an exact constant, but rather,
describes a new dynamical 
degree of freedom (perhaps even a new form of matter). Really, neither
observational data, nor inflationary considerations tell us that a
``cosmological constant'' is constant (though, as discussed above, it
should change sufficiently slowly with time, in particular, slower than 
the Ricci tensor). 
In fact the effective $\l$-term which appears in the inflationary scenario of
the early universe is never an exact constant and rarely even an
approximate constant of motion.
(A recent analysis of observational data in the light of a time dependent $\l$
may be found in ~\cite{wang99,science}.) 

To quantitatively describe this new degree of freedom (or a new form of 
matter), some phenomenological models of a dynamical $\l$-term have to be
introduced. The word ``phenomenological'' means that no attempt to derive
these models
from an underlying quantum field theory is being made, in contrast to
examples discussed in previous sections. 
Historically, many phenomenological $\l$-models were proposed 
since 1986 (not counting the ``C-field'' of Hoyle and Narlikar (1962) 
which was perhaps the earliest, though unsuccessful, 
attempt to introduce a dynamical $\l$-term in cosmology). 
Depending upon their level of ``fundamentality'', these phenomenological
methods may be classified 
into 3 main groups: 

1) Kinematic models. \\
Here $\l$ is simply assumed to be a function of either the cosmic time $t$
or the scale factor $a(t)$ of the FRW cosmological model.

2) Hydrodynamic models. \\
Here a $\l$-term is described by a barotropic fluid with some equation of 
state $p_{\l}(\rho_{\l})$ (dissipative terms may also be present).

3) Field-theoretic models. \\
The $\l$-term is assumed to be a new physical classical field (which we shall 
call a {\em lambda-field}) with some phenomenological Lagrangian.

Of course, models from the last group are in a sense also the 
most fundamental. In 
particular, they may be used in a non-FRW setting. 
Additionally, their quantization is straightforward. However, if we restrict
ourselves to a FRW model with small perturbations, the three different
way of describing a $\l$-term could lead to converging results. Note also, that
it was recently proposed to call a dynamical $\l$-term ``quintessence''~
\cite{caldwell98b} (irrespective of the specifities of modelling, so this
notion is wider than the notion of a {\em lambda-field}), though we don't
consider it to be obligatory.

In the case of field-theoretic models, the most simple and natural 
is the model of a scalar field $\phi$ with some self-interaction potential
$V(\phi)$, minimal coupling to gravity and no (or very weak) coupling to 
other known physical fields. The latter requirement follows not only from
simplicity, but also from observational evidence (see~\cite{car98} for
a recent analysis of upper bounds on coupling of $\phi$ to the electromagnetic 
field). The assumption of minimal coupling to gravity may be relaxed,
but only slightly (see~\cite{chiba99} for constraints on the
coefficient $\xi$ in case of the $\xi R\phi^2/2$ coupling). Since
the minimally coupled scalar field model has proven to be
extremely successfuly in the case of
the inflationary scenario, one might be tempted to use it for the description
of a $\l$-term. As a result, an overwelming part of recent theoretical
activity has focussed on the scalar field model 
(more appropriately, on this class of 
models differing between themselves by the form of the scalar field
potential $V(\phi)$). 

\subsection{When may a $\l$-term be described by a minimally coupled
scalar field?}

Now let us consider the following important question: can a $\l$-term 
{\em always} be described by a minimally coupled scalar field, for any
observed behaviour of $a(t)$ or $H(a)$ ? 
To answer this question let us first consider 
a FRW universe with matter (dust) $\rho_m$ and a homogeneous scalar field
$\phi$ with energy density $\rho$ and pressure $P$ given by
\begin{eqnarray}
\rho &=& \half\dot{\phi}^2 + V(\phi) \nonumber\\
P &=& \half\dot{\phi}^2 - V(\phi).
\label{eq:infl5}
\end{eqnarray}
The evolution of the universe is described by the equations of motion
\ber
&H^2 = \frac{8\pi G}{3}(\rho_m + \frac{\dot \phi^2}{2} + V),
~\rho_m = \frac{3\o_0H_0^2}{8\pi
G}(\frac{a_0}{a})^3,\label{eq:recon0}\\
&{\ddot\phi} + 3H{\dot\phi} + \frac{dV}{d\phi} = 0,\label{eq:recon01}\\
&{\dot H} = -4\pi G(\rho_m + {\dot\phi}^2).
\label{eq:recon1}
\eer
The clue to whether a $\l$-term can be successfully described by
a minimally coupled field
is provided by the background equation for $\dot H$ which
we rewrite in the following form changing the independent variable from
$t$ to $a$ ($\kappa = 0$ is assumed):
\begin{equation}
4\pi G a^2H^2\left({d\phi\over da}\right)^2= - aH{dH\over da} - {3\over 2}
\Omega_mH_0^2\left({a_0\over a}\right)^3~,
\label{dphida}
\end{equation}
where $a_0$ is the present value of the FRW scale factor $a(t)$ and
$\Omega_m$ includes all dust-like matter at present (CDM, baryons,
sufficiently massive neutrinos, etc.). Since the left-hand side
of Eq. (\ref{dphida}) is always non-negative therefore so is the right-hand
side. From this follows a fundamental restriction on the
expansion law for the Universe, which we write in terms of the following
inequality
on the redshift dependence of the Hubble parameter $H(z),~1+z\equiv a_0/a$:
\begin{equation}
{dH^2\over dz}\ge 3\Omega_mH_0^2(1+z)^2~.
\label{ineq}
\end{equation}
Actually, Eq. (\ref{ineq}) is nothing more than the weak energy
condition for a lambda-field: $\rho_{\l}+p_{\l}\geq 0$.

This inequality saturates in the case of a constant $\l$-term (a
cosmological constant). Equation (\ref{ineq}) 
 constitutes the necessary condition for 
an arbitrary $H(z)$ dependence to be physically described by 
a minimally coupled 
scalar field (in the absence of spatial curvature). It will be shown below 
that Eq. (\ref{ineq}) is also a sufficient condition, since a 
knowledge of $H(z)$ and $\Omega_m$ permits a unique reconstruction of the 
self-interaction potential $V(\phi)$ of this scalar lambda-field 
(see section \ref{sec:reconstruct}). Taken
at $z=0$, Eq. (\ref{ineq}) reduces to the following relation between
the acceleration parameter $q_0$ and $\Omega_m$:
\begin{equation}
q_0\ge {3\over 2}\,\Omega_m - 1~.
\label{ineq1}
\end{equation}

It should be emphasized that we have no idea at present whether or not 
Eqs. (\ref{ineq},\ref{ineq1})
are fulfilled. Only future observations will tell us that. 
Moreover, as was explained in previous sections, a constant $\l$-term fits 
existing data very well. Thus, we know already that the inequalities 
(\ref{ineq},\ref{ineq1}) are close to saturation. So, it will be not an easy
observational task. In this case, the presence of even a small spatial 
curvature may dramatically change our conclusions.

In the case of non-zero spatial curvature ($\kappa \not= 0$), 
Eq. (\ref{ineq}) generalizes to:
\begin{equation}
{dH^2\over dz}\ge 3\Omega_mH_0^2(1+z)^2 - {2\kappa\over a_0^2}\,(1+z)~.
\label{ineq2}
\end{equation}
Therefore, if future data show that the inequality (\ref{ineq}) is
not valid, one has either to invoke a {\em positive} spatial curvature
for the Universe ($\kappa =1$), or else to discard this model entirely and to 
consider a more complicated model of a $\l$-term, modelled by, say,
a scalar-field 
non-minimally coupled to gravity. It is easy to verify that in the case of 
$\xi R\phi^2/2$ coupling, no necessary conditions such as (\ref{ineq}) or
(\ref{ineq2}) appear. However, as was mentioned above, this type of coupling
is strongly restricted by observational data~\cite{chiba99}.  

\subsection{Late-time Inflation and $\l$.}
\label{sec:infl}

Conceivably, one might appeal to inflationary mechanisms which are so
successful
at generating a large cosmological constant during an early epoch
to generate a small cosmological constant today.
As pointed out in section \ref{sec:ssb}, effective potentials giving rise to
symmetry breaking generically predict a large negative value for a cosmological
constant
which has to be `regularized' to give the small positive $\l$ observed
today. The problem with these methods is that they usually
prescribe an unevolving cosmological term
whose present value is fixed at the time of
symmetry breaking.
This necessarily implies some fine tuning of
parameters which can be as large as one part in
$10^{123}$ (for symmetry breaking at the
Planck scale) to one part in $10^{53}$ for the electroweak scale.

A different possibility is suggested by
the family of potentials which lead to `chaotic inflation' 
$$V \propto \phi^q, ~~ q \geq 2.$$
From (\ref{eq:infl5}) we find that for a scalar field which
rolls down its potential `slowly' so that ${\dot\phi}^2 \ll V(\phi)$
 will result in the inflationary equation of state $P \simeq
-\rho$ usually associated with a cosmological constant.
For the potential $V = \half m^2\phi^2$ this translates into the constraint
$(m/H_0)^2 \lleq 1$,
in other words,
the Compton wavelength of the inflaton should be larger than the present Hubble
radius $\lambda = {\hbar}/mc {\lower0.9ex\hbox{ $\buildrel
> \over \sim$} } ~cH_0^{-1}$
suggesting an extremely small mass for the inflaton
$m {\lower0.9ex\hbox{ $\buildrel < \over \sim$} }~ 10^{-33}$ eV.
Such a small mass could arise in models with approximate global symmetries
and `lambda-field models' of this kind have been discussed in the context of
pseudo-Nambu-Goldstone bosons \cite{frieman95,hill88}.
One may also be tempted to associate $m$ with the small
mass difference associated
with solar neutrino oscillations
%and the Planck (or GUT) scale:
$m = \Delta m_\nu^2/M_P \simeq 10^{-33}$ eV  where
$\Delta m_\nu^2 \simeq 10^{-5}$ eV$^2$,
an idea which is speculative but not implausible.

It must however be pointed out that models
with the simplest potentials including
$V \propto m^2\phi^2$ run into problems similar to
those encountered by a cosmological constant. The enormous overdamping of the
scalar field equation during radiation and matter dominated epochs causes
$V(\phi)$ to remain unchanged virtually from the Planck
epoch $z_{pl} \sim 10^{19}$ to $z \sim 2$ \cite{frieman95} resulting in an
enormous difference in the scalar field energy density and that of
matter/radiation at early times. This leads to a fine tuning problem:
the relative values of $\rho_\phi$ and $\rho_m$ must be adjusted to very high
levels of accuracy in order to ensure $\rho_\phi/\rho_m \sim 1$ at
precisely the present epoch \cite{sahniwang}. (A very large present value
$\rho_\phi/\rho_m \gg 1$
would conflict with observations of large scale structure since
galaxy formation would be strongly suppressed.
On the other hand a very small value
$\rho_\phi/\rho_m \ll 1$ would not give rise to an accelerating universe.)
Fortunately the fine tuning problem can be significantly reduced in
a class of scalar field models which we consider next.

An interesting example of the dissipationless decay of a lambda-field 
is provided
by Peebles and Ratra (1988) who consider 
a minimally coupled scalar field 
rolling down a potential $$V(\phi) = k/\phi^\alpha$$ 
subject to the equation of motion (\ref{eq:recon01})
($k$ and $\alpha$ are constants, we set $M_P = 1$ for simplicity).
Let us assume that the energy density of the scalar field
\beq
\rho_\phi = \half {\dot \phi}^2 + \frac{k}{\phi^\alpha}
\label{eq:rp2}
\eeq
is subdominant at early epochs (as demanded by CMB and nucleosynthesis 
constraints) so that $\rho_\phi < \rho_{B}$ at $z \gg 1$,
where $\rho_B$ is the density of background matter driving the expansion 
of the universe. 
Assuming a 
general expansion law for the universe $a(t) \propto t^q$ the field equation
of motion (\ref{eq:recon01}) becomes 
\beq
\ddot\phi + \frac{3q}{t}\dot\phi - \frac{\alpha k}{\phi^{1 + \alpha}} = 0
\label{eq:rp3}
\eeq
which has the solution
\beq
\phi \propto t^p, ~~~ p = \frac{2}{2 + \alpha}.
\label{eq:rp4}
\eeq
Substituting $\phi$ 
in (\ref{eq:rp2}) we find $\rho_\phi \propto t^{2p - 2}$, as a result
if $p > 0$ the scalar field density $\rho_\phi$ decreases more slowly than 
the background density of matter or radiation which decreases
as $\rho_{B} \propto t^{-2}$. Consequently we find
\beq
\frac{\rho_\phi}{\rho_{B}} \propto t^\frac{4}{2 + \alpha}
\label{eq:rp5} 
\eeq
\ie for $\alpha > 0$ the scalar field density can dominate the
matter/radiation density at late times even if it was subdominant to begin with
\cite{ratra88a,ratra88b,ratra99}. (This attractive property of scalar fields 
is occasionally referred to as `quintessence'.)
The rate of growth of $\rho_\phi/\rho_{m,r}$ can be modulated by
`tuning' the value of $\alpha$.
Another way of arriving at this conclusion is to examine the equation of
state of the scalar field while the latter is subdominant, this turns out to be
%From (\ref{eq:rp5}) it is easy to show that
%the equation of state of the scalar field has the form
\beq
w_\phi = \frac{\alpha w_B - 2}{\alpha + 2}
\label{eq:rp6}
\eeq
where $w_B$ is the background equation of state.
From (\ref{eq:rp6}) we find $w_\phi < w_B$ \ie the
equation of state of the scalar field is less stiff than that of matter
driving expansion. The conservation condition $\rho_{(\phi, B)}
\propto a^{-3\lbrack 1+w_{(\phi, B)}\rbrack} $
now guarantees that the scalar field
will come to dominate the expansion dynamics of the universe even if
it was initially subdominant.
As a result $\rho_\phi$ can be significantly small during the radiation 
dominated epoch to satisfy nucleosynthesis constraints yet be large 
enough today to
give rise to an accelerating universe in agreement with recent supernovae
results.
(Once $\rho_\phi$ begins to dominate the
energy density, the
universe enters into a period of accelerated expansion driven by 
the scalar field energy density which
begins to mimic an effective $\l$-term.)

A different possibility arises
if we consider a scalar field rolling down
an exponential potential $$V(\phi) = V_0\exp{(-\lambda\phi/M_P)}.$$
%$V(\phi) = M_P^4\exp{(-\lambda\phi/M_P)}$
In the case of a flat universe,
the scalar field density scales exactly like the background density of
matter driving the expansion of the universe so that
the ratio of the scalar field density to the total matter density
rapidly approaches a constant value \cite{ratra88a,wett88,fj97a,fj97b}
\beq
\frac{\rho_\phi}{\rho_{B} + \rho_\phi} = \frac{3(1 + w_B)}{\lambda^2}
\eeq
($w_B = 0, ~1/3$ respectively for dust, radiation).
This `tracker-like' quality whereby the scalar field 
contributes the same fixed amount to the total matter density allows it
to play the role of a form of dark matter. However
strong constraints on this model come from 
cosmological nucleosynthesis which suggests $\Omega_\phi \simeq 
\frac{\rho_\phi}{\rho_{B}} {\lower0.9ex\hbox{ $\buildrel < \over \sim$} } ~0.2$. 
\footnote{In a spatially closed universe the presence of an exponential
potential can give rise to an intermediate `coasting' epoch during which
$a(t) = \alpha t$, where $\alpha \ll 1$. Such a universe bears
great similarity to loitering models considered earlier, density
perturbations grow faster during `coasting' and the `age problem' too
can be resolved \cite{sfs92}.}
As a result the scalar field in these models is forever destined to
remain subdominant, it can neither dominate the matter density of
the universe nor give rise to its accelerated expansion rate. 

A potential which interpolates between an exponential and a power law has been  explored by Sahni \& Wang (1999) 
\beq
V(\phi) = V_0\lbrack \cosh{\lambda\phi/M_P} - 1\rbrack^p.
\label{eq:pot1}
\eeq
Since $V(\phi) \propto \exp{(-p\lambda\phi)}$, 
for $\vert\lambda\phi\vert \gg 1$,
we would expect this potential to reproduce 
features of the exponential potential discussed earlier. As a result 
$\rho_\phi/\rho_B \simeq {\rm constant}$ and $w_\phi \simeq w_B$,
if the scalar field commences rolling from a large initial value.
For small values $\vert\lambda\phi\vert \ll 1$ 
the potential is a power-law 
$V(\phi) \propto (\lambda\phi)^{2p}$ and the scalar field oscillates about
$\phi \simeq 0$ at late times. 
For $p \leq 1/2$ oscillations of $\phi$ can lead to an 
extended period of accelereted expansion since the averaged equation of 
state while the scalar field oscillates is  
$\langle w_\phi\rangle = p-1/p+1$. Thus quintessence models
based on (\ref{eq:pot1}) can play the roll both of cold dark matter (if $p= 1$)
as well as a negative pressure $\Lambda$-field (for $p \leq 1/2$).
An added attraction of this
model is that since $\phi$ rolls to small values at late
times ($\phi \to 0$ as $t \to \infty$) quantum corrections
which could become significant for models having $\phi$ roll to large
values $\phi \geq M_P$ \cite{kolda98} can safely be neglected 
in this case \cite{sahniwang}. 

An unusual potential with interesting features was proposed in
\cite{zws99}
\beq
V(\phi) = V_0\lbrack e^{M_P/\phi} - 1\rbrack.
\label{eq:pot2}
\eeq

A useful property of potentials 
(\ref{eq:pot1}) \& (\ref{eq:pot2}) is that they significantly alleviate
the fine tuning problem associated with generating 
a small cosmological term at {\em precisely} 
the present epoch.  As a result,
$\rho_\phi$ can come to dominate the current cosmological density 
from a fairly general class of initial conditions (see figure \ref{fig:quint}). 
A phase space analysis of scalar field models 
was carried out in \cite{ratra88a,fj97a,fj97b,ls98} 
where it was shown that
both exponential and negative power-law potentials display appealing
attractor-like qualities. 
However, despite the many attractive features of `quintessence' models
a degree of fine tuning does remain in fixing the parameters of
the potential 
and has been commented on in \cite{ls98,kolda98}.
For instance the requirement that $\o_{\phi} \ll 1$ during
the matter dominated epoch, while $\o_{\phi} \sim 1$ now, is fulfilled
for the potential (\ref{eq:pot2})
 only if the present value of $\phi$ is significantly
larger than $M_{Pl}$. Thus, for practical applications in the present
universe, this potential shows little difference from the inverse-power-law
potential $V\propto \phi^{-1}$.

\include{psfig}
\begin{figure}
\centerline{
\psfig{file=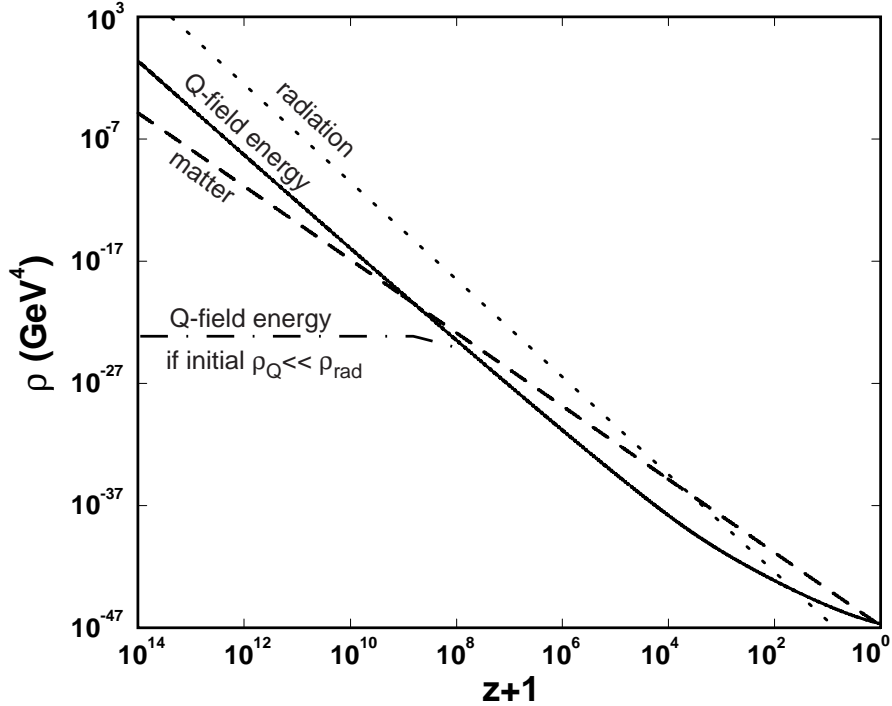,width=12cm}}
\caption{\footnotesize
The `tracker' property of the quintessence field (Q-field) is illustrated
in this figure from Zlatev, Wang and Steinhardt (1999).
The energy density in the quintessence field (solid) initially tracks the 
radiation density (dotted) and then the matter density (dashed) before
finally dominating the total density of the universe 
at the present epoch. If the initial
value of the Q-field is suppressed, then $\rho_Q$ remains
constant until $\rho_Q \sim \rho_{rad}$ after which the $Q$-field
follows the tracker solution converging to the same cosmology today:
$\Omega_m = 0.4$, $w_Q = -0.65$. 
The Q-field evolves according to (\ref{eq:pot2}).}
\label{fig:quint}
\end{figure}

It is worth pointing out in this context that the energy density of relic
gravity waves created during inflation ($\rho_g$) behaves like a tracker
field since $\rho_g/\rho_B \simeq {\rm constant}$, if the expansion factor
grows exponentially during inflation \cite{allen88}. For more realistic 
situations in which the inflaton field rolls down its potential slowly
the ratio $\rho_g/\rho_B$ increases
with time with the result that
the graviton energy density may become comparable to $\rho_B$ 
at very late times provided inflation commenced 
at the Planck epoch \cite{sahni90}.
COBE measurements of the large angle anisotropy of
the cosmic microwave background (CMB) however
ensure that the gravity wave contribution
to the total matter density is
negligibly small today: $\Omega_g 
\lleq 10^{-12}$ \cite{ss92}. However the intriguing
possibility that quanta of a different type
of fundamental field (the dilaton perhaps) may 
come to dominate the energy density of the universe without necessarily
violating CMB bounds remains to be investigated.
(One should also note that good agreement with
supernovae observations can also be obtained by relaxing the constraint
$\rho_\l + P_\l \geq 0$ implicitely assumed in scalar field based
quintessence models \cite{caldwell99,zia00}.)

Some cosmological consequences of scalar field models and models with a 
decaying cosmological term have been analyzed in
\cite{ratra88b,fuji91,moffat95,waga97,coble97,frieman95,frieman98,caldwell98a}
\cite{caldwell98b,wang98,huey99,chiba98,hu99,garn98a,ma99,peebles99,viana98,waga98,waga00,wang99,ht98,cooray99,efst99,ratra99,sahniwang,saini99,zia00}.
Candidates for quintessence based on field theory and 
high energy physics are discussed in 
\cite{albrecht99,binetruy99,choi99,kolda98,bento99,benakli99,sahniwang,brax99a,brax99b,copeland99,cormier00,masiero00,arkani00,branchina00,rosati00}, 
non-minimal scalar field models are treated in 
\cite{morikawa,sahni98,sahni99,parker99a,parker99b,parker99c,amendola99,bertolami99,chiba99,pbm99,uzan99,perrotta99a,pbm99,ritis99}.
Attempts towards obtaining a $\l$-term (or a quintessence-type $\l$-field)
as a relic from inflation  are discussed in \cite{sahni98,sahni99,peebles99,peloso99,rubakov99,rosati00}.
The possibility that the acceleration of the universe is localised within
a bubble of positive vacuum density has been explored in \cite{starkman99}.
Perturbations in quintessence models have been examined in 
\cite{caldwell98b,ma99}. The coupling of the quintessence field
to ordinary matter has been discussed in \cite{car98}.

A brief summary of some models with a decaying
cosmological term is given in Table \ref{table:lambda}
(adapted from \cite{coop98}),
we should stress that most of these models are phenomenological and are
therefore
not necessarily backed by strong physical arguments.

\begin{table}
\begin{center}
\caption{Summary of phenomenological $\Lambda$ models.
Here $a$ is the scale factor, $H$ the Hubble parameter, $T$ the temperature,
$t$ the cosmic time ($A, B, \alpha$ are constants).}
\bigskip
\begin{tabular}{lll}
\tablerule
Evolutionary relation for $\Lambda (t)$ & Reference\\\tablerule
$\l \propto t^{-2} $ & 
\cite{ef77,cha77,bertolami86,bsom90,beesham94,lopna96,coop98}\\
$\l \propto t^{-\alpha}$ & \cite{beesham93,kwe92,kwe95}\\
%$\l \propto t^{-1}(\alpha + t)^{-1}$ & \cite{kwe95}\\
$\l \propto A + B\exp{(-\alpha t)}$ & \cite{beesham93,brout94}\\
$\l \propto a^{-2}$ &
\cite{lopna96,ozer86,ozer87,abdel92,vishwa97,chen90,gott87,kolb89}\\
$\l \propto a^{-\alpha}$ &
\cite{olson87,pavon91,sfs92,maia94,matyj95,waga94,waga97}
\cite{hoyle97,john97,coop98,turner97,caldwell98a,caldwell98b,wang98,hu99,garn98a}\\
%$\l \propto A + \alpha a^{-m}$ & \cite{matyj95}\\
$\l \propto \exp{(-\alpha a)}$ & \cite{rajeev}\\
$\l \propto T^\alpha$ & \cite{cha77,kaz80}\\
$\l \propto H^2$ & \cite{lima94,wett88,wett95,fj97a,clw98}\\
$\l \propto H^2 + A a^{-\alpha}$ & \cite{arbab94,carval92,waga93a,waga93b}\\
$\l \propto f(H)$ & \cite{lima94b,lima96}\\
${\dot\l} \propto g(\l, H)$ & \cite{hisc86,reuter87}\\
\tablerule
\end{tabular}
\label{table:lambda}
\end{center}
\end{table}
\bigskip

Finally one should mention another phenomenological
approach tied to the possibility of
a cosmological term decaying and transferring its energy
into
particles and/or radiation \cite{ozer86,freese87}.
Observationally such an approach can, in principle, be tested:
in the case of dissipative, baryon number
conserving decay of a $\Lambda$-term
into
baryons and antibaryons,
the subsequent annihilation of matter and antimatter would result in a
homogeneous gamma-ray
flux which could be constrained by observations of the
diffuse gamma-ray background in the Universe \cite{freese87,matyj95}.
A decay of the cosmological term directly into radiation could be probed
by cosmic microwave background anisotropies,
cosmological nucleosynthesis etc.
\cite{freese87,sato90,sarkar97,overduin93,matyj95,pavon91,waga94,waga97}.

\subsection{Relation between kinematic and dynamical descriptions of $\l$}
\label{sec:quint}

As pointed out in the previous section, 
although kinematic and dynamical models of $\l$ lie on
completely different levels of fundamentality from the theoretical point
of view, they may be equivalent if a background space-time is described by
a FRW model. In particular, the simplest class of kinematic models
\begin{equation}
\l \equiv 8\pi G\rho_{\l} = f(a)
\label{kin}
\end{equation}
is then equivalent to hydrodynamic models based on an ideal fluid with
the equation of state
\begin{equation}
p_{\l}(\rho_{\l})= - \rho_{\l}(1 + {1\over 3}\, {d\ln \rho_{\l} \over d\ln a})
\label{eqstate}
\end{equation}
(with $a$ being excluded from Eq.~(\ref{eqstate}) using Eq.~(\ref{kin})).

Let us go further and present the correspondence between a popular subclass
of these models where $\l \propto a^{-\alpha}$ (or, equivalently,  
$p_{\l}= \left({\alpha\over 3}-1\right)\rho_{\l}$) and field-theoretical
models for a minimally coupled lambda-field following~\cite{star98c} where
the particular case $\alpha =2$ (i.e. the $\l$-term mimicking temporal
behaviour of spatial curvature or non-relativistic cosmic strings) was
considered. This gives an explicit 
example of the reconstruction of a lambda-field potential from $H(a)$.  
Now $\kappa = 0$ is assumed for simplicity, and we take $0\le \alpha < 3$. 
The left inequality is necessary for the condition (\ref{ineq}) to be 
satisfied, while the right inequality guarantees that $\rho_{\l} \ll \rho_m$
during the matter-dominated stage while $z\gg 1$ 
(in addition, this condition makes $p_{\l}$
negative).
In this case, the Hubble parameter $H(a)$ is given by
\begin{equation}
{H^2\over H_0^2}=\Omega_m\left({a_0\over a}\right)^3+(1-\Omega_m)
\left({a_0\over a}\right)^{\alpha}~.
\label{H2}
\end{equation}

Using the $0-0$ background Einstein equation and Eq.~(\ref{dphida}), the
lambda-field potential $V(\phi)$ can be expressed in terms of $H(a)$:
\begin{equation}
8\pi GV(\phi)= aH{dH\over da} +3H^2 - {3\over 2}\,\Omega_mH_0^2
\left({a_0\over a}\right)^3
\label{vphi}
\end{equation}
which reduces to 
\begin{equation}
V = {3-{\alpha\over 2}\over 8\pi G}\, H_0^2(1-\Omega_m)\left({a_0\over a}
\right)^{\alpha}
\label{vphi1}
\end{equation}
for the case under consideration.

Now Eq.~(\ref{dphida}) may be integrated for the given $H(a)$ dependence
to obtain
\begin{equation}
{a\over a_0} ={\Omega_m\over 1-\Omega_m}\sinh^{{2\over 3-\alpha}}\left((3-
\alpha)\sqrt{{2\pi G\over \alpha}}\,(\phi-\phi_0+\phi_1)\right)
\label{aphi}
\end{equation}
where $\phi_0$ is the present value of the lambda-field and
\begin{equation} 
\exp\left((3-\alpha)\sqrt{{2\pi G\over \alpha}}\,\phi_1\right)=
\left({1-\Omega_m\over \Omega_m}\right)^{3-\alpha\over 2} +
\sqrt{1+ \left({1-\Omega_m\over \Omega_m}\right)^{3-\alpha}}~.
\end{equation}
Finally, combining Eqs. (\ref{vphi1}, \ref{aphi}) we get an explicit
expression for the interaction potential:
\begin{equation}
V(\phi)={(3-{\alpha\over 2})(1-\Omega_m)^{1+\alpha}H_0^2\over 8\pi
\Omega_m^{\alpha}G}\, \sinh^{-{2\alpha \over 3-\alpha}}\left((3-
\alpha)\sqrt{{2\pi G\over \alpha}}\,(\phi-\phi_0+\phi_1)\right)~.
\label{vphifin}
\end{equation}

At early times during the matter-dominated stage, this potential is an inverse
power-law ($V(\phi) \propto  (\phi - \phi_0 +\phi_1)^{-{2\alpha \over 
3-\alpha}}$) (we do not consider here what happens with $V(\phi)$ even 
earlier, during the radiation-dominated stage). 
While during the current, $\l$-dominated epoch,
it changes its form to an exponential. This shows why the 
assumptions of a purely power-law dependence of $\l$ on $a$ or, equivalently, 
of a linear equation of state $p_{\l}=w_{\l}\rho_{\l},~w_{\l}=const$ are not 
``natural'': they require fine-tuning between the present value of the 
lambda-field $\phi_0$ and the value of $\phi$ where the potential changes its 
form. 
On the other hand, neither can this possibility be ruled out
completely.

In addition, this example of reconstruction of $V(\phi)$ shows that, in 
field-theoretic models of $\l$ based on a minimally coupled scalar field, 
there is no lower limit on the present value of $w_{\l}$ other than $-1$ 
(which follows from the weak energy condition (\ref{ineq})). The opposite
statement in~\cite{zws99,stein98} is a consequence of a number of additional
assumptions (equipartition of energy densities of all fields including
the lambda-field at the end of inflation, use of a subclass of possible
initial contitions whose solutions for $\phi$ have reached 
an intermediate asymptote 
which they call the ``tracker'' solution by the present time, 
consideration of some special classes of potentials), none of which is 
obligatory. In particular, a ``tracker'' solution may have $w_{\l}$ 
arbitrarily close to $-1$ at present, if an inverse power-law
potential with a small exponent is used. 
 
\subsection{Reconstructing the effective potential $V(\phi)$
for a time-dependent
$\l$-term.}
\label{sec:reconstruct}

In view of the large number of models capable of
predicting a small cosmological constant at the present epoch, it is
necessary to ask whether cosmological observations themselves may be
used to
determine model parameters uniquely.
The answer to this question
is (fortunately) in the affirmative, at least for the class of minimally
coupled scalar field models discussed earlier. 
This is easily demonstrated by considering Equations (\ref{vphi})
and (\ref{dphida}) which
express $V(\phi)$ and $\phi$ in terms of the Hubble parameter $H$
and its first derivative $dH/dz$. Consequently one can determine the
form of the potential $V(z)$ (mimicking the $\l$-term)
if the Hubble parameter $H(z)$ is
known from observations. There are two independent methods for
determining
$H(z)$.
The first is related to the luminosity distance $d_L$,
discussed in section \ref{sec:lens} \cite{star98b,saini99}.
From (\ref{eq:age11d}) we easily find
\beq
H(z) = c\bigg\lbrack\frac{d}{dz}\frac{d_L(z)}{1+z}\bigg\rbrack^{-1}.
\label{star3}
\eeq
Thus the luminosity distance $d_L(z)$ determines the Hubble parameter
$H(z)$ uniquely ! Now (\ref{dphida},\ref{vphi}) can be used to reconstruct the
form of the potential $V(z)$ (or $V(\phi)$) and the equation of state
$w_\phi(z)$ in a model independent manner \cite{saini99}. 
(Note that 
for an unambiguous determination of $V(\phi)$ one also needs to
know the present matter
density
$\o_m$\cite{star98b}.)
Formula (\ref{star3}) can also be used for an unambiguous
determination
of $H(z)$ from the angular-size distance $d_A(z)$ introduced in section
\ref{sec:angle}, if we
use the relation $d_A(z)=d_L(1+z)^{-2}$~\cite{star98c}.

Another means of
determining $H(z)$ is associated with the growth of linear density
fluctuations responsible for the formation of large scale structure~
\cite{star98a,star98b}.
The growth of linearized perturbations in a collisionless medium has the
well known form
\beq
{\ddot \delta} + 2H{\dot\delta} - 4\pi G\rho_m\delta = 0
\label{eq:recon3}
\eeq
where the value of $H$ is determined from (\ref{eq:recon1}).
(On scales $\ll 200$h$^{-1}$ Mpc the $\l$-field is practically
unclustered and can be treated as a smooth component if $|m_\phi^2|
\equiv
|d^2V/d\phi^2| {\lower0.9ex\hbox{ $\buildrel < \over \sim$} }~ H_0^2$).
Although it is not
possible to solve (\ref{eq:recon3}) analytically for an arbitrary
potential $V(\phi)$, the inverse problem of determining $H$ once
$\delta$ is known is exactly solvable ! We demonstrate this by first
performing
a change of variables $t \rightarrow a,~ d/dt \rightarrow aHd/da$ which
reduces (\ref{eq:recon3}) to a first order linear differential equation
for $H^2$:
\beq
a^2\frac{d\delta}{da}\frac{dH^2}{da} + 2\bigg(a^2\frac{d^2\delta}{da^2}
+
3a\frac{d\delta}{da}\bigg)H^2 = 3\o_0H_0^2(\frac{a_0}{a})^3\delta.
\label{eq:recon4}
\eeq
Equation (\ref{eq:recon4}) has the exact solution
\beq
H^2 =
\frac{3\o_0H_0^2a_0^3}{a^6}\bigg(\frac{d\delta}{da}\bigg)^{-2}\int_0^a
a\delta\frac{d\delta}{da}da = 3\o_0H_0^2\frac{(1 +
z)^2}{\delta'^2}\int_z^\infty\frac{\delta\vert\delta'\vert}{1 + z}dz,
\label{eq:recon5}
\eeq
where $\delta' = d\delta/dz$. Setting $z = 0$ in this expression, we
arrive at a very interesting relationship between $\o_0$ and
$\delta(z)$:
\beq
\o_0 = \delta'^2(0)\bigg(3\int_0^\infty\frac{\delta\vert\delta'\vert}{1
+
z}dz\bigg)^{-1}.
\label{eq:recon6}
\eeq
Substituting this relationship in (\ref{eq:recon5}) we finally obtain
\beq
H(z)= H(0) \bigg\lbrack\frac{(1 + z)^2\delta'^2(0)}{\delta'^2(z)} -
3\o_0\frac{(1 + z)^2}{\d'^2(z)}\int_0^z
\frac{\delta\vert\delta'\vert}{1 + z}dz\bigg\rbrack^\half.
\label{eq:recon7}
\eeq
Clearly knowing $\d$ and $\d'$ we can determine
$H(z)$ and hence
$V(z)$.
For very large $z$, $z \gg 1$, observational difficulties
make it unlikely that $\d$ will be known to great accuracy, at least
in the near future. However in this regime the flat matter dominated
solution $\d \propto (1 + z)^{-1}$ provides a very good approximation
since $\o_m \rightarrow 1$ for $z \gg 1$.
%book as demonstrated in  section \ref{sec:omega}.

It should be pointed out that the above method of
reconstructing the $\l$-term potential from observations
is complementary to that used to reconstruct the inflaton potential
\cite{lidsey97}.
Whereas the
luminosity distance $d_L$ or the growth rate of the
linearized density contrast $\d(z)$ can be used to
reconstruct $V(\phi)$,
the inflaton potential is reconstructed on the basis of
the primordial amplitude and
spectrum of relic density perturbations and gravity waves
created during inflation
(also see \cite{nakachiba98}).

\subsection{Accelerated expansion and topological defects}

Phenomenological $\Lambda$ models
usually belong to the
general category of models in which matter
either violates or marginally satisfies the strong energy condition (SEC)
$\rho + 3P \geq 0$. Scalar fields driving inflation as well as the models
discussed earlier in this section furnish examples
of matter which can violate the SEC. Other
examples of such `strange' or `exotic' forms of
matter include cosmic strings and domain walls
\cite{spergel97,spergel98,spergel99}.
The field configuration
within a string is in the false vacuum state leading to $P = -\rho$
along the string length. A tangled network of random non-intercommuting
strings therefore possesses
the average equation of state $P = -\rho/3$ which marginally satisfies
the SEC \cite{vilenk85}. The
mean energy density of a string network dominated by straight strings
decays as $\rho \propto a^{-2}$
leading to the linear expansion law $a \propto t$
\cite{vilenk85,gott87}. Similarly
$P = -\rho$ is satisfied along any two orthogonal directions
within a domain wall
leading to $P = -2\rho/3$ for a network of walls \cite{vilenk85}
and resulting in `mild' inflation $\rho \propto a^{-1}$,
$a \propto t^2$. In the standard
defect scenario the system of topological defects
looses energy as it untangles, as a result the energy density decreases
faster than the `inflationary' law above. However topological obstructions
to untangling can arise in some field-theoretic models
as a result
of which the evolution of defects is `frustrated' and the decrease in energy
is $\rho \propto a^{-2}$ for strings and $\rho \propto a^{-1}$ for walls.
The presence of a `frustrated network' of
strings and/or domain walls can be tested by measurements
sensitive to the expansion dynamics of the universe.
For instance recent
supernovae results strongly suggest $w \lleq -2/3$ which severely
constraints the string network for which $w \simeq -1/3$. Thus it
appears that a frustrated network of strings is ruled out by current
observations (see section \ref{sec:sn}). Other aspects of defect models
including effects on the CMB have been discussed in \cite{spergel99}.

%\section{Anthropic arguments for a small cosmological constant.}
\section{Universality of ${\l}$ and anthropic arguments for its small value.} 

In this, the final section of our review, we must ask 
the following question:
%The following question naturally arises: 
do we expect the present value of $\l$ to be
fundamental (= defined by the parameters of a physical theory)
 or accidental (= determined by initial conditions in the
early Universe)? At present, we have no answer to this question. Models for 
$\l$ considered in previous sections admit both possibilities.
For instance, in the class of minimally-coupled lambda-field models
with an inverse power-law potential, the present value of $\l$ is fundamental
(= defined by the parameters of $V(\phi)$ only) if ``initially'' (at the end 
of inflation or at a later moment when the lambda-field becomes a separate
degree of freedom of matter) $\phi$ was sufficiently small, so that the
corresponding solution for $\phi(t)$ had time to reach a future attractor
(the ``tracker'' solution of~\cite{zws99,stein98}) by the present epoch.
On the other hand, if the initial value $\phi_{in}$ is large,
then the present value $\phi_0 \approx \phi_{in}$, and the current
value of the
$\l$-term
is accidental. Note that in the latter case $\l$ is practically time
independent now. Such a large value of $\phi_{in}$ may, for instance, be 
generated during an early inflationary stage, in which case stochastic 
methods~\cite{st82,st86,vil83} may be used to derive probability 
distributions for $\phi_{in}$ and $\l$. As a byproduct of such a mechanism,
small quasi-static inhomogeneous perturbations of $\l$ will also be generated.
\footnote{Previous discussions involving quantum cosmology also held the
possibility that the value of $\l$ is not determined uniquely. For
instance
Hawking (1984)
showed that the wave function for the universe could contain a
superposition of terms with different values for the cosmological
constant.
Investigating the effect of wormholes on quantum gravity, Coleman
(1988a,b)
subsequently showed that coupling constants whose values were not fixed
by
symmetries in the Lagrangian could take on all possible values in
the superposition
of terms describing
the state
vector in quantum cosmology.}

If $\l$ is accidental, then a wide range of ``explanations'' for its currently
(small) value can be given, based on 
the most reliable form of the anthropic principle - the 
weak anthropic principle. However, even if $\l$ is fundamental and can be
expressed through other microphysical constants, one may still try to use 
a more controversial form of this principle - the strong anthropic principle.
\footnote{The weak anthropic principle in the narrow sense states that
our location in space and time 
should be such that it admits the existence of intelligent life.
An extension of this principle is that initial conditions allow
the existence of such a region in space-time.
On the other hand the strong anthropic principle 
states that laws of nature should permit the existence of intelligent
life. It may be noted that the border between these two versions of the
anthropic principle is not absolutely rigid. Namely, by generalizing a
physical
theory (say, the electroweak model) with fixed constants into a more
general 
theory where these constants may have arbitrary values depending upon
initial 
conditions, we make a step from the strong to the weak anthropic principle
(also see \cite{barrow86}).}

An anthropic argument for $\l \neq 0$, originally suggested by Banks (1985) and
Weinberg (1987), bases itself on the observation that,  
in the absence of a
fundamental symmetry which set the value of $\l$ to precisely zero,
the extraordinary difference between likely values of the
vacuum energy $\rho_\l \sim \rho_m \sim 10^{-29}\hbox{g/cm}^3$ 
and the expected value (from a 
consideration of Planck scale physics) $\rho_P \sim 10^{93}\hbox{g/cm}^3$ 
could 
only be understood through anthropic arguments, since
it would be extremely 
fortuitous if particle physics determined a value for $\rho_\l$ 
which was comparable to
the matter density at {\em this precise} moment in
the history of the universe.
The case for the anthropic principle 
as a viable means for understanding properties of
the universe has received a strong measure of
support from recent developments in inflationary cosmology.
A self-consistent treatment of quantum effects in inflationary models
has shown that the entire universe may consist of an ensemble
of sub-universes (separated from each other by particle horizons) having
`all possible types of vacuum states and all possible types of 
compactification'
of extra space-time dimensions \cite{linde85}. According to this picture
our observable universe is but one of an infinite
number of universes each having its own set of conserved quantities and
dimensions. 
%book (see Sec \ref{sec:anthro}).
Since in each sub-universe physical fields determining the value of $\l$ 
have distinct values it is reasonable to expect that the value of
$\l$ varies from one sub-universe to another.

Weinberg (1987) showed that large values of $\l$ were unlikely to be
`observed' since the presence of observers demanded the existence of galaxies
and galaxy formation was strongly suppressed if the energy in the cosmological
constant greatly exceeded the matter density 
(also see
\cite{efst95,vilenk95}).
Martel, Shapiro \& Weinberg (1998) have suggested that the
probability that observers living in a given sub-universe will measure
a value $\rho_\l$ for the `vacuum energy' be given by the expression
\beq
{\cal P}(\rho_\l) = \frac{F(\rho_\l)}{\int_0^\infty F(\rho_\l) d\rho_\l}
\eeq
where $F(\rho_\l)$ is the fraction of matter in galaxies in a sub-universe
with vacuum energy $\rho_\l = \l/8\pi G$ \cite{wein98}. 
The value of $F(\rho_\l)$ is
calculated assuming Gaussian initial fluctuations at recombination,
with a COBE-normalized
cold dark matter spectrum with a cosmological
constant ($\l{\rm CDM}$).
The requirement that the observed value $\o_{\l,*}$ in our sub-universe equal
the statistical mean or median evaluated over 
all sub-universes (\ie  $\o_{\l,*} = \langle \o_\l\rangle $, where $\o_\l = 
\l/3H_0^2$) gives
a value which peaks in the region 
$\o_{\l,*} \sim 0.6 - 0.9$ 
for a broad region of parameter space and
assuming fairly reasonable conditions for galaxy formation \cite{wein98,wein00}.
Thus small observed values of $\o_\l$ appear to be
strongly disfavoured by the anthropic argument ! Other aspects of the anthropic
argument have been discussed in \cite{vilenk99}.

\section{The future of our universe in the presence of $\l$}
\label{sec:future}

The conventional viewpoint concerning the future of our universe is one of the 
following: (i) if the universe is spatially open or flat then it will expand
for ever, alternatively (ii) if the universe is spatially closed then its 
expansion will be followed by recollapse. Both these scenario's are based on
the premise that the universe contains `normal' matter, one that satisfies
the strong energy condition (SEC) $\rho + 3P \geq 0$. A cosmological $\l$-term
violates the SEC and it is easy to see that in its presence postulates
(i) \& (ii) become: 

[A] A spatially open or flat universe ($\kappa = 0, -1$)
will recollapse if the $\l$-term is 
constant and negative;

[B] A closed universe ($\kappa = +1$)
with a constant positive $\l$-term can, under certain
circumstances, expand forever. 

If the $\l$-term decays faster than $a^{-2}$ then, in the case of [B],
the universe will eventually recollapse if $\kappa = +1$, alternatively
if $\kappa = 0$ then the universe may still recollapse in some regions
due to the influence of local inhomogeneities in the distribution of matter. 
In the case of a $\l$-term based on a scalar field Lagrangian, the decay
of $\l$ is linked to the decrease of $V(\phi)$ with time. Existing observational
data seem to indicate that $\l$ is decreasing rather slowly, if at all. 
For instance if one assumes $P_\l = w\rho_\l$, with $w = $ constant,
 then $w < -0.6$ is indicated \cite{efst99,saini99}.
Since $\rho_\l \propto a^{-3(1+w)}$ this corresponds to $\rho_\l$
decreasing less rapidly than $a^{-1.2}$ at present. However, this behaviour
might change in the future. 

Consider next what happens if the $\l$-term is really a cosmological
{\em constant}.
For $\l > 0$ the expansion of the universe will rapidly approach the
de Sitter value $H = H_\infty = \sqrt{\l/3} = H_0\sqrt{1 - \o_m}$, even as
the density of matter asymptotically declines to zero $\rho_m \propto a^{-3}
\to 0$. Density perturbations will freeze to a constant value 
$\delta\rho/\rho \to constant$ if they are still in the linear regime
and angular anisotropies in the CMB ($\Delta T/T_l$), 
particularly the quadrupole, will
also freeze to some constant value. However the large scale acceleration of
the universe will not affect gravitationally bound systems on present scales
of $R < 10$h$^{-1}$ Mpc (which includes our own galaxy as well as galaxy
clusters). As a result the universe will soon consist of islands of matter 
immersed in an accelerating sea of vacuum energy: `$\l$'. 

An interesting issue in a $\l$ driven universe concerns how the redshift
of a given object changes with time. The present redshift of an object
$z \equiv z(t_0)$ located at a coordinate distance $r$ is given by
\beq
1 + z = \frac{a(\eta_0)}{a(\eta_{em})}, ~~\eta_{em} = \eta_0 - r,
\label{eq:time}
\eeq
$\eta_{em} = \eta(t_{em})$ is the moment when the object
emitted the light which we are observing now. The physical distance to such an 
object is $R = ar$. Let us differentiate (\ref{eq:time}) with respect to $t_0$,
to get ${\dot z}$ -- the time rate of change of the redshift of the object.
It is easy to see that if $\l = 0$, then  ${\dot z} < 0$ $\forall z$.
In fact $z(t)$ monotonically decreases with time tending to 0 as $t \to \infty$.
On the other hand, if $\l > 0$, $z(t)$ stops decreasing at some instant of time
and then begins to increase due to the acceleration of the universe.
As a result ${\dot z} > 0$ if $z < z_c$. 
For a flat universe the present value of
$z_c$ for which ${\dot z}_c(t_0)=0$ can be determined from the equation
\beq
(1 + z_c)\bigg (\o_m + \frac{1 - \o_m}{(1 + z_c)^3}\bigg ) = 1.
\eeq
Substituting $\o_m = 0.3$ we get $z_c = 2.09$. (Note that $z_c$ decreases
with increasing $\o_m$.) Thus we find that there is a reversal in the 
sign of ${\dot z}$ for sufficiently close objects, an effect that may even
be observable in the not-too-distant future \cite{loeb98,star99} !

Another interesting aspect of a $\l$-dominated universe is that, provided
the universe is accelerating over a large enough region, the local neighborhood
of an observer from which he/she is able to receive signals will eventually
contract and shrink. As a result even those regions of the universe which
are observable to us at present will eventually be hidden from view. 
Conversely, the coordinate volume of space which may be affected by our 
civilization in the infinite future is finite and its boundary is given by
$r_{H} = \eta(t = \infty) - \eta_0$. 
These features are related
to the appearance of a de Sitter-like future event horizon in a universe
which inflates in the future. This reasoning applies to galaxies
and other high redshift objects which we are currently observing: if the
redshift of an object is $z > z_{H}$ then such an object will remain
`out of bounds' to our civilization forever ! The redshift $z_{H}$ can
be determined from the following considerations.
Light propogation in a FRW universe is described by setting $ds^2 = 0$ in
(\ref{eq:lam1a}). Consider an event at ($r_1,t_1$) we wish to observe
at our location at $r = 0$, then
\beq
\int_0^{r_1}\frac{dr}{\sqrt{1-\kappa r^2}} = \int_{t_1}^{t}\frac{dt'}{a(t')}.
\label{eq:future1}
\eeq
If the integral in the RHS of (\ref{eq:future1}) diverges (as $t \to \infty$)
then an observer at $r=0$ will be able to receive signals from any event
provided he/she waits long enough. For $a \propto t^p$,
this implies $p < 0$, {\em i.e.} the universe is decelerating. In an 
accelerating universe the integral converges which means that one 
can only receive signals from those events which satisfy \cite{wein}
\beq
\int_0^{r_1}\frac{dr}{\sqrt{1-\kappa r^2}} \leq \int_{t_1}^{\infty}\frac{dt'}{a(t')}.
\label{eq:future2}
\eeq
(Taking the equality sign and assuming the expansion is de Sitter-like, so that
$a \propto \exp{Ht}, H = \sqrt{\l/3}$, we get $r_1 \equiv r_H$, 
$R = a_0r_H = H^{-1}$, which
is the proper distance to the event horizon in de Sitter space.)
Next consider an event ($r_H,t_1$) whose signals are being received by
us today ($t=t_0$) so that 
$a_0/a(t_1) - 1 = z_H$ is its redshift. 
Substituting $dz/dt = -(1+z)H(z)$ with $H(z)$ given by (\ref{eq:age4})
in (\ref{eq:future1}) we obtain (assuming $\kappa = 0$)
\beq
H_0a_0r_H = \int_0^{z_H} \frac{dz}{\lbrack \o_\l + \o_m (1+z)^3\rbrack^\half}.
\label{eq:future3}
\eeq
We now assume that $r_1=r_H$ is arbitrarily close to the horizon
so that light emitted today ($t_1=t_0$)
will reach us at $t \to \infty$, from
(\ref{eq:future2}) we obtain
\beq
H_0a_0r_H = \int_{-1}^{0} \frac{dz}{\lbrack \o_\l + \o_m (1+z)^3\rbrack^\half}.
\label{eq:future4}
\eeq
Equating (\ref{eq:future3}) \& (\ref{eq:future4}) we get
\beq
\int_1^{1+z_H} \frac{dx}{\sqrt{ 1 - \o_m + \o_mx^3}} = \int_0^1
\frac{dx}{\sqrt{ 1 - \o_m + \o_mx^3}}
\eeq
which can be solved for $z_H$.
If $\o_m = 0.3$, then $z_{H} = 1.80$, a rather small redshift 
since we see many galaxies and QSO's at much higher redshifts. 
(Note: $z_{H} \to \infty$ as $\o_m \to 1$).
For $\l$ = constant,
$z < z_H$ is the `sphere of influence of our civilization'
since celestial objects with $z > z_{H}$ will always remain
 inaccessible to signals emitted by us either now or in the future.
Thus comoving observers once visible to us will gradually disappear from view
as light emitted by them gets redshifted and declines in intensity.
(This is analogiues to what is observed for an object falling through the
horizon of a black hole.)
More discussion on these
and related issues may be found in \cite{loeb98,starkman99,star99}

\section{Summary and Discussion}
\label{sec:discussion}

In the absence of a symmetry in Nature which would set the value of the
cosmological constant to precisely zero, one is forced to either set $\l = 0$
by hand, or else look for mechanisms that can generate $\l = \l_{\rm obs} > 0$,
where $\l_{\rm obs} \sim 10^{-29}$g cm$^{-3}$ is the value of the $\l$-term
inferred from recent supernovae observations. 
We have discussed several mechanisms which could, in principle,
give rise either to
a time independent cosmological {\em constant}, or else a time dependent 
$\l$-term. To the former category primarily 
belong models which associate $\l$ with a
property of the vacuum such as the vacuum energy associated with symmetry
breaking, or vacuum polarization and particle production effects in curved
space-time. 
Mechanisms predicting a time dependent $\l$ take their
cue from inflation and generate a time varying $\l$ out of scalar fields
rolling down a potential. 
Models with a fixed $\l$ run into 
fine-tuning problems since the ratio of the energy
density in $\l$ to that of matter/radiation must be tuned to better
than one part in $10^{60}$ during the early universe 
in order that $\l/8\pi G \simeq \rho_{\rm matter}$ today. Scalar field models
considerably alleviate this problem though some fine-tuning does remain 
in determining the `correct choice' of parameters in the scalar field 
potential. 

It has been known for several years that the flat FRW $\l$CDM
cosmological model with an approximately flat spectrum of initial
adiabatic perturbations fits observational data better and has
a larger admissible region of parameters ($H_0,\Omega_m$)
than any other cosmological model with both inflationary and
non-inflationary initial conditions 
(see, e.g.,~\cite{ks85,kgb93,os95,star96,bagla96}). 
For instance according to a typical expert opinion made 
several years ago ``for $H_0>60$ km\,s$^{-1}$\,Mpc$^{-1}$,
this model is probably the {\em only} feasable model'' 
~\cite{star96}). Now, with new data on high redshift type Ia 
supernovae becoming available, we are closer than ever to concluding that
this is {\em the right} cosmological model (at least to a first
approximation) even if $H_0<60$.
Moreover, using type Ia supernovae data and with
improved data
on gravitational clustering at high redshifts soon expected, we may
progress 
further and investigate whether $\l$ depends weakly on time.

Turning to the observational situation,
constraints on the cosmic equation of state arise from
observations at: low redshifts (age of universe, cluster abundances, 
baryon fraction, velocity fields, etc.), intermediate redshifts
(ages of distant galaxies \& QSO's, angular size vs. redshift, gravitational 
lensing, Type Ia supernovae, 
the Lyman $\alpha$ forest etc.) and high redshifts (cosmic microwave 
background).
Each set of observations has its own systematic errors and although considerable 
progress has been made in trying to understand systematics it is safe to say
that at any given time at least one set of observations is likely to be 
well off the mark !

Of the low redshift tests, the age of the universe, cluster abundances and 
baryon fraction all appear to favour a low density universe, with $\o_m \lleq 
0.3$ 
in clustered matter. 
A tone of dissonance is however provided by
recent observations of the angular size of compact radio sources 
which seem to suggest a
critical density matter dominated universe,
although evolutionary effects clearly need to be better understood
before a strong case for $\o_m \simeq 1$ is made based on these results alone.

The strongest support for an accelerating universe comes from intermediate
redshift results for Type Ia supernovae. At the time of writing close to
a hundred supernovae
have been analyzed by two teams: The Supernova Cosmology Project and the High-Z
Supernova Search Team, both teams getting mutually consistent results
for $\lbrace\o_m,\o_\l\rbrace$.
It should be pointed out that the supernovae results do not by themselves
pick out a flat universe from other possibilities; a cursory look at
fig. (\ref{fig:sn}) shows that a closed universe with $\o_m + \o_\l > 1$
appears preferred although a flat universe is also accommodated
by current observations.
However the combined likelihood analysis of SnIa +
CMB observations strongly supports a flat universe with $\o_m + \o_\l \simeq 1$,
primarily due to the presence of a Doppler peak in the CMB data at intermediate
angular scales $\theta \sim 1^\circ$. Thus although observations do seem
to suggest that the universe may be spatially flat with a large fraction
of its density in the form of a cosmological $\l$-term, it may be 
premature to rule out, on the basis of current data alone, models that are
spatially open or even matter dominated and flat.

Great progress is however expected on the observational front in the coming 
5 - 10 years. Conservative estimates suggest that one should expect over
$\sim 50$ new
Type Ia events to be added to the supernovae inventory every year (including
several at significantly higher redshifts than $z \sim 1$).
Thus by the time of the launch of the MAP and PLANCK 
satellites (during 2001 \& 2007 respectively) one 
would expect our understanding of supernovae related parameter estimation
to have improved by over an order of magnitude.
In  particular more insight 
would have been gained on current sources of uncertainty including:
extinction caused by host galaxy and/or intergalactic
dust; composition of SnIa progenitors;
impact of rise times on peak supernova luminosity etc.
The possibility that extinction by `grey dust' consisting of
elongated dust grains could be responsible 
for the faintness of high redshift supernovae has recently been examined in 
\cite{aguirre99a,aguirre99b,simonsen99,riess00}.
(The flat extinction curve of elongated grains does not lead to reddening
which distinguishes `grey' dust from standard reddening dust characteristic
of small spherical grains.)
Evolutionary effects (if any) are of singular importance
since they have in the past severly restricted the potency of tests aimed
at determining the deceleration parameter and hence the equation of state.
In the case of
SnIa \cite{riess99} ``a luminosity evolution of $\sim 25\%$ over a
lookback time of $\sim 5$Gyr would be sufficient to nullify the cosmological
conclusions''  and therefore the $\l$ hypothesis.
One must therefore keep in mind the possibility that progenitor evolution
\cite{drell99,dominguez99,hoeflich99,riess99,suntzeff00,wang99a}
and the inhomogeneity of the SnIa sample \cite{li99}
might lead to systematic biases in the high z SnIa data.

Since both MAP and PLANCK missions
are expected to pinpoint the location and amplitude of the first
Doppler peak at the level of a few percent accuracy, they should provide
a decisive answer to the question of whether or not
we live in a critical density
universe.
The definitive answer to the question of 
whether the universe is flat and accelerating may therefore have to wait just
a few more years !

\vfill\eject

\begin{ack}
The authors acknowledge stimulating discussions with 
Somak Raychaudhuri, Tarun Souradeep Ghosh, Tarun Deep Saini and Ewan D. Stewart. They also thank
Neta Bahcall, Richard Ellis, John Peacock, Saul Perlmutter, Robert A. Schommer,
Max Tegmark and Limin Wang
for generously supplying some of the figures shown in this manuscript.
VS acknowledges support from the Indo-Russian Integrated Long Term Programme 
of cooperation in science and technology
(ILTP).
AS was partially supported by the Russian Foundation for Basic
Research, Grant 99-02-16224, and by the Russian Research Project
``Cosmomicrophysics''. This review was finished during the visit of
AS to the Institute of Theoretical Physics, ETH, Zurich.
\end{ack}

\end{document}